\documentclass[article]{aa}  
\usepackage{graphicx}
\usepackage{txfonts}

\usepackage{xfrac, amsmath, bm, multirow}
\usepackage[dvipsnames]{xcolor}

\newcommand{\figpath}{}

\usepackage{empheq}

\graphicspath{{figures/}}
\usepackage{hyperref}
\hypersetup{colorlinks, linkcolor={blue},citecolor={blue},urlcolor={magenta}}  

\usepackage[labelfont=bf]{caption, subcaption}
 
\usepackage[compact]{titlesec}
 
 \allowdisplaybreaks

\begin{document} 

\title{Improved source classification and performance analysis using Gaia DR3}

\author{
	Sara Jamal \inst{1}
  	\and
	Coryn A.\ L.\ Bailer-Jones \inst{1}
  }

\institute{
	Max-Planck Institute f\"ur Astronomy,
	K\"onigstuhl 17, 69117 Heidelberg, Germany\\
	\email{jamal@mpia.de}
	}

\date{Received xxx. Accepted xxx}

\abstract 
{
The Discrete Source Classifier (DSC) provides probabilistic classification of sources in Gaia Data Release 3 (GDR3) 
using a Bayesian framework and a global prior. 
The DSC \texttt{Combmod} classifier in GDR3 achieved for the extragalactic classes (quasars and galaxies) a high completeness of 92\%, 
but a low purity of 22\% (all sky, all magnitudes)
due to contamination from the far larger star class. 
However, these single metrics mask significant variation in performance with magnitude and sky position. Furthermore, a better combination of the individual classifiers that comprise \texttt{Combmod} is possible.
Here we compute two-dimensional (2D) representations of the completeness and the purity as a function of Galactic latitude and source brightness, and also exclude the Magellanic Clouds where stellar contamination significantly reduces the purity. 
Reevaluated on a cleaner validation set and without introducing changes to the published GDR3 DSC probabilities themselves, we here achieve for \texttt{Combmod}
	average 2D completenesses of 92\% and 95\% and average 2D purities of 55\% and 89\% for the quasar and galaxy classes, respectively.
Since the relative proportions of extragalactic objects to stars in Gaia is expected to vary significantly with brightness and latitude, we then introduce a new prior that is a continuous function of brightness and latitude, and compute new class probabilities
from the GDR3 DSC component classifiers, \texttt{Specmod} and \texttt{Allosmod}.
%
Contrary to expectations, this variable prior only improves the performance by a few percentage points, mostly at the faint end.
Significant improvement, however, is obtained by a new additive combination of \texttt{Specmod} and \texttt{Allosmod}. This classifier, \texttt{Combmod}-$\alpha$, achieves
average 2D completenesses of 82\% and 93\% and average 2D purities of 79\% and 93\% 
for the quasar and galaxy classes, respectively, when
using the global prior. 
Thus, we achieve a significant improvement in purity for a small loss of completeness.
The improvement is most significant for faint quasars 
(${G}$$\geq$20) where the purity rises from 20\% to 62\%. 

}


\maketitle
\section{Introduction}
The Gaia ESA mission \citep{prusti_gaia_2016, vallenari_gaia_2023} maps with {micro-arcsecond astrometry} more than $10^9$ sources in the sky into a 6D phase-space of positions and velocities, allowing us to 
	trace the Galactic acceleration field 
			\citep{malhan_ghostly_2018,ibata_charting_2021},
	infer the 3D distribution of interstellar matter 
			\citep{dharmawardena_three-dimensional_2022,dharmawardena_three-dimensional_2023},
	characterise the kinematics of stellar clusters and associations 
			\citep{soubiran_open_2018, kuhn_kinematics_2019},
	and improve the calibration of the cosmic distance scale ladder 
	\citep{riess_milky_2018,riess_cosmic_2021}, to mention just a few examples.
Through Gaia, the distribution and the kinematics of matter (dark+visible) 
provides important insights into the formation history and the structure of the Milky Way components, from the Galactic centre extending to the outermost parts of the 
discs and into the halo.
The distances derived from the Gaia parallaxes \citep{bailer-jones_estimating_2021}, complemented with photometry, allow us to populate the {Hertzsprung-Russell} diagram highlighting different {stellar populations} \citep{babusiaux_gaia_2023, creevey_gaia_2023}.
Primarily designed to map the stars in the Milky Way, Gaia observes a large scope of extragalactic sources such as quasars and galaxies \citep{bailer-jones_gaia_2023}.

{
Large samples of galaxies and quasars are used in statistical studies to constrain key cosmological parameters.
Quasars are extremely luminous sources powered by a supermassive black hole at the centre of their host galaxy. 
The most distant quasars at redshifts z$\gtrsim$6 are thought to have formed before the epoch of reionisation and are therefore useful probes of the early Universe and the epoch of reionisation \citep{fan_constraining_2006,mortlock_luminous_2011,banados_800-million-solar-mass_2018}.
Quasars are also used to investigate supermassive black hole formation and evolution history \citep{volonteri_quasars_2006}. 
Cosmological probes such as the baryon acoustic oscillations (BAOs) and the redshift distortions space (RSD) are key observables for constraining different cosmological models and for investigating the role of the dark energy in the late-time acceleration of the expansion of the Universe. 
The distributions of galaxies and quasars over large volumes across the Universe at different cosmic times carry an imprint on the growth rate of cosmic structures and emerge as an anisotropic clustering in the redshift space. The measurement of the RSD is used to constrain dark energy models through measurements of the linear growth rate of large-scale structures $f(z)\sigma_8(z)$ \citep{beutler_6df_rsd_2012,zarrouk_clustering_2018}.
Likewise, signatures of baryons, seen in the large-scale distribution of galaxies at low redshifts, are used to trace the expansion history through measurements of the Hubble distance $H(z)$ and the co-moving angular distances $D_{A}(z)$ \citep{beutler_6df_bao_2011}.
Together with direct measurements from galaxy samples, BAOs can also be measured from the spatial distribution of the neutral hydrogen in the Lyman-$\alpha$ forest in quasars spectra \citep{busca_baryon_2013, font-ribera_quasar-lyman_2014, delubac_baryon_2015, ata_clustering_2018}. 
}

{
The construction of catalogues of quasars and galaxies with a high level of purity is therefore essential to cosmological studies aiming at a better understanding of the evolution history of the Universe and probing the role of the dark components, namely the dark energy and dark matter, in the late acceleration of the expansion.
}
Classification of sources relies on information from spectra, photometry, astrometry, and image reconstructions to discriminate between different types. 
Quasar spectra are characterised by a featureless continuum with noticeable emission lines such as the Balmer lines and Ly$\alpha$ \citep{berk_composite_2001}, 
while galaxy spectra display a strong continuum component with absorption lines indicative of the stellar population as well as emission lines from the ionised gas in galaxies with an active star formation.  
{
Galaxy spectra are a composite of different contributions, from stellar populations, HII regions, active galactic nuclei, as well as contributions from the interstellar medium such as reddening} \citep{conroy_modeling_2013}.
%
%
In stellar spectra, the shape of the continuum is a direct proxy to the star's temperature and the characteristics of the absorption lines (e.g. ratio, strength and broadening) inform about the evolutionary phase of the star.
%
Colour information and astrometry are also used to identify a source as a quasar, a galaxy, or a star. 

Historical methods for the quasar-galaxy-star classification employ direct approaches such as colour cuts to optical or infrared photometry to identify aggregates of sources sharing similar properties \citep{newberg_three-dimensional_1997,fan_simulation_1999}.
For instance, quasars appear bluer compared to stars in the optical, and point-like sources such as single stars are fairly concentrated along a stellar locus.
However, despite their simplicity, such methods are hampered by the reduced dimensionality of the segmentation and the (inherent) subjectivity in selecting the separating planes through visual inspection, especially down to fainter magnitudes. 
{
With the rapid growth of astronomical data sets from large-scale surveys, traditional classification methods are unable to efficiently process the information space. Therefore, we use automated techniques such as machine learning (ML) for a repeatable, robust, and efficient way to classify the data and dive into finer details at levels beyond visual inspection or colour cuts.
%
%
ML has been applied to the quasar-galaxy-star classification 
using a diverse range of techniques such as 
	random forests \citep{weir_automated_1995, bai_machine_2018, nakazono_discovery_2021, zhang_classification_2021,rimoldini_gaia_2023},
	extremely randomised trees \citep{baqui_minijpas_2021, delchambre_gaia_2023},
	support-vectors machines \citep{peng_selecting_2012, malek_vimos_2013, nakazono_discovery_2021,wang_j-plus_2022}, 
	Bayesian kernel density estimators \citep{peters_quasar_2015, bailer-jones_quasar_2019}, 
	convolutional neural networks \citep{kim_stargalaxy_2017, burke_deblending_2019, he_deep_2021, stoppa_autosourceid-classifier_2023, merz_detection_2023, rodrigues_minijpas_2023, chaini_photometric_2023}, 
	and dense neural networks \citep{martinez-solaeche_minijpas_2023}. 
Unsupervised approaches have also been proposed in the literature such as the Hierarchical Density-Based Spatial Clustering of Applications with Noise (HDBSCAN) algorithm \citep{logan_unsupervised_2020} and contrastive learning networks \citep{guo_unsupervised_2022}. 
More recently, several contributions have exploited boosted trees classifiers such as extreme gradient boosting (XGBoost) for classification\citep{golob_classifying_2021, li_identification_2021, hughes_quasar_2022, stoppa_autosourceid-classifier_2023, rimoldini_gaia_2023} 
that proved to be highly performing in separating point-like sources (quasars and stars) from extended sources (galaxies).
}

{By scanning the entire sky down to faint magnitudes, Gaia observes millions of extragalactic sources that require classification methods to identify quasar and galaxy candidates} \citep{bailer-jones_gaia_2023}.
Within the Gaia Data Processing and Analysis Consortium (DPAC), several modules are responsible for the classification and the characterisation of the Gaia sources. 
Among the classification modules, the Discrete Source Classifier (DSC) provides probabilistic classifications into five classes, namely the quasar, galaxy, star, white dwarf, and binary star classes \citep{creevey_gaia_2023}.
DSC exploits a decision tree-based classifier and a density-based model in a supervised approach trained on Gaia astrometry and spectro-photometric data \citep{delchambre_gaia_2023}. 
%
In the latest Gaia Data Release 3 (GDR3; \citeauthor{vallenari_gaia_2023}, \citeyear{vallenari_gaia_2023}), DSC classification reports for the extragalactic classes a high completeness {($\sim$92\%)} owing to the classifiers' ability in identifying the true classes, but a low purity {($\sim$22\%)} due to large numbers of stellar detections erroneously classified as a quasar or a galaxy candidate. 
However, single metrics to summarise the performance for all types of sources do not constitute a fair assessment. 
For instance, low performance is unsurprisingly correlated to fainter magnitudes and crowded regions. 
{
Moreover, a reduced purity implies a limited use of the catalogue for cosmological studies.
}
In this work, {our goal is two-fold, 
	first} to quantify DSC performance as a function of brightness and Galactic latitude in order to identify the range of the best results and limitations, 
	and second to improve the purity of the extragalactic classes.
To assess the performance, we define a two-dimensional (2D) representation of the purities and the completenesses at different magnitudes across the sky. 
To improve the purity, we apply to the {GDR3} DSC probabilities a continuous variable prior based on the expected distribution of brightness and Galactic latitude and provide a new approach for combining the DSC classifiers. 
In this work, we concentrate our efforts in assessing the performance of the extragalactic classes and briefly discuss the white dwarf class.

The paper is structured as follows. 
Section \ref{sec:gaiadata} succinctly presents the data.
Section \ref{sec:DSCresultsGDR3} summarises the classification results from DSC in GDR3 and introduces the 2D assessment of the performance. 
In Section \ref{sec:DSCcombiner}, we present the continuous variable prior function of brightness and Galactic latitude 
and outline a new approach for combining DSC classifiers to improve the purity of the extragalactic classes.
We finally conclude in Section \ref{sec:conclusions}, and provide an Appendix with additional information.

\section{{Data}} \label{sec:gaiadata}

The Gaia data collected by the satellite between the period of 2014 and 2017 was published in the latest catalogue on June 2022, the Gaia Data Release 3 (GDR3; \citeauthor{vallenari_gaia_2023}, \citeyear{vallenari_gaia_2023}), for $\sim$1.9 billion sources down to faint magnitudes ${G}$$\simeq$20, providing $\sim$1.6 billion sources with classifications and 470 million sources with astrophysical parameters such as effective temperatures, surface gravity, metallicity, extinction, and abundances.
Gaia provides measurements of astrometry, photometry, and spectroscopy for billions of sources. 
The low-resolution spectro-photometry, namely the XP spectra, cover a wavelength range in the blue BP band at 330-680 nm and in the red RP band at 640-1050 nm ({$\sim$220 million mean spectra in GDR3}; \citeauthor{angeli_gaia_2023} \citeyear{angeli_gaia_2023}).
%
In GDR3, DSC computes probabilistic classification into the quasar, galaxy, star, white dwarf, and binary star classes using the parallaxes, the proper motions, the Galactic latitude, the colours, the magnitudes in the G-band, and the low-resolution XP spectra \citep{bailer-jones_quasar_2019,delchambre_gaia_2023}. 
The main source table (\texttt{gaiadr3.gaia\_source}) reports the posterior probabilities of the quasar, galaxy, and star classes from the DSC {combined} classifier, 
	while the astrophysical parameters table (\texttt{gaiadr3.astrophysical\_parameters}) reports the posterior probabilities of all classes from the DSC classifiers. 
	The GDR3 extragalactic tables (\texttt{gaiadr3.qso\_candidates}, \texttt{gaiadr3.galaxy\_candidates}) provide a list of quasar and galaxy candidates identified by different 
		classification modules within DPAC including DSC candidates. 
		A purer sub-sample of the GDR3 extragalactic tables, presented in \cite{bailer-jones_gaia_2023}, is obtained by imposing a joint condition on the probabilities of the DSC {classifiers}. 
In this work, we use the DSC posterior probabilities from the astrophysical parameters table in GDR3.

\section{Reassessment of {classification} results in GDR3} \label{sec:DSCresultsGDR3}

\subsection{An outline of DSC in GDR3}\label{subsec:DSC}

\begin{figure*}[htp!]
	\centering	
	\includegraphics[scale=0.22]{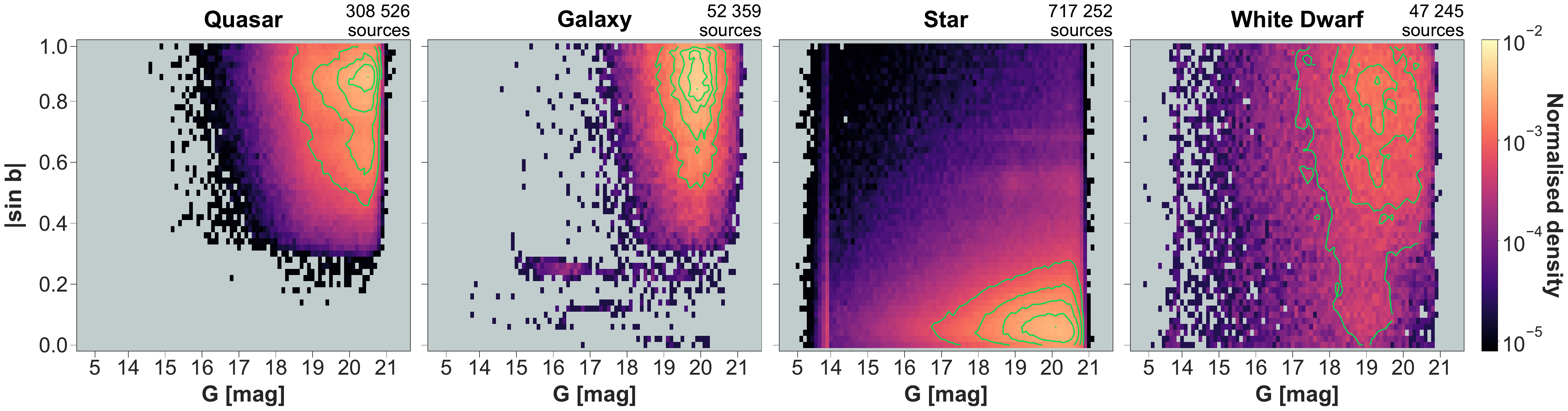}

	%
	\caption{
		Representation in 2D of the distribution of sources as a function of Galactic latitude and brightness of the original validation data set. 
		{Each distribution is normalised by the total number of sources in that panel.}
		The colour scale is set such that bright colours indicate a higher density of sources compared to dark regions. 
		The 2D representation of the density of sources is defined on a grid in $|\sin b|$ and ${G}$ of 50$\times$91 bins. 
		As a design choice, the binning in magnitude bins compresses the range at ${G}$<14.
		{Contours (green lines) are used in the regions of highest density to indicate the normalised density in log scale.} 
		}
	\label{fig:2d_alldata_combmod_counts}
\end{figure*} 

\begin{table*}[htp!]
   \caption{Summary of the overall classification performance evaluated on the validation data set. 
   }               
    \label{table:tab_GDR3_and_newcombmod_summary}  
    \small \centering    
    \setlength{\tabcolsep}{7pt}                        
    \begin{tabular}{l  | l| c c c | c c c }         
        \hline\hline                  
        		\multirow{2}{5em}{\centering Prior}
			&\multirow{2}{4em}{\centering Class}
        			& \multicolumn{3}{c|}{Overall completeness} 
    			& \multicolumn{3}{c}{Overall purity}\\
    			\cline{3-8}
    			&	& \it original & \it {{cleaned}} &  \it \texttt{Combmod}-${\alpha}$
    				& \it original & \it {{cleaned}} &  \it \texttt{Combmod}-${\alpha}$ \\
    	\hline\hline  
	\multirow{5}{5em}{\centering global prior}              
           &quasar	 	&	\bf 0.916	&	\bf 0.916	&	\bf 0.819	&	\bf 0.240	&	\bf 0.489	&	\bf 0.749	\\	 
           &galaxy		&	\bf 0.935	&	\bf 0.953	&	\bf 0.931	&	\bf 0.219	&	\bf 0.353	&	\bf 0.590	\\	 
           &star		&	0.996	&	0.998	&	0.999	&	0.990	&	0.990	&	0.990	\\	 	
           &white dwarf 	&	0.432	&	0.430	&	0.395	&	0.250	&	0.406	&	0.429	\\	 	
           &binary		&	0.002	&	0.002	&	0.0014	&	0.075	&	0.062	&	0.079	\\	 
        \hline   
    \end{tabular}
    \tablefoot{
    Predicted labels are obtained from the class with the maximum posterior probability. 
   The labels `original' and `{{cleaned}}' refer to the results evaluated on the full validation data set and the {{cleaned}} validation data set, respectively.
   \texttt{Combmod}-${\alpha}$ is the parametric combination of the GDR3 DSC \texttt{Allosmod} and \texttt{Specmod}, where $\alpha_{\rm glob}$=[1.0, 0.6, 0.5].
    Results of the extragalactic classes are highlighted in boldface.
    }
\end{table*}

%
DSC architecture in GDR3 is composed of three classifiers computing posterior probabilities: two baseline classifiers trained using a set of labelled sources, \texttt{Specmod} and \texttt{Allosmod}, complemented with a third classifier, \texttt{Combmod}. 
\texttt{Specmod} is trained using low-resolution XP spectra and computes posterior probabilities across the five classes, while \texttt{Allosmod} is trained using discretised features from the Gaia astrometry and photometry to discriminate between the quasar, galaxy, and anonymous classes, with the latter referring to the stellar types of the single star, white dwarf and binary star classes combined. 
\texttt{Combmod} combines the probabilities of the baseline classifiers and provides posterior probabilities over the five classes.
The baseline classifiers are trained on the same sources, thus requiring the availability of all input data for the joint training. However, at the prediction stage, the architecture with separate classifiers offers more flexibility in processing sources with missing information. For instance, sources without parallax measurements but with a valid XP spectrum are classified through \texttt{Specmod} and not processed through \texttt{Allosmod}.

\texttt{Specmod} is a decision tree-based classifier Extratrees \citep{geurts_extremely_2006} and \texttt{Allosmod} is a Gaussian Mixture Model \citep{reynolds_gaussian_2015}.
The Extratrees algorithm is an ensemble-based algorithm combining class predictions from a collection of randomised decision trees trained on bootstrap samples of the data. 
Compared to the Random Forest algorithm, Extratrees is similar in the randomisation of the input features at the nodes level, but differs by adding a randomisation of the thresholds applied to the selected features.
The Gaussian Mixture Model is a probabilistic model assuming that the data are a mixture of Gaussian distributions. Training the algorithm aims to find the best model parameters, the mean and variance of the gaussian components.
DSC classifies sources using a Bayesian framework and a global prior to account for the rarity of the extragalactic classes. The global prior $\bm{\pi}^{\rm glob}$ is defined such that for one source to be identified as a star, the probabilities to observe a quasar and a galaxy are, respectively, one over 1000 and one over 5000. 

\subsection{Performance metrics}

To assess the performance of a classifier in predicting the correct labels, we compute the completeness and the purity from the confusion matrix. 
By definition, the completeness defines the ability of the classifier to retrieve the true labels while the purity depicts its ability to filter out false detections. The two metrics are defined as  
\begin{equation}
	\rm completeness = \frac{TP}{TP+FN},  
	\qquad
	purity = \frac{TP}{TP+FP},  
	\label{eq:metrics}
\end{equation}
where TP, FN, and FP refer to true positives, false negatives and false positives, respectively.
A high completeness and a high purity are indicative of a satisfactory performance. However, a trade-off between completeness versus purity is a known compromise.
A low completeness and a high purity suggest that the classifier is able to reject false positives (i.e. sources from other classes erroneously classified as the target class) at the cost of missing true predictions, while a high completeness coupled to a low purity emphasises the ability of the classifier to correctly predict the true labels to the expense of introducing large numbers of false detections.
DSC assigns class labels by identifying the class with the largest probability above a fixed threshold.
A higher threshold entails a more stringent condition to assign a source to a target class, otherwise the source is marked as unclassified. 
A high threshold typically produces a purer sample from rejecting more false detections, to the detriment of a reduced completeness from rejecting the true positives with probabilities below the threshold.

Classification results, when evaluated on a limited number of test objects, are not indicative of the performance on the overall distribution across the sky as the number of stars overly exceeds the number of extragalactic sources. 
Therefore, if stars are underrepresented in the test set compared to reality, there will not be a correct estimate of the fraction of potential stellar contaminants and the purity will be underestimated.
To illustrate this point, we consider a simple example of a 2-class classifier able to correctly predict class-1 but completely misclassifies the class-2. 
	For a balanced test set, the completeness of class-1 is 100\% and the purity is 50\%. 
	For an imbalanced test set where class-2 is at least 100 times larger, the completeness of the class-1 remains 100\% but its purity is $\sim$1\% (=1/101).
For DSC, the problem of the class imbalance was addressed in \cite{bailer-jones_quasar_2019} in two folds, firstly by applying the class prior to compute posterior probabilities, and secondly by applying an adjustment to the confusion matrix to scale the test set to the overall class distribution in the sky. The adjustment, applied per target class to the confusion matrix, affects only the purities and not the completenesses.
Throughout this work, the classification metrics are the metrics after adjustment.

\subsection{Performance of DSC {as} reported in GDR3}

In \cite{delchambre_gaia_2023}, the classification performance of DSC in GDR3 is evaluated on a validation data set. 
	\texttt{Specmod} reports a completeness of 41\% and 83\% and a purity of 23\% and 40\%, 
	and \texttt{Allosmod} reports a higher completeness of 84\% and 92\% and a purity of 41\% and 30\%, 
	for the quasars and the galaxies, respectively.
	\texttt{Combmod} combines the probabilities of the baseline classifiers and reports a high completeness of 92\% and 94\% 
	and a low purity of 24\% and 22\% for the quasars and the galaxies respectively, 
	showing the ability of the classifier to achieve a higher prediction rate (higher completeness) 
	to the disadvantage of introducing the false detections (lower purity).
	Low purity values result from increased numbers of the stellar false positives, a contribution largely inflated due to the strong prior of the star class.
At higher latitudes, farther from the Galactic plane where most single star contaminants lie, the purity of the extragalactic classes increases by $\sim$20\% for all classifiers, which pinpoints anew to the contamination from the star class as the main factor affecting the purity.
In GDR3, DSC \texttt{Combmod} reports a completeness of 99\%, 43\%, and 0.2\%, and a purity of 99\%, 25\%, and 7.5\% for the stars, white dwarfs, and physical binaries, respectively.
For the extragalactic classes, a conditional conjunction of \texttt{Specmod} and \texttt{Allosmod} probabilities is introduced through the label \texttt{classlabel\_dsc\_joint} such that the two classifiers jointly identify the same class with a maximum probability larger than 0.5. This combination reports the highest purities ($\sim$62\%) for the extragalactic classes but the lowest completeness for the quasar class (38\%). 

In this work, we use the same validation data set of DSC in GDR3, which comprises
	308\,526 quasars, 
	52\,359 galaxies, 
	717\,252 single stars, 
	47\,245 white dwarfs 
	and 331\,526 physical binary stars.
This validation set was built from SDSS spectroscopic candidates for the extragalactic classes, from the Montreal White Dwarf database\footnote{\url{https://www.montrealwhitedwarfdatabase.org/}} for the white dwarfs, and from a selection of Gaia data for the single star and physical binary star classes. 
In this work we omit the analysis of the binary class, due to the poor performance of DSC in GDR3 on this class.  
Figure \ref{fig:2d_alldata_combmod_counts} shows the distribution of the number of sources in the validation data set per target class as a function of the Galactic latitude and magnitude. The relative density of quasars and galaxies peaks at higher latitudes at the faint end while the density of stars is higher close to the Galactic plane.
The density of the white dwarfs across the 2D space peaks at higher latitudes at the faint end, but such agglomeration could be due to a selection effect from the constructed sample.

Using the validation data set, we first reproduce the GDR3 results (the overall completeness and purity for all three classifiers and the posterior probabilities from \texttt{Combmod} computed from \texttt{Allosmod} and \texttt{Specmod}).
%
The overall performance evaluated on the full validation data set is summarised in Table \ref{table:tab_GDR3_and_newcombmod_summary} under the label `original'.
{The confusion matrix is provided in the Appendix Figures \ref{fig:confmat_alldataset_globalprior} for reference.}

\begin{table*}[htp!]
\caption{Summary of the average 2D classification performance evaluated on the validation data set. 
   }  
    \label{table:tab_GDR3_and_newcombmod_2dsummary}   
    \small \centering    
    \setlength{\tabcolsep}{7pt}                        
    \begin{tabular}{l | l | l| c c c | c  c c}         
        \hline\hline                  
        		\multirow{2}{7em}{\centering Magnitude range}      
        		&\multirow{2}{5em}{\centering Prior}
		& \multirow{2}{4em}{\centering Class}
        			& \multicolumn{3}{c|}{Weighted average 2D completeness} 
    			& \multicolumn{3}{c}{Weighted average 2D purity}\\
    			\cline{4-9}
    			&&	& \it original & \it {{cleaned}} &  \it \texttt{Combmod}-${\alpha}$
    				& \it original & \it {{cleaned}} &  \it \texttt{Combmod}-${\alpha}$ \\
    	\hline\hline                
	\multirow{10}{6em}{\centering All magnitudes}
	&\multirow{5}{5em}{\centering global prior}
	&	quasar		&	0.916	&	\bf 0.916	&	\bf 0.819	&	0.344	&	\bf 0.553	&	\bf 0.786	\\
	&&	galaxy		&	0.935	&	\bf 0.953	&	\bf 0.931	&	0.641	&	\bf 0.884	&	\bf 0.926	\\
	&&	star			&	0.996	&	0.998	&	0.999	&	0.994	&	0.994	&	0.994	\\
	&&	white dwarf	&	0.432	&	0.430	&	0.395	&	0.901	&	0.950	&	0.954	\\
	&&	binary 		&	0.002	&	0.002	&	0.001	&	0.325	&	0.278	&	0.718	\\
	\cline{2-9}
	&\multirow{5}{5em}{\centering 2D variable prior}
	&	quasar		&	-	&	\bf 0.953	&	\bf 0.885	&	-	&	\bf 0.570	&	\bf 0.737	\\
	&&	galaxy		&	-	&	\bf 0.957	&	\bf 0.939	&	-	&	\bf 0.864	&	\bf 0.896	\\
	&&	star			&	-	&	0.998	&	0.999	&	-	&	0.994	&	0.994	\\
	&&	white dwarf	&	-	&	0.422	&	0.401	&	-	&	0.947	&	0.953	\\
	&&	binary		&	-	&	0.002	&	0.001	&	-	&	0.281	&	0.717	\\
	 \hline   
    \end{tabular}
    \tablefoot{
   Predicted labels are obtained from the class with the maximum posterior probability. 
   The labels `original' and `cleaned' refer to the results evaluated on the full validation data set and the {{cleaned}} validation data set, respectively.
   \texttt{Combmod}-${\alpha}$ is the parametric combination of the GDR3 DSC \texttt{Allosmod} and \texttt{Specmod}, where $\alpha_{\rm glob}$=[1.0, 0.6, 0.5] and $\alpha_{\rm vari}$=[0.6, 0.6, 0.5].
   Results of the extragalactic classes are highlighted in boldface.
   } 
\end{table*}

\subsection{{An {{cleaned}} validation data set}} \label{subsec:cleanedvalset}

The reported performance assumes that the labels in the validation data set are correct. However, through cross-matches using public tools\footnote{\url{https://skyserver.sdss.org/dr17/}}$^{,}$\footnote{\url{https://www.star.bristol.ac.uk/mbt/topcat/}} with external non-Gaia catalogues {at a maximum radius of 1 arcsecond}, we identify, in the stellar classes of the validation data set, quasar and galaxy candidates from Milliquas v7.2-v8 \citep{flesch_half_2015, flesch_million_2023}, LAMOST DR7 \citep{luo_vizier_2022}, SDSS DR 17 photometry \citep{abdurrouf_seventeenth_2022}, 
UKIDSS DR9 \citep{lawrence_ukirt_2007}, DES DR2 \citep{abbott_dark_2021}, Assef R90 and C75 WISE AGN catalogues \citep{assef_wise_2018}, and SIMBAD \citep{wenger_simbad_2000}.
{We further filter out the sources with unreliable classifications in SIMBAD}.
%
%
We remove these sources from the stellar classes to build a cleaner validation data set and assess the performance of the classifiers without a bias.  
The filtering amounts {in total} to 1.6\% removal of sources (22\,912 sources) applied to the star, white dwarf, and binary classes. 
{The Appendix Table \ref{table:ref_xmatch} summarises the selection criteria applied to the results in order to filter contaminants from the stellar classes in the validation data set.}
We also remove the sources located in the Magellanic Clouds\footnote{excluding sources with coordinates (\texttt{ra},\texttt{dec}) within a radius of 9° and 6° around the LMC and the SMC, respectively.} from the stellar classes, 
about 0.8\% of sources (11\,837 sources), 
and filter the galaxy class from sources {that deviate from the expected distribution of galaxies in the colour-colour diagram and} fall in the stellar locus as defined in \citeauthor{bailer-jones_quasar_2019} \citeyear {bailer-jones_quasar_2019} ($\sim$10$^{3}$ sources). 
After accounting for duplicates, the total removal from the original validation data set amounts to 2.5\% (35\,753 sources).
%
The {{cleaned}} validation data set contains 
	308\,526 quasars, 
	51\,347 galaxies,         
	702\,450 stars,            
	45\,005 white dwarfs  
	and 313\,827 binaries. 
{The Appendix Table \ref{table:ref_counts} summarises the total number of sources in the validation data set.}

\subsection{{Two-dimensional representation of results}} \label{section25}
Classification performance typically summarises results averaged over a diverse validation data set. 
However, the performance of a classifier is expected to vary as a function of source parameters, such as magnitude and Galactic latitude that are strongly linked to the signal-to-noise (S/N) level and possible crowding.
For instance, in crowded regions, classification is limited either by a reduced data quality or by a higher density of stellar contaminants. 

Similar to \cite{hughes_quasar_2022}, we explore a 2D representation of the classification performance. 
We compute a 2D grid of a total of 4550 bins (50$\times$91) in $|\sin b|$ and ${G}$. The magnitude binning is composed of a coarse grid for ${G}$<14 and a finer grid for 14$\leq$${G}$<22, while the binning in $|\sin b|$ is {defined} on a uniform grid {between 0 and 1}.
Using the prior and the number of sources per bin, we apply the adjustment to the confusion matrices in each bin and compute the completeness and the purity.
We intuitively expect a better performance for brighter sources, especially in low-density regions farther from the Galactic plane.

\begin{figure*}[htp!]
    \centering
    \begin{subfigure}{1\textwidth}
        \centering
        \includegraphics[scale=0.21, trim={0 0 0 0}, clip]{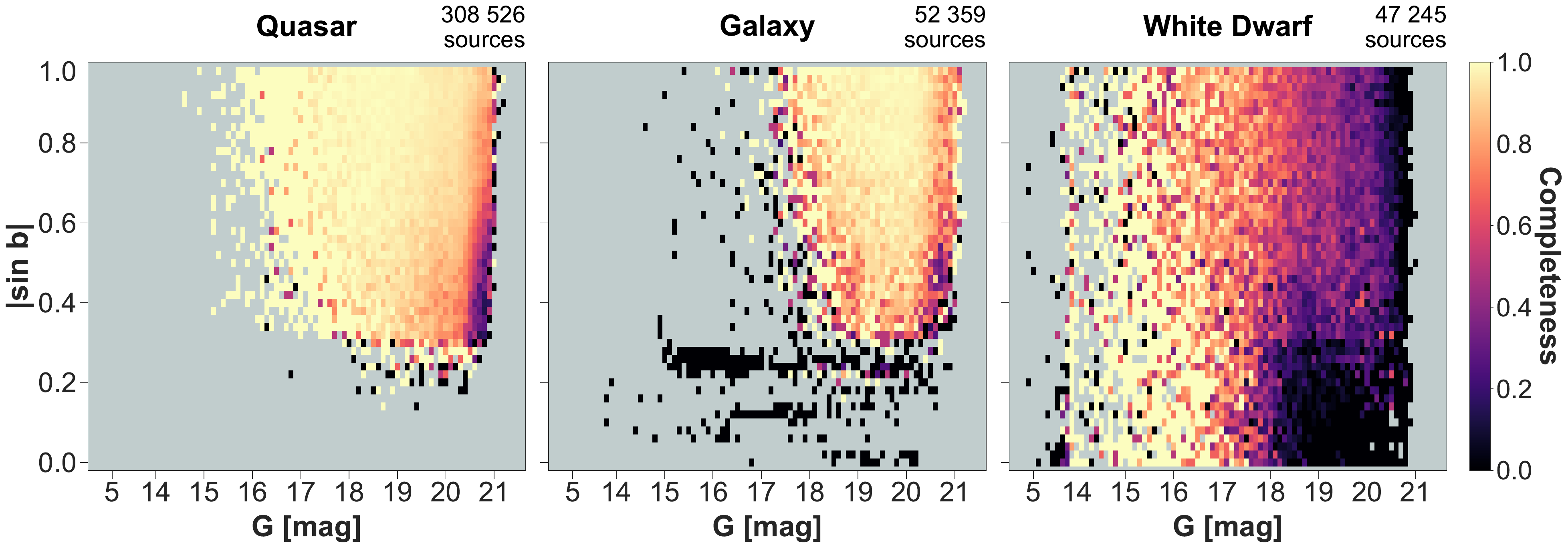}
         \caption{}
        \label{fig:2d_alldata_combmod_globalprior_completeness}
    \end{subfigure}
    
      \begin{subfigure}{1\textwidth}
        \centering
        \includegraphics[scale=0.21, trim={0 0 0 0}, clip]{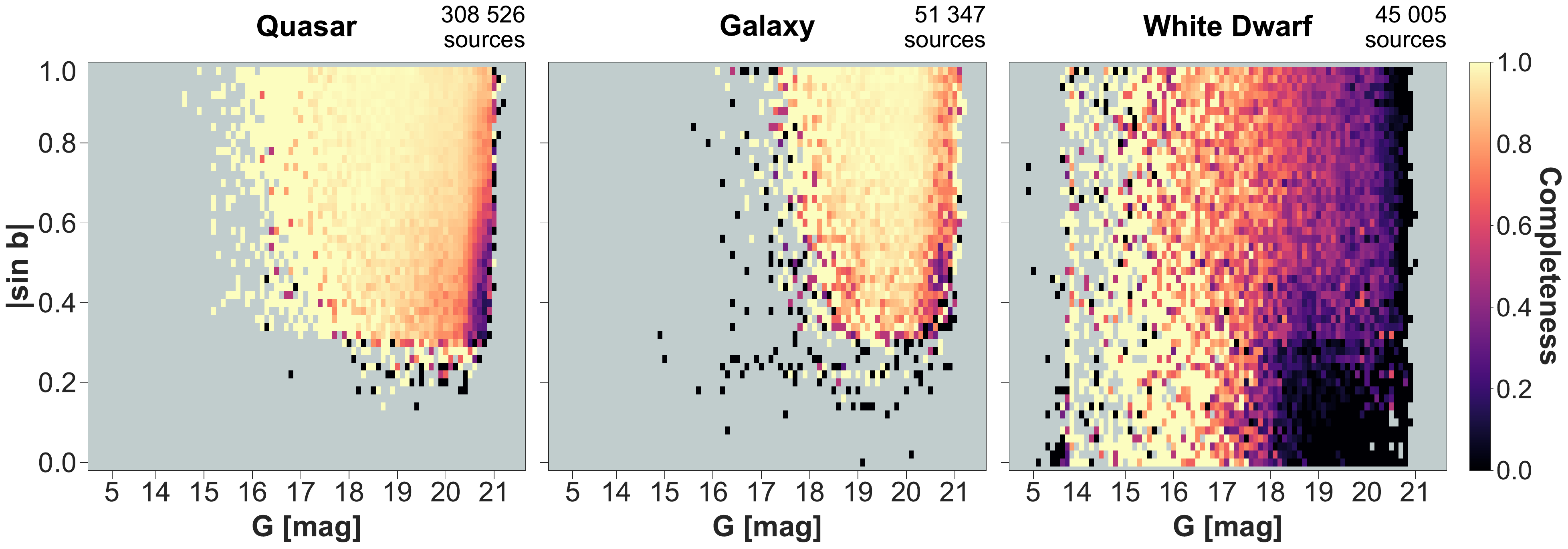}
         \caption{}
        \label{fig:2d_alldata_combmod_globalprior_noMCcontaminantsExt_completeness}
    \end{subfigure}
        
        \begin{subfigure}{1\textwidth}
           \raggedleft
            \includegraphics[scale=0.28, trim={0 0 0 0}, clip]{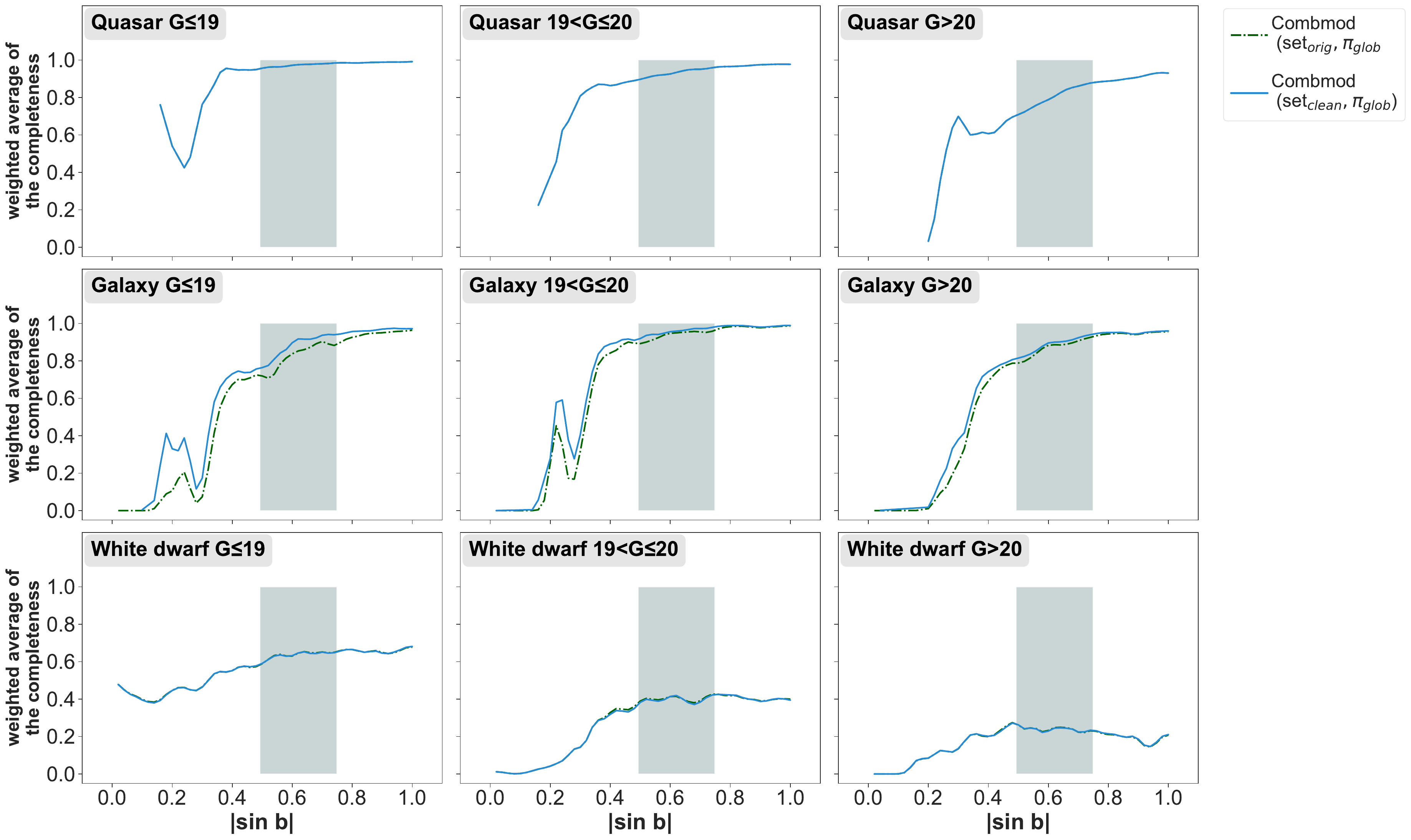}
            \caption{}
	\label{fig:1dsmooth_0a_global}
    \end{subfigure}
    
    \vspace{-0.20cm} 
    \caption{
        Completeness of DSC \texttt{Combmod}. 
        Predicted labels are obtained from the class with the largest posterior probability using the global prior.
	\textit{(a)} Variation of completeness with magnitude and absolute Galactic latitude for each of the three classes using the original validation data set. 
	\textit{(b)} As (a), but using the cleaned validation set (i.e. after removal of the Magellanic Clouds and mislabelled sources).
	\textit{(c)} Completeness as a function of absolute Galactic latitude for three brightness ranges (columns) for each class (rows). 
	In each panel the dashed line shows the performance on the original validation data set and the solid line the cleaned validation data set. 
	These one-dimensional plots are essentially a marginalisation of the 2D representations in panels (a) and (b) 
	but weighting by the number of sources in each bin. 
	The curves have been smoothed with a Gaussian filter to remove noise. 
	The shaded region encompasses the LMC at $|\sin b|$=--0.54 and the SMC $|\sin b|$=--0.70 (which are excluded from the cleaned set).
	}
        \label{fig:1dsmooth_0_global}
\end{figure*}

\begin{figure*}[htp!]
	\centering
	\begin{subfigure}{1\textwidth}
		\centering
		\includegraphics[scale=0.21, trim={0 0 0 0}, clip]{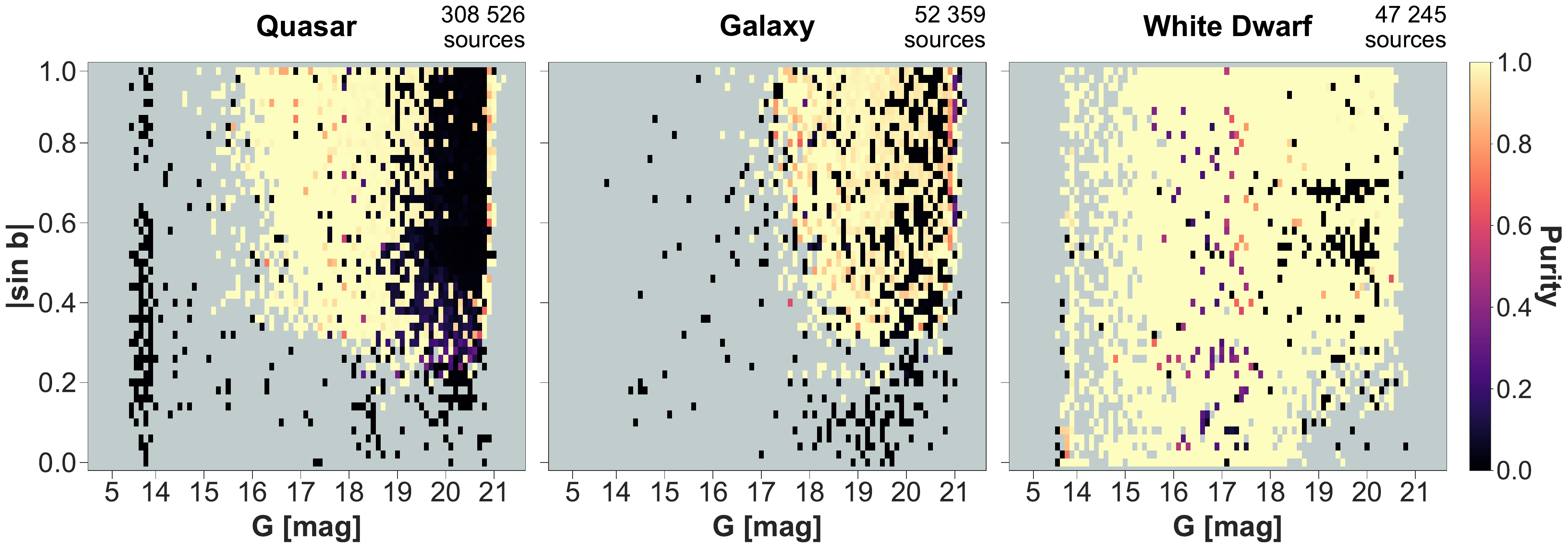}
		 \caption{}
		\label{fig:2d_alldata_combmod_globalprior_purity}
	\end{subfigure}
	
     \begin{subfigure}{1\textwidth}
		\centering
		\includegraphics[scale=0.21, trim={0 0 0 0}, clip]{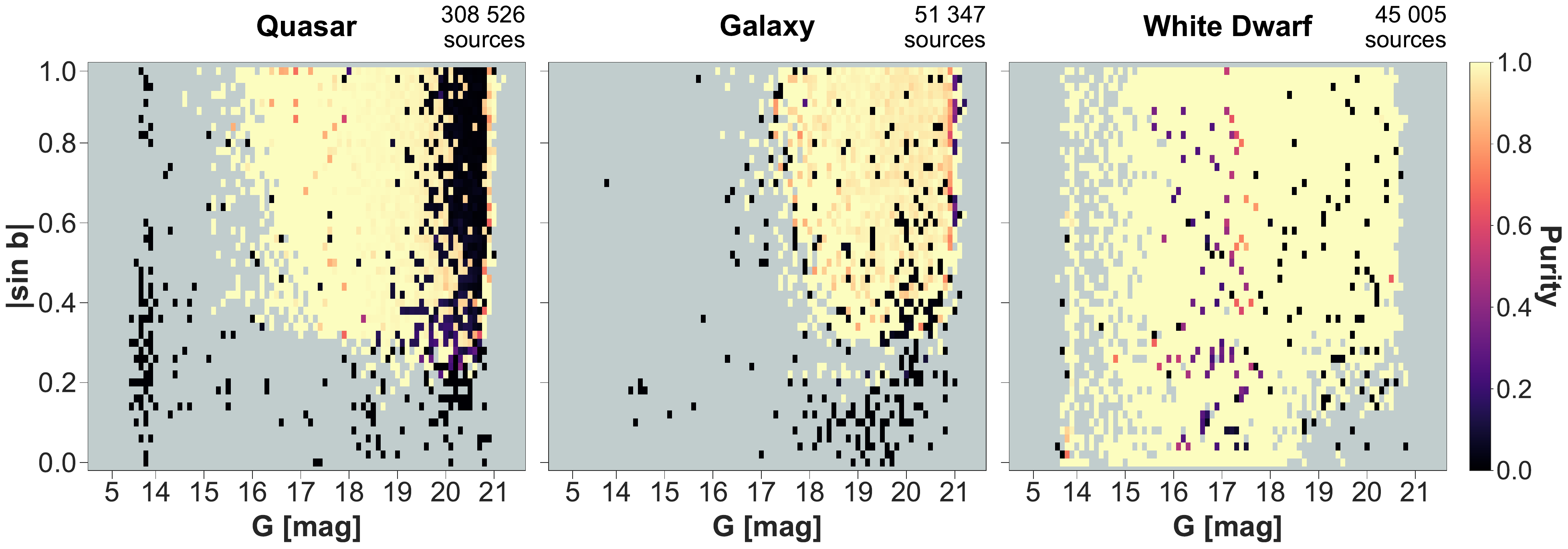}
		 \caption{}
		\label{fig:2d_alldata_combmod_globalprior_noMCcontaminantsExt_purity}
	\end{subfigure}
		
    	\begin{subfigure}{1\textwidth}
		\raggedleft 
    		\includegraphics[scale=0.28, trim={0 0 0 0}, clip]{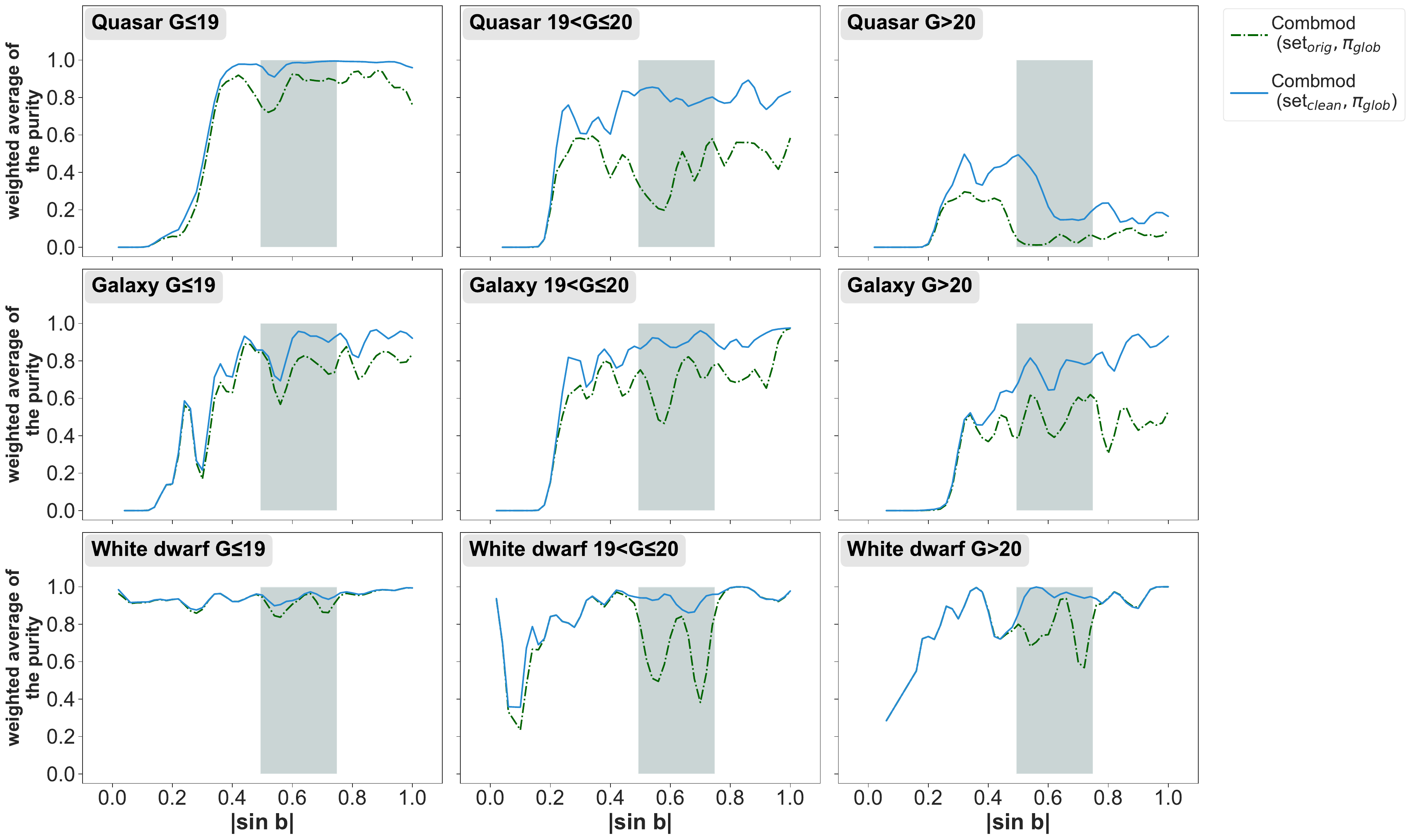}
	 \caption{}
    \label{fig:1dsmooth_0b_global}
    \end{subfigure}
    
    \vspace{-0.15cm} 
    \caption{
        As Figure \ref{fig:1dsmooth_0_global}, but for the 2D purities.
        \vspace{2cm}
	}
\end{figure*}

Representations of the 2D completenesses and 2D purities of the DSC \texttt{Combmod} classifier evaluated on the original validation data set are provided in Figures \ref{fig:2d_alldata_combmod_globalprior_completeness}-\ref{fig:2d_alldata_combmod_globalprior_purity}, respectively. 
Similar representations evaluated on the cleaned validation data set are reported in Figures \ref{fig:2d_alldata_combmod_globalprior_noMCcontaminantsExt_completeness}-\ref{fig:2d_alldata_combmod_globalprior_noMCcontaminantsExt_purity}.
{To assess the performance, we also compute 1D representations of the completenesses and purities as a function of Galactic latitude in Figures \ref{fig:1dsmooth_0a_global}-\ref{fig:1dsmooth_0b_global}.}
%
We exclude from our assessment the star class as its purity and completeness remain very close to 1.
For the quasar, galaxy, and white dwarf classes, 
	the completenesses and the purities decrease with decreasing magnitudes.
	From the 2D representations, we identify regions with zero completenesses (TP=0) indicative of the inability of the classifier to correctly identify sources. 
	However, these regions have low occupancy (i.e. few sources per bin), and thus do not affect the overall assessment of the performance.
	%
	We also identify regions with an undefined purity (TP=0 and FP=0) and a zero completeness (TP=0) due to the absence of 
	correct predictions and false positives (no source is mislabelled but no true class is correctly identified).
	Such regions correspond to the gaps in the 2D representations of the purity, as identified for 
	the white dwarf class in the strip at $\rm {G}$>20.5 across all latitudes 
	and the bulk region close to the Galactic plane at $|\sin b|$<0.2 and $\rm {G}$>18.5, totalling 4\,742 sources.

Overall, the 2D completenesses indicate a good performance across all magnitudes (Figures \ref{fig:2d_alldata_combmod_globalprior_completeness}-\ref{fig:2d_alldata_combmod_globalprior_purity}).
The 2D representations of the purities show that the false positives of the quasar class are mainly localised at the faint end with noticeable aggregates at $|\sin b|$=--0.54 and $|\sin b|$=--0.70, in addition to a few bright stellar contaminants at ${G}$<14. 
The false detections of the galaxy class are distributed across the 2D space mostly at fainter magnitudes $\rm {G}$>19 and at lower latitudes (i.e. close to the Galactic plane). 
The false positives of the white dwarf class are similarly distributed across the 2D space with noticeable aggregates at the fainter magnitudes at $|\sin b|$=--0.54 and $|\sin b|$=--0.70.
These aggregates of false positives are traced back to the Magellanic Clouds\footnote{Coordinates of the Large Magellanic Cloud (LMC) and the Small Magellanic Cloud (SMC) are (RA 80.8942\textdegree, DEC --69.7561\textdegree) and (RA 13.1583\textdegree, DEC --72.8003\textdegree), respectively.}, identifiable in the distribution of the star class in Figure \ref{fig:2d_alldata_combmod_counts}. 
The performance of DSC in the LMC and SMC is dominated by the stellar contaminants that strongly affects the overall purity. 
{To assess the completeness and purity outside these regions}, we chose to remove the LMC and SMC regions from the validation data set (Figures \ref{fig:2d_alldata_combmod_globalprior_noMCcontaminantsExt_completeness}-\ref{fig:2d_alldata_combmod_globalprior_noMCcontaminantsExt_purity}).

\begin{table*}[htp!]
\caption{Summary of the average 2D classification performance evaluated on the validation data set at different magnitude limits.}            
    \label{table:tab_GDR3_and_newcombmod_2dsummary_faint_bright}  
    \small \centering    
    \setlength{\tabcolsep}{7pt}                        
    \begin{tabular}{l | l | l| c c c | c  c c}         
        \hline\hline  
       		\multirow{2}{7em}{\centering Magnitude range}      
        		&\multirow{2}{5em}{\centering Prior}
		& \multirow{2}{4em}{\centering Class}
        			& \multicolumn{3}{c|}{Weighted average 2D completeness} 
    			& \multicolumn{3}{c}{Weighted average 2D purity}\\
    			\cline{4-9}
    			&&	& \it original & \it {{cleaned}} &  \it \texttt{Combmod}-${\alpha}$
    				& \it original & \it {{cleaned}} &  \it \texttt{Combmod}-${\alpha}$ \\
	
    	\hline\hline                	
	\multirow{10}{6em}{\centering \\ ${G}<$19}
	&\multirow{5}{5em}{\centering global prior}
	&	quasar		&	\bf 0.982	&	\bf 0.982	&	\bf 0.967	&	\bf 0.883	&	\bf 0.984	&	\bf 0.982	\\
	&&	galaxy		&	\bf 0.840	&	\bf 0.928	&	\bf 0.891	&	\bf 0.797	&	\bf 0.917	&	\bf 0.967	\\
	&&	star			&	0.999	&	0.999	&	0.999	&	0.991	&	0.991	&	0.991	\\
	&&	white dwarf	&	0.594	&	0.592	&	0.551	&	0.938	&	0.952	&	0.956	\\
	&&	binary		&	0.002	&	0.002	&	0.001	&	0.380	&	0.310	&	0.718	\\
	\cline{2-9}
	&\multirow{5}{5em}{\centering 2D variable prior}
	&	quasar		&	-	&	\bf 0.987	&	\bf 0.973	&	-	&	\bf 0.980	&	\bf 0.988	\\
	&&	galaxy		&	-	&	\bf 0.921	&	\bf 0.881	&	-	&	\bf 0.893	&	\bf 0.951	\\
	&&	star			&	-	&	0.999	&	0.999	&	-	&	0.991	&	0.991	\\
	&&	white dwarf	&	-	&	0.592	&	0.562	&	-	&	0.948	&	0.954	\\
	&&	binary		&	-	&	0.002	&	0.001	&	-	&	0.311	&	0.718	\\

	\hline\hline
	\multirow{10}{6em}{\centering \\19$\leq$${G}$<20}
	&\multirow{5}{5em}{\centering global prior}	
        &	quasar		&	\bf 0.955	&	\bf 0.955	&	\bf 0.904	&	\bf 0.470	&	\bf 0.801	&	\bf 0.865	\\
	&&	galaxy		&	\bf 0.964	&	\bf 0.975	&	\bf 0.961	&	\bf 0.747	&	\bf 0.916	&	\bf 0.955	\\
	&&	star			&	0.997	&	0.999	&	1.000	&	0.998	&	0.998	&	0.998	\\
	&&	white dwarf	&	0.345	&	0.340	&	0.314	&	0.835	&	0.944	&	0.948	\\
	&&	binary		&	0.000	&	0.000	&	0.000	&	0.000	&	0.000	&	N/A		\\
	\cline{2-9}
	&\multirow{5}{5em}{\centering 2D variable prior}
	&	quasar		&	-	&	\bf 0.973	&	\bf 0.938	&	-	&	\bf 0.814	&	\bf 0.884	\\
	&&	galaxy		&	-	&	\bf 0.976	&	\bf 0.965	&	-	&	\bf 0.902	&	\bf 0.911	\\
	&&	star			&	-	&	0.999	&	0.999	&	-	&	0.998	&	0.998	\\
	&&	white dwarf	&	-	&	0.328	&	0.315	&	-	&	0.945	&	0.948	\\
	&&	binary		&	-	&	0.000	&	0.000	&	-	&	0.000	&	0.000	\\
 
	\hline\hline
	\multirow{10}{6em}{\centering \\${G}$<20} 
	&\multirow{5}{5em}{\centering global prior}	
        &	quasar		&	\bf 0.964	&	\bf 0.964	&	\bf 0.926	&	\bf 0.610	&	\bf 0.864	&	\bf 0.905	\\
	&&	galaxy		&	\bf 0.941	&	\bf 0.967	&	\bf 0.949	&	\bf 0.760	&	\bf 0.918	&	\bf 0.958	\\
	&&	star			&	0.998	&	0.999	&	0.999	&	0.993	&	0.993	&	0.993	\\
	&&	white dwarf	&	0.487	&	0.485	&	0.450	&	0.910	&	0.952	&	0.955	\\
	&&	binary		&	0.002	&	0.002	&	0.001	&	0.332	&	0.281	&	0.718	\\
	\cline{2-9}
	&\multirow{5}{5em}{\centering 2D variable prior}
	&	quasar		&	-	&	\bf 0.978	&	\bf 0.950	&	-	&	\bf 0.873	&	\bf 0.921	\\
	&&	galaxy		&	-	&	\bf 0.966	&	\bf 0.950	&	-	&	\bf 0.902	&	\bf 0.918	\\
	&&	star			&	-	&	0.999	&	0.999	&	-	&	0.993	&	0.993	\\
	&&	white dwarf	&	-	&	0.480	&	0.457	&	-	&	0.949	&	0.954	\\
	&&	binary		&	-	&	0.002	&	0.001	&	-	&	0.284	&	0.717	\\
	
        \hline\hline
	\multirow{10}{6em}{\centering \\ ${G}\geq$20}
        &\multirow{5}{5em}{\centering global prior}			
	&	quasar		&	\bf 0.864	&	\bf 0.864	&	\bf 0.706	&	\bf 0.065	&	\bf 0.196	&	\bf 0.616	\\
	&&	galaxy		&	\bf 0.919	&	\bf 0.930	&	\bf 0.898	&	\bf 0.483	&	\bf 0.832	&	\bf 0.877	\\
	&&	star			&	0.991	&	0.997	&	0.999	&	0.999	&	0.999	&	0.999	\\
	&&	white dwarf	&	0.206	&	0.206	&	0.169	&	0.845	&	0.931	&	0.924	\\
	&&	binary		&	0.000	&	0.000	&	0.000	&	0.000	&	0.000	&	N/A		\\
	\cline{2-9}
	&\multirow{5}{5em}{\centering 2D variable prior}
	&	quasar		&	-	&	\bf 0.927	&	\bf 0.816	&	-	&	\bf 0.247	&	\bf 0.508	\\
	&&	galaxy		&	-	&	\bf 0.940	&	\bf 0.917	&	-	&	\bf 0.814	&	\bf 0.862	\\
	&&	star			&	-	&	0.995	&	0.998	&	-	&	0.999	&	0.998	\\
	&&	white dwarf	&	-	&	0.184	&	0.169	&	-	&	0.922	&	0.929	\\
	&&	binary		&	-	&	0.000	&	0.000	&	-	&	0.000	&	N/A		\\
        \hline   
    \end{tabular}
    \tablefoot{
   Predicted labels are obtained from the class with the maximum posterior probability. 
   The labels `original' and `cleaned' refer to the results evaluated on the full validation data set and the {{cleaned}} validation data set, respectively.
   \texttt{Combmod}-${\alpha}$ is the parametric combination of the GDR3 DSC \texttt{Allosmod} and \texttt{Specmod}, where $\alpha_{\rm glob}$=[1.0, 0.6, 0.5] and $\alpha_{\rm vari}$=[0.6, 0.6, 0.5].
   Results of the extragalactic classes are highlighted in boldface.
   \textit{N/A} values correspond to TP=0 (absence of correct predictions) and FP=0 (absence of false detections).
   }           
\end{table*}

Reevaluated on the cleaned validation set and without introducing changes to the published GDR3 DSC probabilities, the overall performance set is summarised in Table \ref{table:tab_GDR3_and_newcombmod_summary} under the label `cleaned'. 
{The confusion matrix is provided in the Appendix Figure \ref{fig:confmat_updateddataset_globalprior} for reference.}
Figures \ref{fig:2d_alldata_combmod_globalprior_noMCcontaminantsExt_completeness} and \ref{fig:2d_alldata_combmod_globalprior_noMCcontaminantsExt_purity} report the 2D performance of \texttt{Combmod} evaluated on the {{cleaned}} validation data. 
{Comparing Figures \ref{fig:2d_alldata_combmod_globalprior_purity} and \ref{fig:2d_alldata_combmod_globalprior_noMCcontaminantsExt_purity}}, the absence of the contaminants is easily noticeable in the 2D purities. This is also apparent in the smoothed representations of the purities in Figure \ref{fig:1dsmooth_0b_global} at different magnitude limits.

To summarise the 2D representations, we compute the average 2D metrics over magnitude and latitude bins, defined as the weighted sum across all bins. Per target class, the average 2D completenesses are weighted by the number of true sources per bin and the average 2D purities are weighted by the number of predictions per bin.
%
Table~\ref{table:tab_GDR3_and_newcombmod_2dsummary} reports the summary of the 2D performance evaluated on the cleaned validation set showing higher 2D purities for the quasars, galaxies and white dwarfs, compared to the original validation data set.
In Table \ref{table:tab_GDR3_and_newcombmod_2dsummary_faint_bright}, the comparison at different magnitude limits shows that the improvement in purity is more pronounced at fainter magnitudes, 
	by 33, 17 and 11 percentage points at magnitudes 19$\leq$${G}$<20,     
	and 13, 35, and 9 percentage points at magnitudes ${G}$$\geq$20, 
	for the quasars, galaxies and white dwarfs, respectively.

\begin{figure*}[htp!]
\begin{minipage}[t]{1.\textwidth}  
	\centering
	\hspace{-.25cm}
	\begin{subfigure}{1\textwidth}
	\centering
	\includegraphics[scale=0.470]{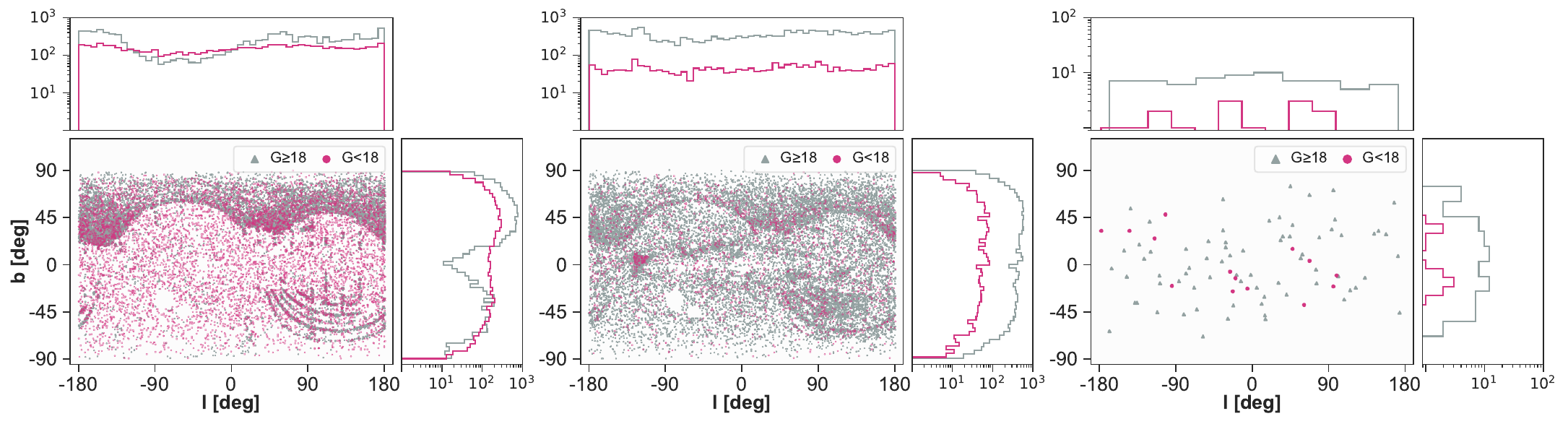}	
	\includegraphics[scale=0.470]{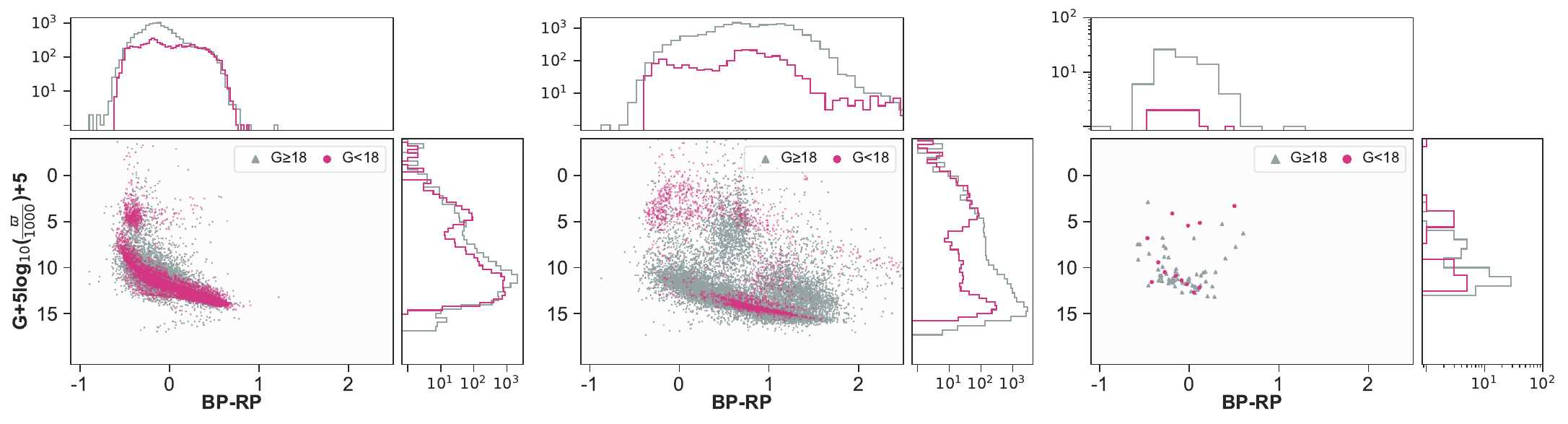}	
	\caption{} 
	\label{fig:WD_preds_all}
	\end{subfigure}

	\vspace{0.50cm} 
	\begin{subfigure}{1\textwidth}
	\centering	
	\includegraphics[scale=0.470]{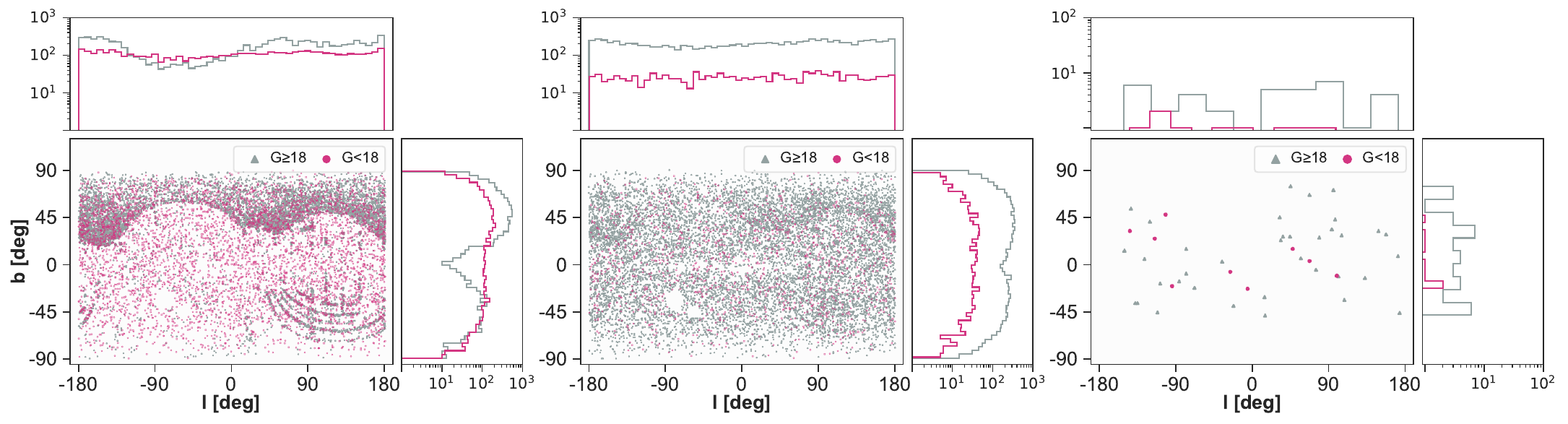} 
	\includegraphics[scale=0.470]{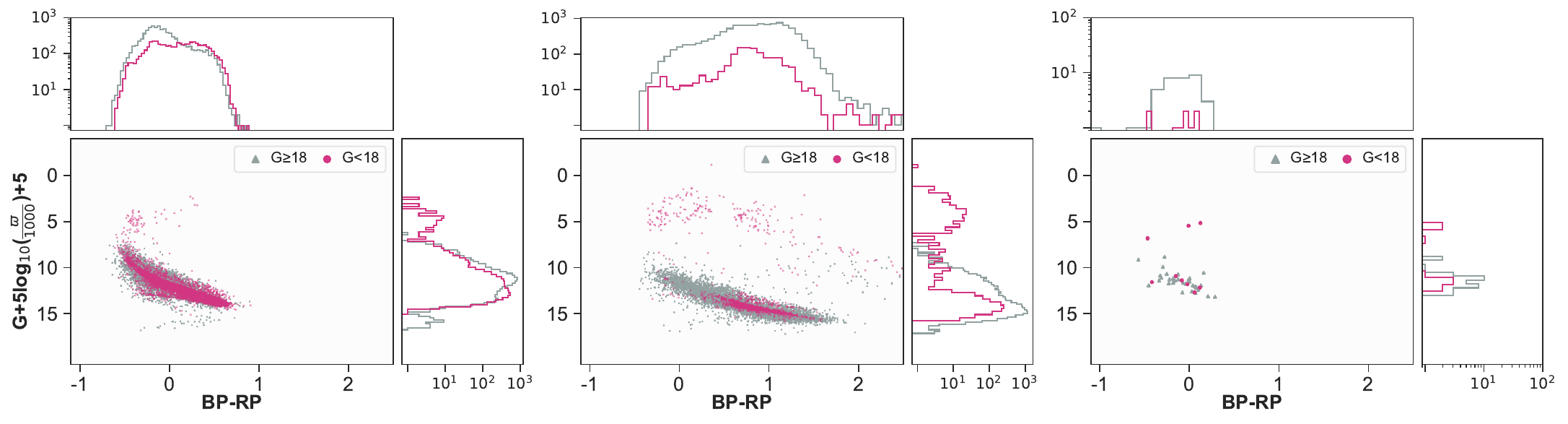} 
	\caption{} 
			%
	\label{fig:WD_preds_qualcutsel}
	\end{subfigure}
	
	\caption{
		DSC \texttt{Combmod} classifications for the white dwarf class on the {{cleaned}} validation data set using the global prior. 
		\textit{(a)} DSC WDs predictions on the full sample.
		\textit{(b)} DSC WDs predictions after the quality cut on Gaia data.
		Each panel shows, from left to right, 
			the true positives (i.e. correct predictions),
			the stellar false negatives (i.e. sources from the white dwarf class classified 
			as stars), 
			and the stellar false positives (i.e. sources from the star class 
			classified as white dwarfs).
		Faint sources ${G}$$\geq$18 (grey) and 
			bright sources ${G}$<18 (magenta) are plotted separately. 
		}
	\label{fig:WD_preds}
\end{minipage} 
\end{figure*} 

 \subsection{{A closer look at white dwarfs}} \label{section_wd}
We now examine the distribution of the white dwarfs (WDs) predictions.
Figure \ref{fig:WD_preds_all} shows the distributions of the correct predictions (TPs), the WDs falsely classified as stars (FNs), and the stars falsely classified as WDs (FPs). 
We build the Gaia Hertzsprung-Russell diagram (HRD) by estimating the absolute magnitude from the $G$ band and the parallax in millarcseconds. The absolute magnitude is plotted against the colour $G_{\rm BP}-G_{\rm RP}$.
On the HRD, most TPs lie in the expected WDs sequence.
Within the TPs, a subset of hot subdwarfs can be identified.
However, since hot subdwarfs and white dwarfs share similar characteristics in their black-body emission spectra, these sources can be misclassified.
{In} GDR3, the DSC training of the white dwarf class used the MWDD that contains hot subdwarf candidates in addition to white dwarfs.
The stellar FNs are distributed over the full sky with small clusterings of faint sources at higher latitudes or close to the Galactic plane.
On the HRD, the faint FNs show a large scatter and the bright FNs extend towards redder colours (higher $G_{\rm BP}-G_{\rm RP}$) in comparison to the TPs. 
The distribution of the few FPs from the star class follows the WD sequence on the HRDs. These sources can possibly be true WDs candidates, although misclassified in the validation data set as stars.  

To investigate the stellar FNs we apply a quality cut on Gaia data similar to \cite{gentilefusillo_catalogue_2021} {using features from the Gaia main source table (\texttt{gaiadr3.gaia\_source})} as 
\begin{align*}
	&& \texttt{parallax\_over\_error} Eq{\geq4}, \\
	&  \rm and & \texttt{ruwe}<1.1,						\\
	&  \rm and & \texttt{astrometric\_sigma5d\_max}<1.5,	\\
    	& \rm and  & \texttt{ipd\_gof\_harmonic\_amplitude}<1,	\\
        &  \rm and & \texttt{phot\_[bp,rp]\_n\_obs}>2, 	
	\tag{\stepcounter{equation}\theequation}
\end{align*}                    
 where
 \texttt{parallax\_over\_error} is the parallax divided by its error,
 \texttt{astrometric\_sigma5d\_max} the longest semi-major axis of the 5-d error ellipsoid, 
 \texttt{ruwe} the renormalised unit weight error, 
 \texttt{ipd\_gof\_harmonic\_amplitude} the amplitude of the goodness-of-fit of the Image Parameter Determination (IPD) versus the position angle of the scan direction,
 \texttt{phot\_[bp,rp]\_n\_obs} the number of observations in BP and RP, respectively.
On Figure \ref{fig:WD_preds_qualcutsel}, the quality cut rejects the bright subdwarfs and the faint sources from the TPs, and the faint sources from the FNs. 
The clusterings of the faint FNs are rejected and the large dispersion on the HRD is corrected. This shows that the data quality contributes to the misclassification. 
After the quality cut, the remaining FNs mislabelled by DSC as stars result from inherent limitations of the GDR3 DSC classifiers.
In GDR3, \texttt{Combmod} achieves for the white dwarf class average 2D completenesses and 2D purities of 43\% and 95\%, respectively (Table \ref{table:tab_GDR3_and_newcombmod_2dsummary}).
At different magnitude limits, the average 2D completenesses and 2D purities are  
	59\% and 95\% at ${G}$<19, 
	34\% and 94\% at 19$\leq$${G}$<20, 
	and 20\% and 93\% at ${G}$$\geq$20, respectively
(Table \ref{table:tab_GDR3_and_newcombmod_2dsummary_faint_bright}).
Although incomplete, the GDR3 DSC WD candidates constitute a pure set of candidates that can be useful for follow-up studies. 
We advise to apply quality cuts and discard DSC predictions in the LMC and SMC regions.

\section{{New classifications based on the published DSC probabilities}}\label{sec:DSCcombiner}

\subsection{{A magnitude- and latitude-dependent prior}}\label{sec:varprior}

\subsubsection{Method}

DSC exploits a Bayesian framework to compute posterior probabilities per target class, and uses a global prior to account for the overall distribution of classes over the entire data set \citep{delchambre_gaia_2023}. The global prior is set to $\bm{\pi}^{\rm glob}$=[0.000989, 0.000198, 0.988728, 0.000198, 0.009887] for the quasar, galaxy, star, white dwarf, and physical binary classes, respectively.
Following the work by \cite{hughes_quasar_2022}, we compute a variable prior $\bm{\pi}^{\rm vari}$ function of Galactic latitude and brightness from the expected class distribution across the sky.
Stars dominate over the whole sky, but their number density decreases towards higher latitudes and fainter magnitudes where the contribution from the quasars and galaxies increases.
Ideally, a deeper and complete survey would provide the relative numbers of stars, quasars, galaxies as a function of Galactic latitude and brightness. However, we do not have access in reality to a deep survey covering the whole sky at all magnitudes. 
We compute the prior from a selected data set and assume that the ratios computed in the constructed 2D space are representative of the expected class distributions across all magnitudes and latitudes. 
The variable prior $\bm{\pi}^{\rm vari}$ for class $k$ (quasar, galaxy, stellar class) is defined as 
\begin{equation}
	\begin{split}
		&\pi_{k}^{\rm vari} (x) 
            = \pi_{k}^{\rm glob}  \, \frac{F_{k}(x)}{N_{k}},
	\end{split}
	\label{eq:newprior_definition}
\end{equation}
where 
	$x$ refers to the parameters (Galactic latitude and brightness), 
	$k$ the class,
	$F_{k}(x)$ the total number of sources per class $k$ evaluated at $x$,
	$N_{k}$ the total number of sources per class $k$,
	and $\pi_{k}^{\rm glob} $ the global prior at class $k$.
We expand the variable prior from the three classes to the five classes by adjusting the fraction of the stellar types as
\begin{equation}
	\begin{split}
		\pi_{j}^{\rm vari} 
			&= \pi_{\rm stellar}^{\rm vari} \left(\frac{\pi_{j}^{\rm glob} }{\pi_{\rm stellar}^{\rm glob} }\right),
			\end{split}
	\label{eq:newprior_frac_stellar}
\end{equation}
where $j$ refers to the star, white dwarf and binary classes and $\pi_{\rm stellar}^{\rm glob}  = \pi_{\rm star}^{\rm glob}+\pi_{\rm white dwarf}^{\rm glob}+\pi_{\rm binary}^{\rm glob}$.\\
Finally, the variable prior is normalised across the five classes using 
\begin{equation}
		\pi_{k}^{\rm vari} =  \left.\pi_{k}^{\rm vari}  \right/{\sum_{k' \rm{classes}} \pi_{k'}^{\rm vari}}.
	\label{eq:newprior_normalise}
\end{equation}	
To build the variable prior, we select quasar and galaxy sources from SDSS DR17 with spectral types \texttt{class=`GALAXY'} or \texttt{class=`QSO'} and a redshift quality flag \texttt{zWarning=0} indicative of a reliable redshift estimation and spectroscopic classification. 
The stellar class is a random selection from Gaia cleaned from the selected extragalactic sources and any match to quasar and galaxy candidates from cross-matches with external non-Gaia catalogues (Milliquas v7.2-v8 \citeauthor{flesch_half_2015} \citeyear{flesch_half_2015},
\citeauthor{flesch_million_2023} \citeyear{flesch_million_2023}; 
SDSS DR 17 photometry \citeauthor{abdurrouf_seventeenth_2022} \citeyear{abdurrouf_seventeenth_2022};  
and the Assef R90 and C75 WISE AGN catalogues \citeauthor{assef_wise_2018} \citeyear{assef_wise_2018}).
Since SDSS does not cover the full sky nor the full magnitude and latitude range of Gaia, we extrapolate the prior outside this magnitude range to the values at the boundary. 
The selected extragalactic classes cover a magnitude range of 14$\leq$${G}$$\leq$21. Priors at magnitudes fainter than 21 are extrapolated to the prior at ${G}$=21, and at magnitudes brighter than 14 extrapolated to the prior at ${G}$=14. Moreover, to compensate for the disparity in the SDSS footprint between the northern versus the southern Galactic hemispheres, we build the prior in latitudes using the absolute values in $|\sin b|$.

We compute the variable prior from a discretised 10$\times$10 grid across magnitudes and latitudes using Eq.\ \ref{eq:newprior_definition}, compute the variable prior for the stellar classes using Eq.\ \ref{eq:newprior_frac_stellar} and then normalise using Eq. \ref{eq:newprior_normalise}.
We expand the discretised mapping to a continuous representation $\hat{\bm\pi}^{\rm vari}$ by fitting a bivariate spline to the prior of the quasar and galaxy classes. 
To ensure the sum of the prior across all classes to be equal to 1 at every point, the continuous variable prior for the stellar class is obtained by subtracting the priors of the extragalactic classes 
$\hat{\pi}_{\rm stellar}^{\rm vari}=1-\hat{\pi}_{\rm quasar}^{\rm vari}-\hat{\pi}_{\rm galaxy}^{\rm vari}$.
The individual stellar contributions are obtained using Equation \ref{eq:newprior_frac_stellar}.
The fitting is performed using the implementation of the \texttt{RectBivariateSpline} from the \texttt{python} package \texttt{scipy.interpolate}. 
As previously stated, priors outside the grid are extrapolated from the values estimated at the boundary.
Computing DSC probabilities using a new prior does not require re-training nor re-applying DSC to the data \citep{bailer-jones_quasar_2019}.
The posterior probability using the variable prior $P_{k}^{\rm vari}$ for class $k$ is defined as
\begin{equation}
		P_{k}^{\rm vari} 
			= \left.{P_{k} \frac{\hat{\pi}_{k}^{\rm vari}}{{\pi}_{k}^{\rm glob}} }\right
				/ 	{ \sum_{k' \rm{classes}} \left( P_{k'} \frac{\hat{\pi}_{k'}^{\rm vari}}{{\pi}_{k'}^{\rm glob}} \right) },
	\label{eq:newprior}
\end{equation}
where $P_{k}$ refer to the posterior probability using the global prior, $\pi_{k}^{\rm glob}$ the global prior at class $k$ and $\hat{\pi}_{k}^{\rm vari}$ the variable prior at class $k$. 
We apply Eq. \ref{eq:newprior} to compute new posterior probabilities for all classifiers in DSC.
{For reference, the Appendix Figure \ref{fig:varprior} reports the variable prior for the quasar, galaxy, and stellar classes.}

\begin{figure*}[htp!]
    \centering
        \begin{subfigure}{1\textwidth}
            \raggedleft
            \includegraphics[scale=0.28, trim={0 0 0 0}, clip]{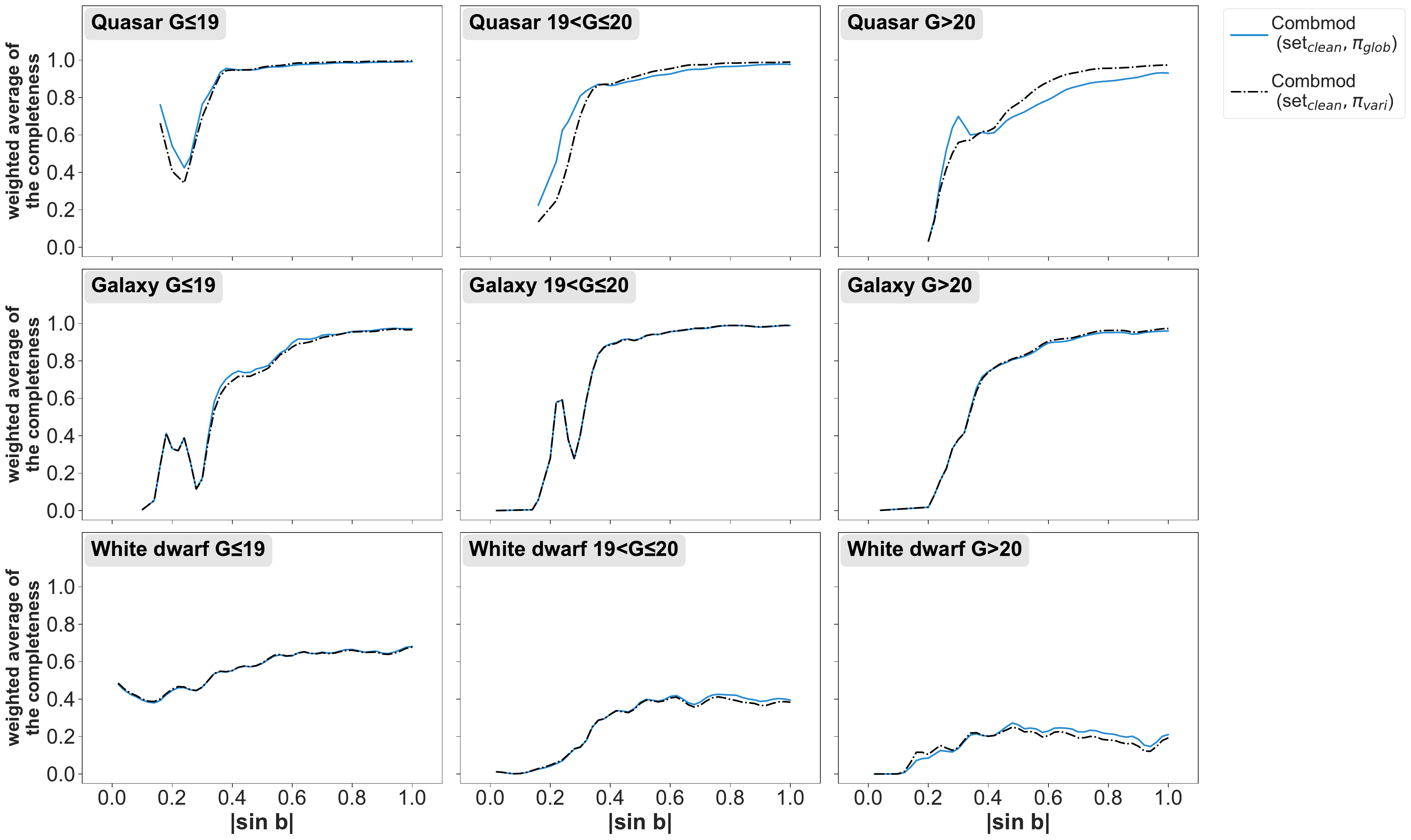}
             \caption{}
    \label{fig:1dsmooth_1a_variable}
    \end{subfigure}
    
     \vspace{0.5cm}
    \begin{subfigure}{1\textwidth}
		\raggedleft
    		\includegraphics[scale=0.28, trim={0 0 0 0}, clip]{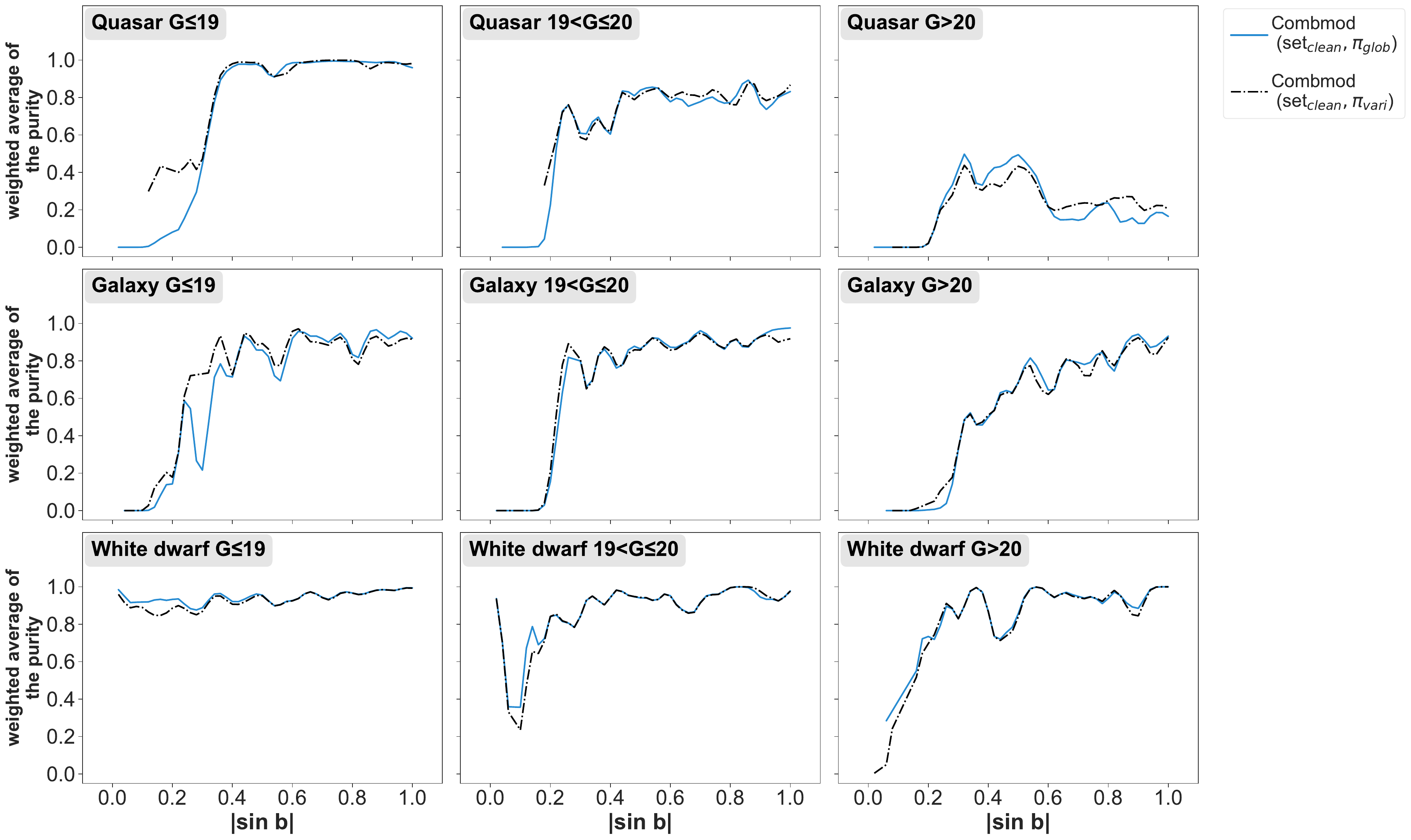}
	 \caption{}
	\label{fig:1dsmooth_1b_variable}
    \end{subfigure}
    
    \caption{
        As Figures \ref{fig:1dsmooth_0a_global} and \ref{fig:1dsmooth_0b_global}, but comparing the DSC \texttt{Combmod} using the variable prior and the global prior on the clean validation data set.
        In each panel the solid line shows the performance of the \texttt{Combmod} using the global prior and the dash dot line using the variable prior. 
        \textit{(a)} Smoothed average of the 2D completenesses.
        \textit{(b)} Smoothed average of the 2D purities.
        }
\end{figure*} 

\subsubsection{Performance}

{We compute 1D representations of the completenesses and purities as a function of Galactic latitude in Figures \ref{fig:1dsmooth_1a_variable}-\ref{fig:1dsmooth_1b_variable}}.
%
Results of the \texttt{Combmod} classifier using the variable prior show a slightly higher purity and completeness at fainter magnitudes at high latitudes. 
{
Figure \ref{fig:2d_updateddata_combmod_variableprior} reports, in the 2D performance of the extragalactic classes, the removal of the few bright contaminants in the quasar class.
Overall, the application of the variable prior shows overall} no significant improvement in the completeness and the purity of any of the three classes. 

{The confusion matrix is provided in the Appendix Figure \ref{fig:confmat_updateddataset_variableprior_combmod} for reference.}
Table \ref{table:tab_GDR3_and_newcombmod_2dsummary} summarises the 2D performance evaluated on the {{cleaned}} validation data set (as described in Section \ref{subsec:cleanedvalset}).
{Using the} variable prior does not strongly modify (improve nor decrease) the average 2D completenesses and purities compared to the GDR3 \texttt{Combmod} using the global prior.
The comparison at different magnitude limits (Table \ref{table:tab_GDR3_and_newcombmod_2dsummary_faint_bright}) shows the highest improvement at magnitudes ${G}$$\geq$20, although small (under $\sim$7 percentage points in the 2D completenesses and $\sim$2-3 percentage points in the 2D purities).
%
%

To summarise, the application of a continuous variable prior on \texttt{Combmod} improves by a small fraction the average 2D completenesses and purities of the target classes compared to the GDR3 results using the global prior. 
The highest improvement is seen at fainter magnitudes ${G}$$\geq$20 for quasars, where the average 2D completenesses increases from 86\% to 93\% and the average 2D purities from 19\% to 24\%. The effect of the variable prior on the galaxies and the white dwarfs remains not significant ($\sim$2 percentage points).

\subsection{{ \texttt{Combmod}-$\alpha$: A new combination of classifiers}}

\subsubsection{Method}

In GDR3, \texttt{Combmod} is obtained from a weighted combination of the DSC classifiers \texttt{Specmod} and \texttt{Allosmod} to compute probabilities for the five classes. 
Since \texttt{Specmod} and \texttt{Allosmod} have the same prior, the combination of the posterior probabilities removes the prior from one contribution \citep{ulla_gaia_2022}.
The posterior probability of \texttt{Combmod} for class $k$ is defined as
\begin{equation}
	\begin{split}
	&P_{k}^{c} =
	\begin{cases}
		P_{k}^{s} P_{k}^{a} \dfrac{1}{\pi_k^{s}}
			& \text{\footnotesize for the quasar and galaxy classes}, \\
		P_{k}^{s} P_{k}^{a} \dfrac{\pi_k^{s}}{(\pi_{star}^{a})^2}
			& \text{\footnotesize {for the stellar classes}},
	\end{cases} 
	\end{split}
	\label{eq:combmod_gdr3}
\end{equation}
and normalised across all five classes
	\begin{equation}
	\begin{split}
		P_{k}^{c} = \left.{P_{k}^{c}} \right/{\sum_{k'} P_{k'}^{c}},
	\end{split}
\end{equation}
where
$P_k^s$ and $P_k^a$ are the posterior probabilities for \texttt{Specmod} and \texttt{Allosmod} respectively,  
and the priors $\{\pi_k^{s},\pi_k^{a}\}$ refer, respectively, to the 5-class prior used in \texttt{Specmod} and the 3-class prior used in \texttt{Allosmod}, 
such that $\pi_{\rm star}^{a}$=$\pi_{\rm star}^{s}$+$\pi_{\rm white dwarf}^{s}$+$\pi_{\rm binary}^{s}$.

We investigate how a different combination of the DSC baseline classifiers performs in terms of completeness and purity.
We report in this work results of a parametric weighted sum combination.
%
We compute a new classifier, \texttt{Combmod}-$\alpha$, from a weighted sum of the GDR3 \texttt{Allosmod} and \texttt{Specmod} posterior probabilities.
The posterior probability of \texttt{Combmod}-$\alpha$ for class $k$ is defined as
\begin{equation}
	\begin{split}
	&P_{k}^{c_{\alpha}} =
	\begin{cases}
		(1- \alpha_k) P_{k}^{s} + \alpha_k P_{k}^{a}
			& \text{\footnotesize for the quasar and galaxy classes}, \\
		(1- \alpha_k) P_{k}^{s} + \alpha_k P_{k}^{a} \dfrac{\pi_k^{s}}{\pi_{star}^{a}}
			& \text{\footnotesize {for the stellar classes}},
	\end{cases} 
	\end{split}
	\label{eq:combmod_alpha}
\end{equation}
and normalised across all five classes 
\begin{equation}
	\begin{split}
		P_{k}^{c_{\alpha}} = \left.{P_{k}^{c_{\alpha}}} \right/ {\sum_{k'} P_{k'}^{c_{\alpha}}},
	\end{split}
\end{equation}
where the parameters $\{\alpha_k\}$ control the contribution of each classifier, subject to $\alpha_k$$\geq$0. 
We explore the optimal values of these parameters over a uniform grid between 0 and 1, and set the contribution of the stellar classes to $\alpha_{\rm star}$=$\alpha_{\rm white dwarf}$=$\alpha_{\rm binary}$. 

For different configurations of the $\{\alpha_k\}$, we compute the posterior probabilities of the classifier and assess its performance through the geometric score. {We define} the geometric score as the geometric mean of the completeness and purity of the extragalactic classes, as follows :
\begin{equation}
	\begin{split}
		&\rm score_{\rm geometric} 
			&= \Big( \rm completeness_{\rm quasar} 
					\times\,  \rm completeness_{\rm galaxy}  \\
				&&    \times\, \rm purity_{\rm quasar}
					\times\, \rm purity_{\rm galaxy} \Big)^{1/4}.
		\end{split}	
	\label{eq:geoscore}
\end{equation}
The extreme case of $\alpha_{k}$=0 for all $k$ has only \texttt{Specmod} contributing to the posterior probabilities, while the case of $\alpha_{k}$=1 has only \texttt{Allosmod} contributing.
When $\alpha_{\rm star}$>0.5, the contribution from \texttt{Allosmod} is increased compared to \texttt{Specmod}. Since the GDR3 \texttt{Allosmod} did not discriminate between the three stellar types, we limit the contribution $\alpha_{\rm star}$ to values below 0.5, in order to balance out the contributions of \texttt{Specmod} and \texttt{Allosmod} across the five classes. 

\begin{figure*}[htp!]

	\begin{subfigure}{1.0\textwidth}
		\centering
		\includegraphics[scale=0.21, trim={0 0 5cm 0},clip]{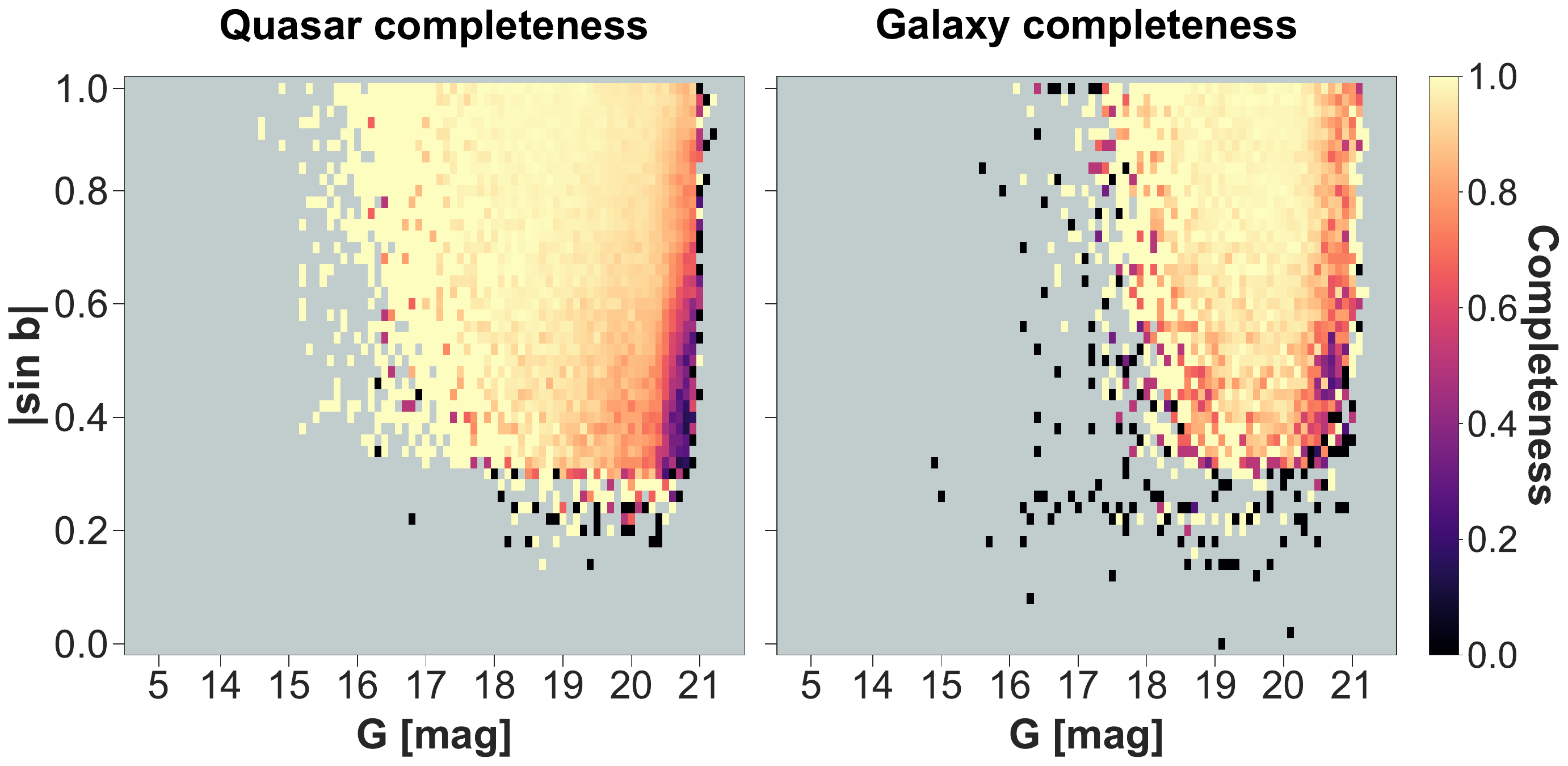}
		\hspace{.1cm}
		\includegraphics[scale=0.21, trim={3.5cm 0 1.5cm 0},clip]{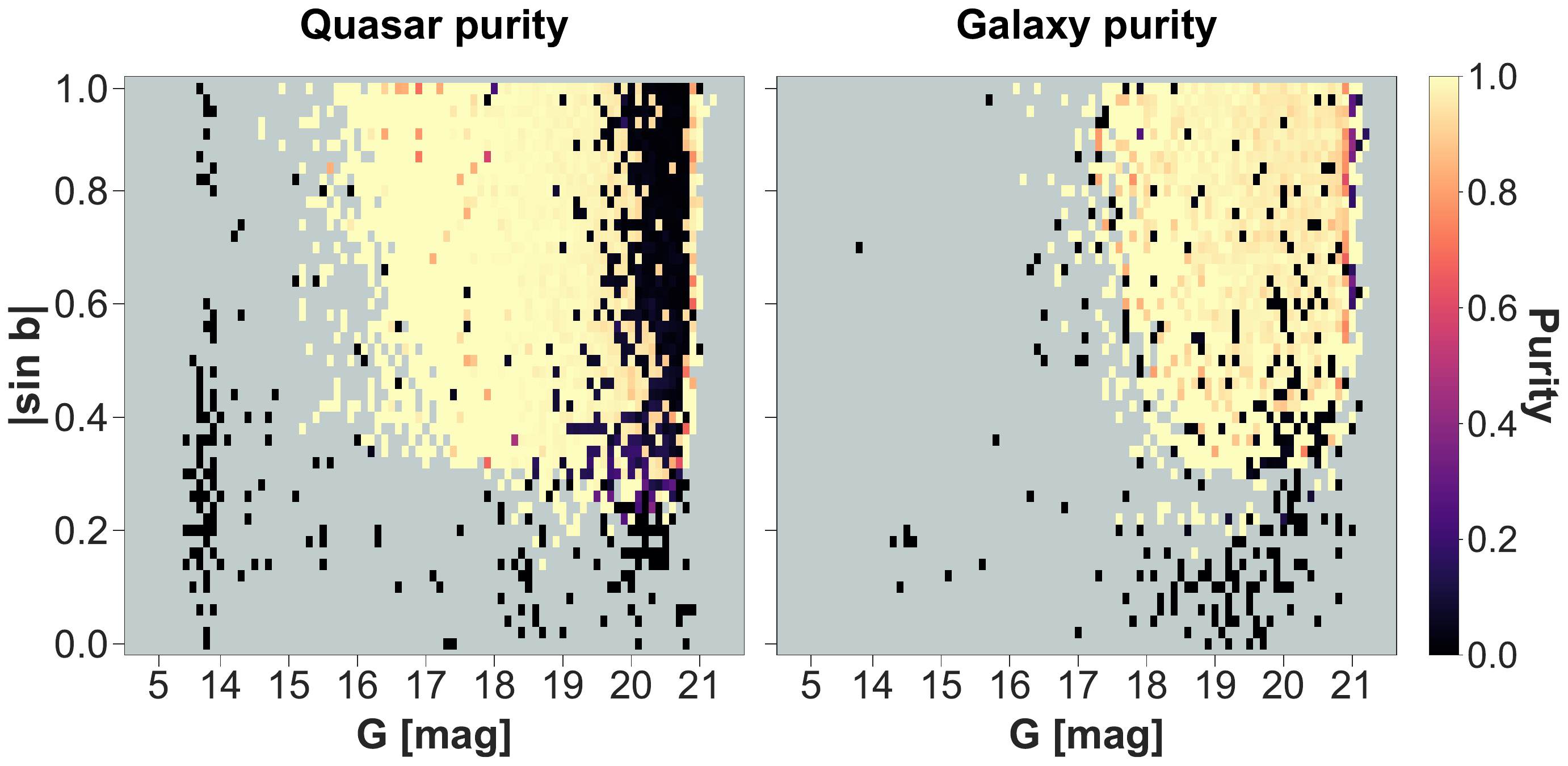}
		\caption{} 
		\label{fig:2d_updateddata_combmod_globalprior_2}
	\end{subfigure}
	
	\vspace{.1cm} 
	\begin{subfigure}{1\textwidth}
		\centering
    		\includegraphics[scale=0.21, trim={0 0 5cm 0},clip]{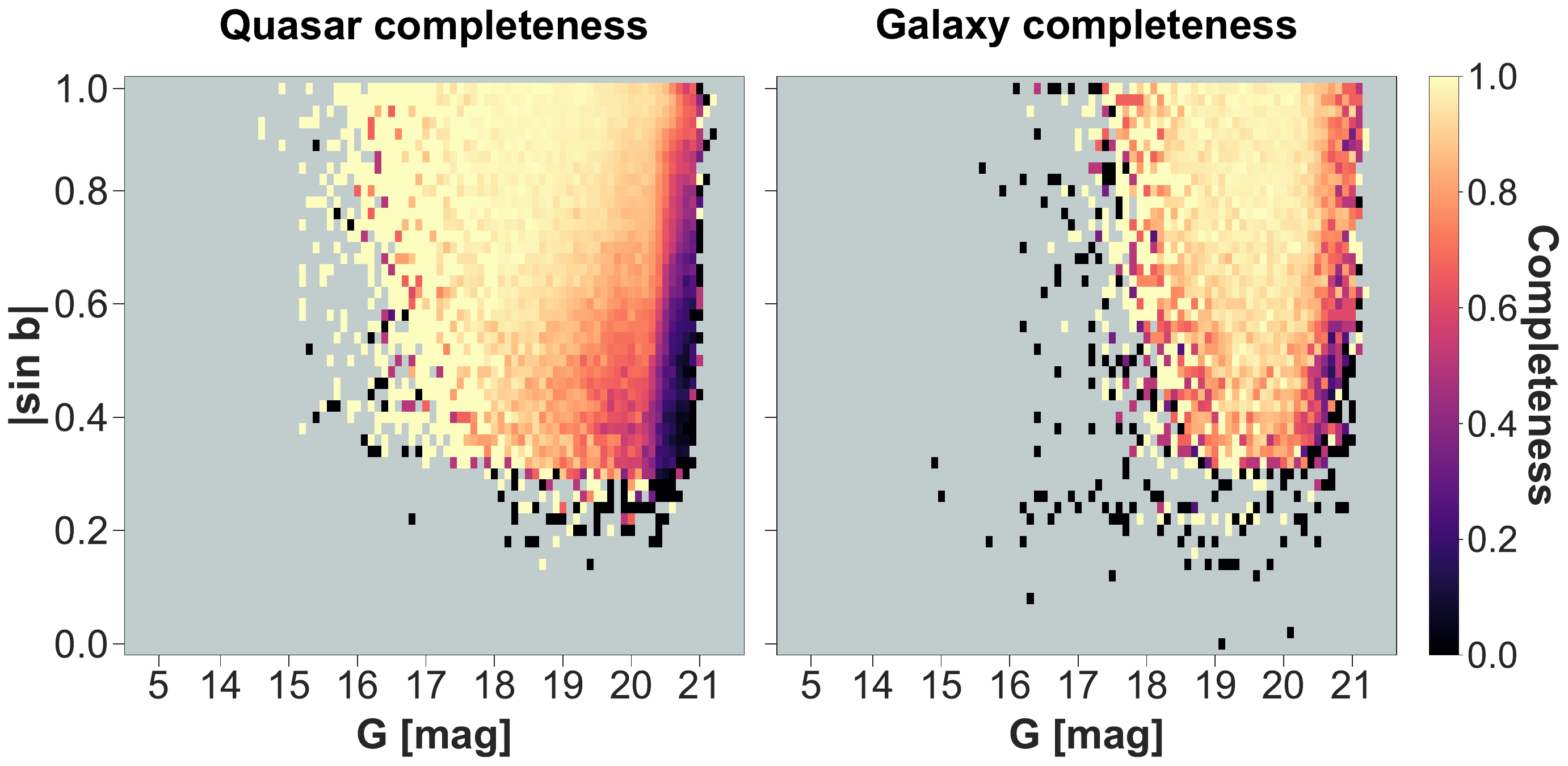}
		\hspace{.1cm}
    		\includegraphics[scale=0.21, trim={3.5cm 0 1.5cm 0},clip]{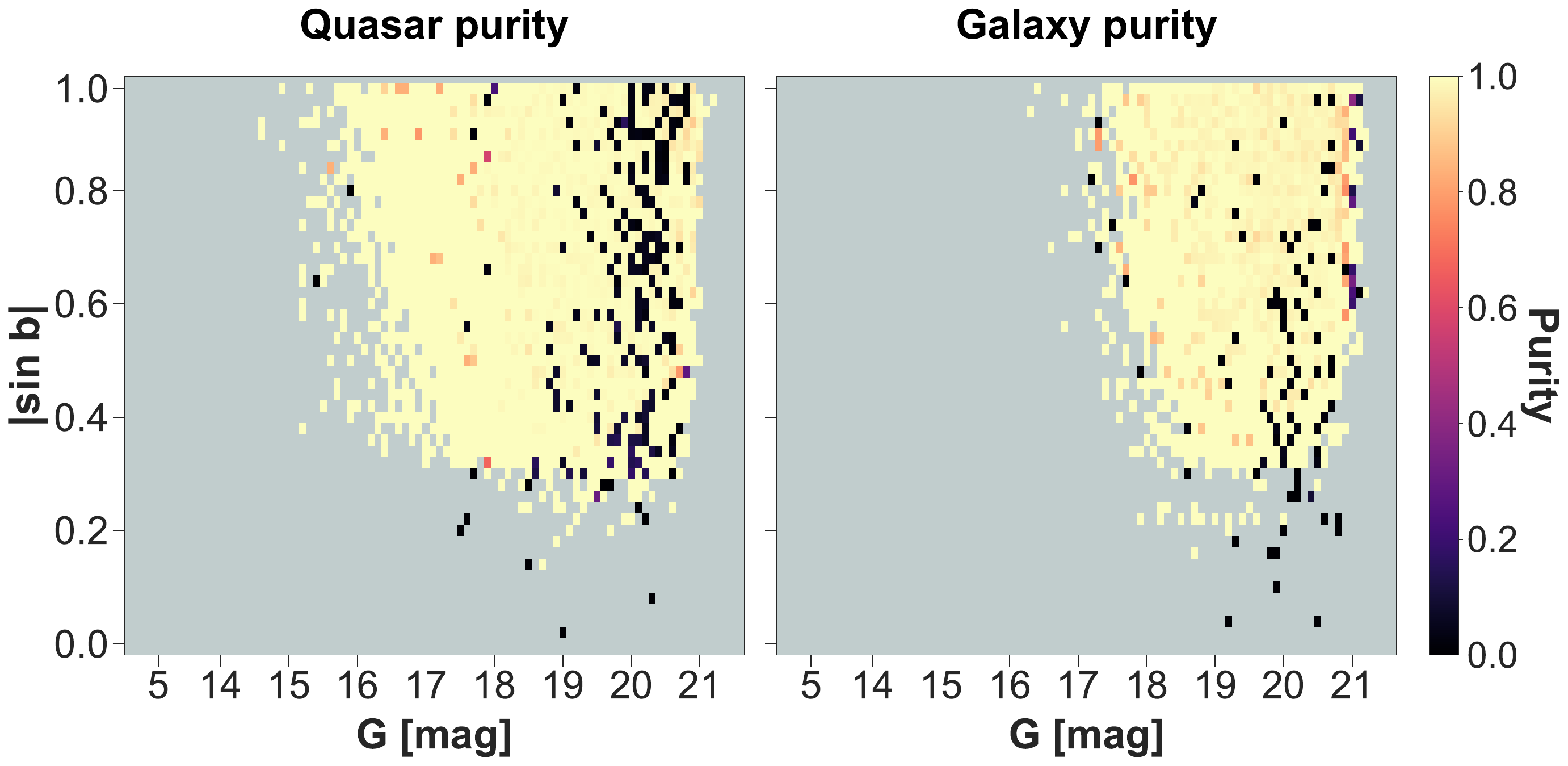}
    		\caption{} 
    		\label{fig:2d_updateddata_newcombmod2_globalprior}
	\end{subfigure}
	
	\vspace{.1cm} 
	\begin{subfigure}{1.0\textwidth}
		\vspace{.1cm} 
		\centering
		\includegraphics[scale=0.21, trim={0 0 5cm 0},clip]{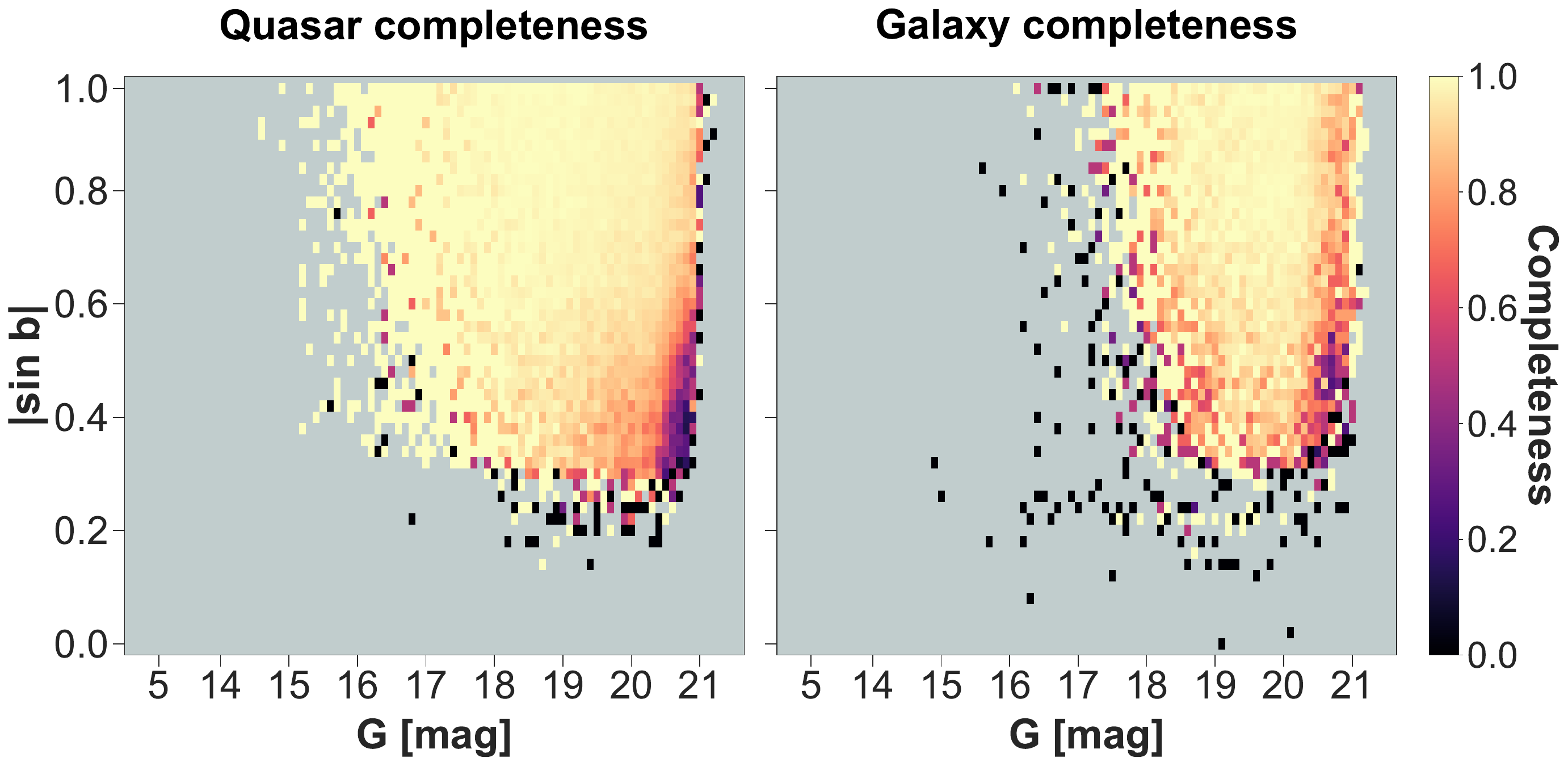}
		\hspace{.1cm}
		\includegraphics[scale=0.21, trim={3.5cm 0 1.5cm 0},clip]{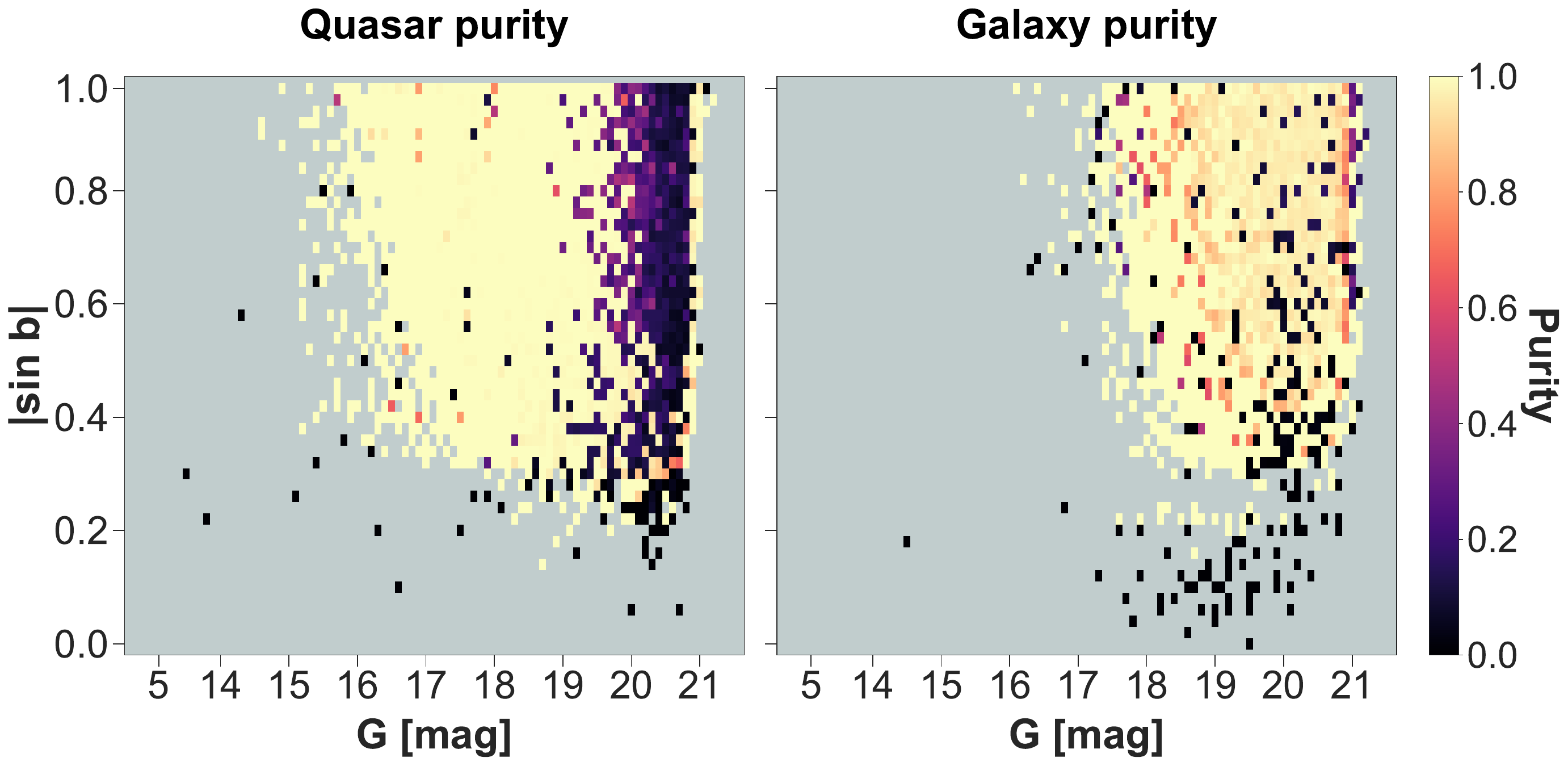}
		\caption{} 
		\label{fig:2d_updateddata_combmod_variableprior}
	\end{subfigure}

	\vspace{.1cm} 
	\begin{subfigure}{1\textwidth}
		\centering
    		\includegraphics[scale=0.21, trim={0 0 5cm 0},clip]{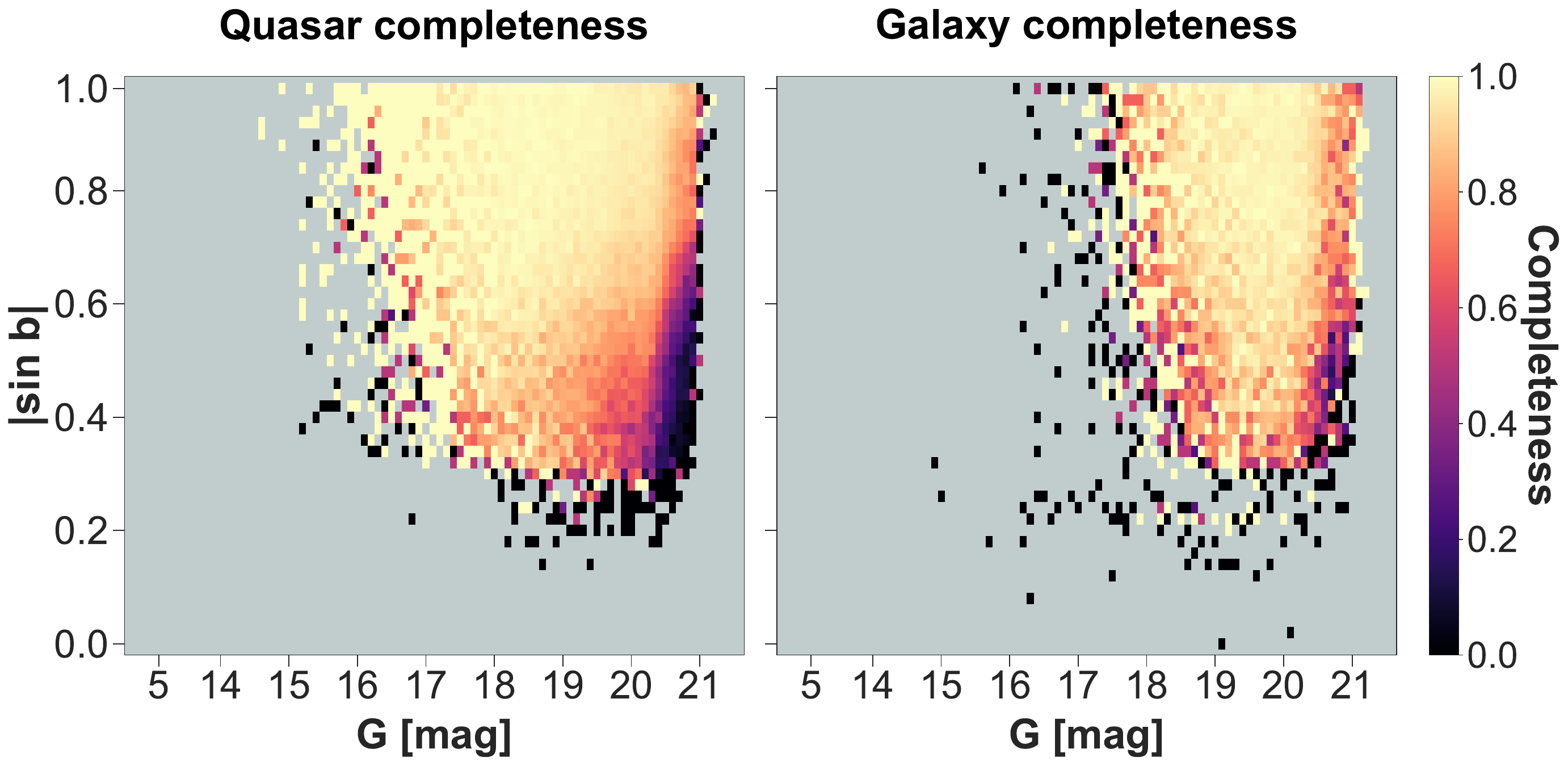}   
    		\hspace{.1cm}
    		\includegraphics[scale=0.21, trim={3.5cm 0 1.5cm 0},clip]{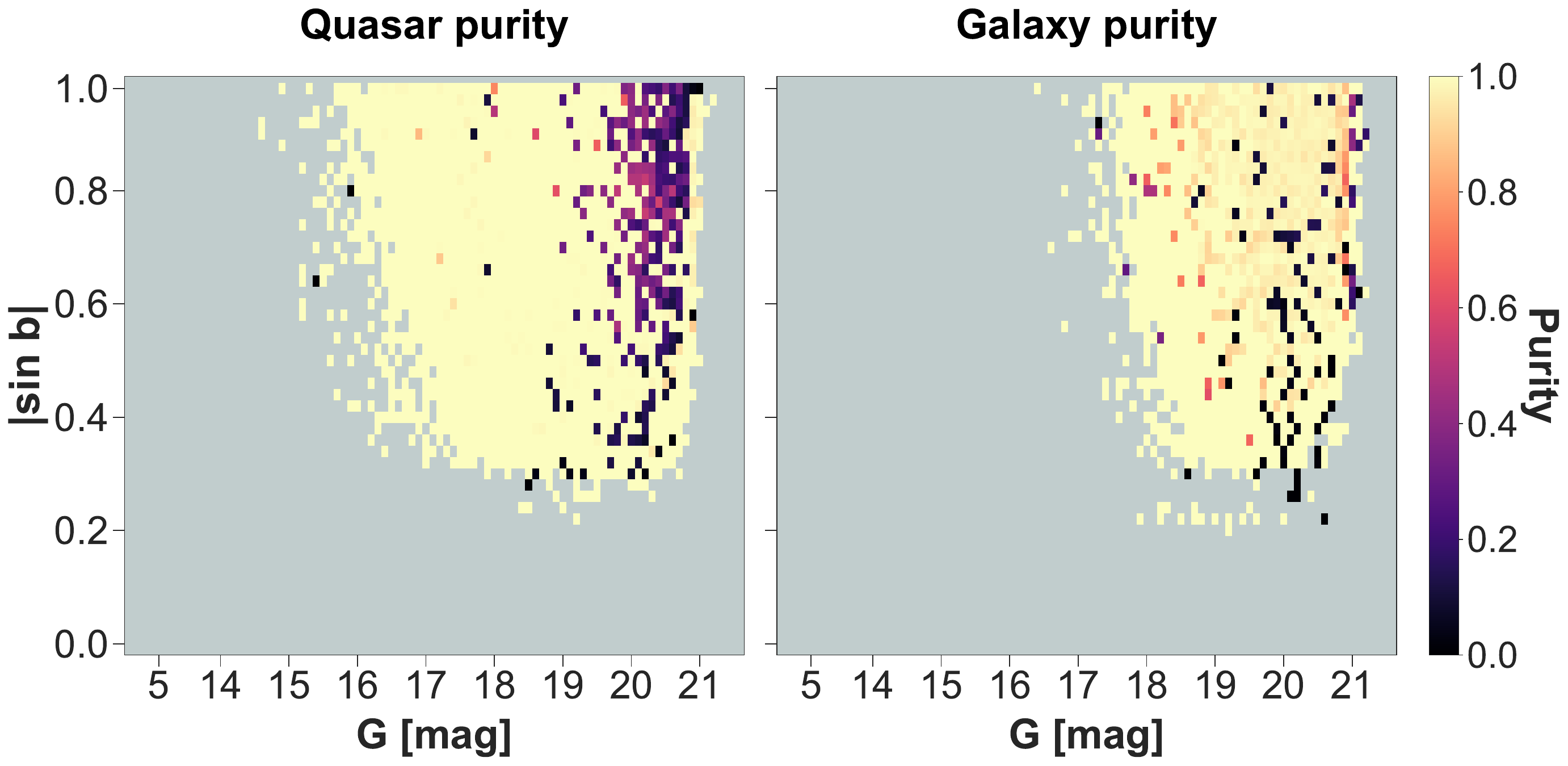}
    		\caption{} 
    		\label{fig:2d_updateddata_newcombmod2_variableprior}
	\end{subfigure}
	
	\caption{
		Representation in 2D of DSC classification results of the \texttt{Combmod} classifier compared to new combination of the DSC classifiers \texttt{Combmod}-$\alpha$ evaluated on the {{cleaned}} validation data set as a function of Galactic latitude and brightness. In each subplot, left and right two panels refer to the completenesses and the purities of the extragalactic classes, respectively. 
	{The number of sources per target class in the cleaned validation set are reported in Appendix Table \ref{table:ref_counts}}.
	\textit{(a)} \texttt{Combmod} using the global prior
			(average 2D completenesses of 92\% and 95\%, 
			and 2D purities of 55\% and 88\%, for the quasar and galaxy classes, respectively).
	\textit{(b)} \texttt{Combmod}-$\alpha$ using the global prior
			(average 2D completenesses of 82\% and 93\%, 
			and 2D purities of 79\% and 93\%).
	\textit{(c)} \texttt{Combmod} using the variable prior 
			(average 2D completenesses of 95\% and 96\%, 
			and 2D purities of 57\% and 86\%).
	\textit{(d)} \texttt{Combmod}-$\alpha$ using the variable prior 
			(average 2D completenesses of 89\% and 94\%, 
			and 2D purities of 74\% and 90\%).
	}

	\label{fig:2d_alldata_comparison}

\end{figure*}

\begin{figure*}[htp!]
	\centering
	\begin{subfigure}{1\textwidth}
	\raggedleft 
    	\includegraphics[scale=0.28, trim={0 0 0 0}, clip]{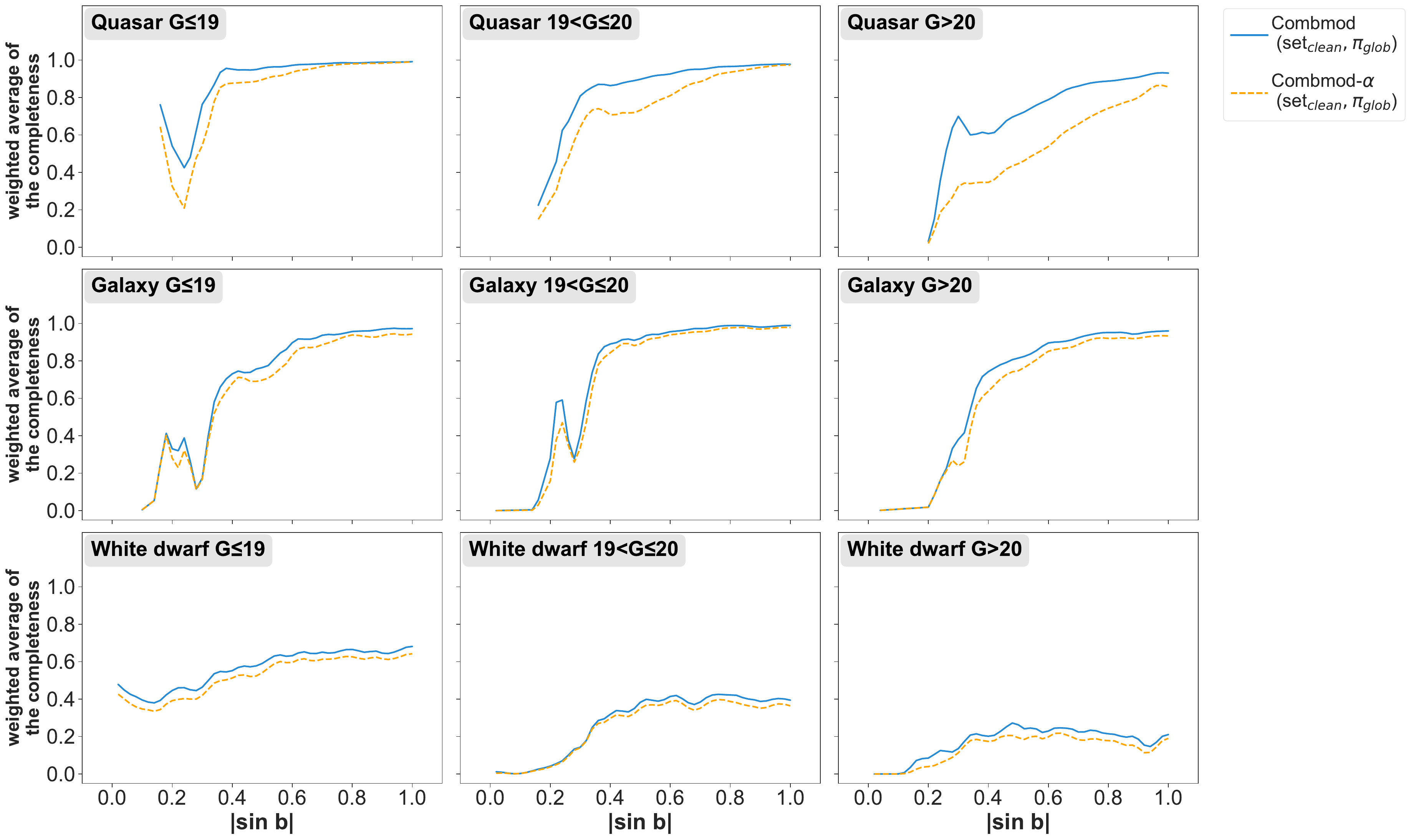}
	\caption{}
	\end{subfigure}
	
	\vspace{0.5cm} 
	\begin{subfigure}{1\textwidth}
	\raggedleft
    	\includegraphics[scale=0.28, trim={0 0 0 0}, clip]{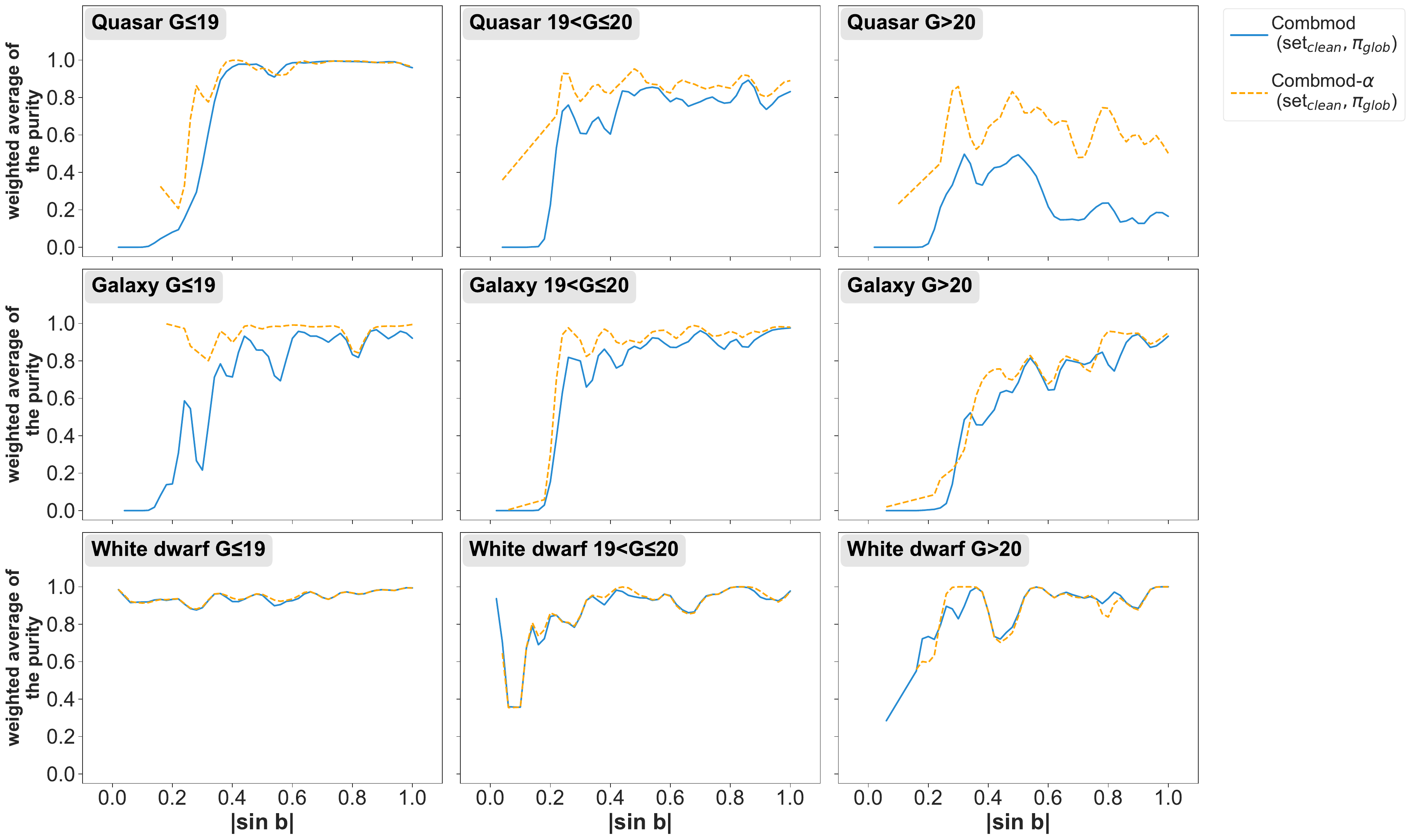}
	\caption{}
	\end{subfigure}
	
	\caption{
		As Figures \ref{fig:1dsmooth_0a_global} and \ref{fig:1dsmooth_0b_global}, but comparing the DSC \texttt{Combmod} and the new classifier \texttt{Combmod}-$\alpha$ using the global prior on the clean validation data set.
		In each panel the solid line shows the performance of the \texttt{Combmod} and the dash line the \texttt{Combmod}-$\alpha$, using the global prior. 
		\textit{(a)} Smoothed average of the 2D completenesses.
		\textit{(b)} Smoothed average of the 2D purities.
	}
	\label{fig:1dsmooth_2_global}
\end{figure*} 
\begin{figure*}[htp!]
	\centering
	\begin{subfigure}{1\textwidth}
	\raggedleft 
    	\includegraphics[scale=0.28, trim={0 0 0 0}, clip]{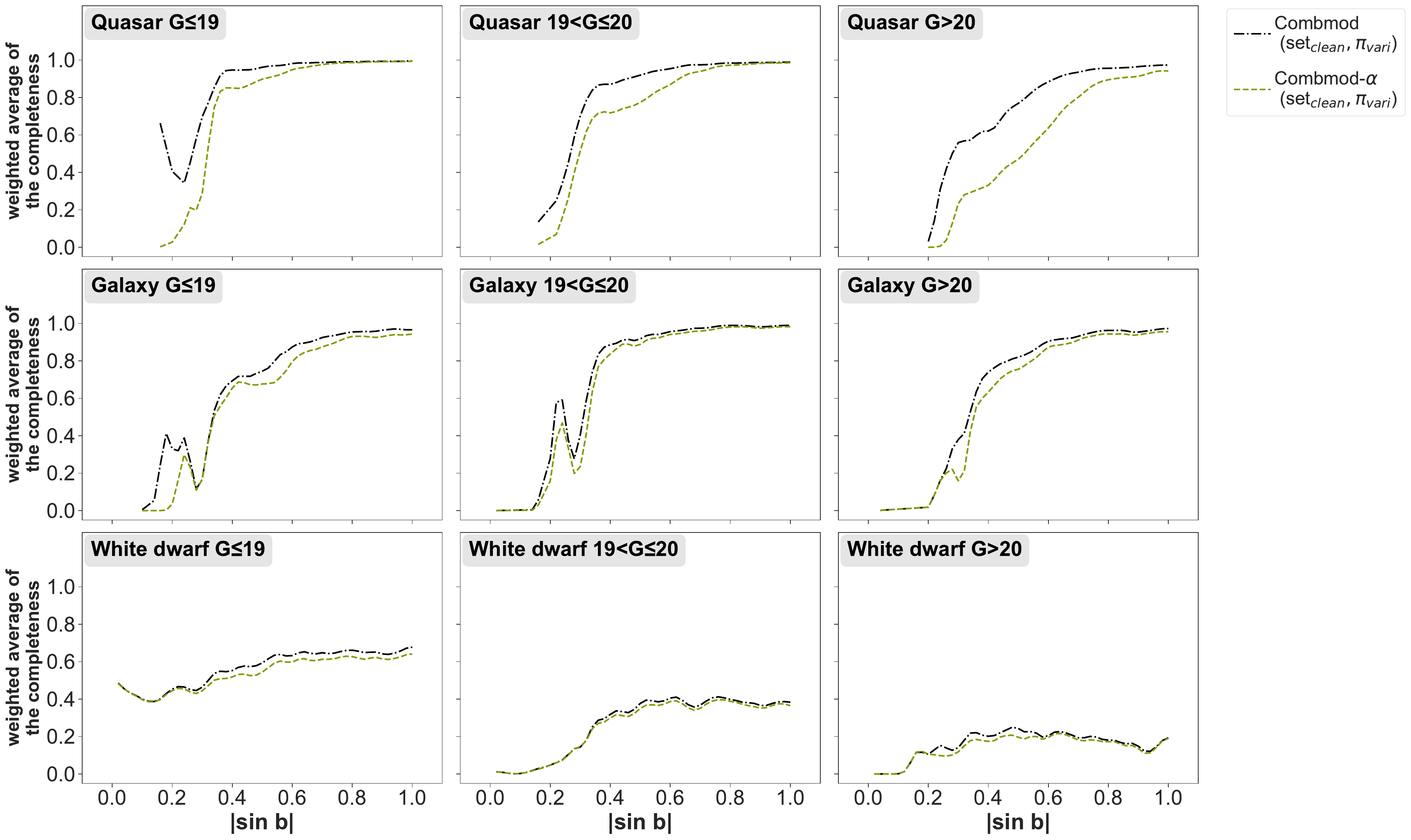}
	\caption{}
	\end{subfigure}
	
	\vspace{0.5cm} 
	\begin{subfigure}{1\textwidth}
	\raggedleft 
    	\includegraphics[scale=0.28, trim={0 0 0 0}, clip]{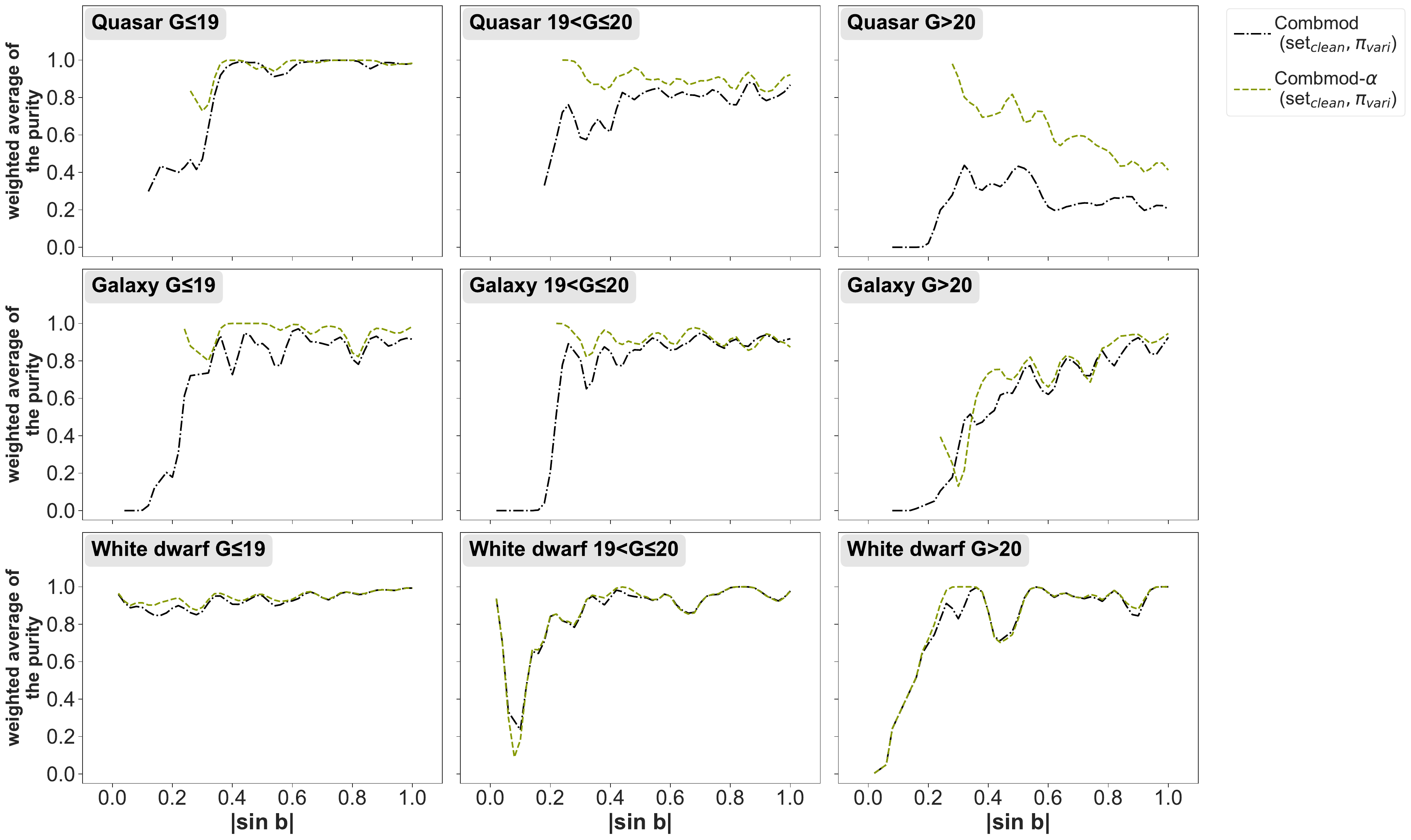}
	\caption{}
	\end{subfigure}
	
	\caption{
	As Figure \ref{fig:1dsmooth_2_global}, but using the variable prior instead of the global prior.
	In each panel the dash dot line shows the performance of the \texttt{Combmod} and the dash line the \texttt{Combmod}-$\alpha$, using the variable prior. 
		\textit{(a)} Smoothed average of the 2D completenesses.
		\textit{(b)} Smoothed average of the 2D purities.
	\vspace{1cm}
	}
	\label{fig:1dsmooth_2_variable}
\end{figure*}

\subsubsection{Performance}\label{seclperformance_alpha}

For different configurations of the $\{\alpha_k\}$, the configuration with the highest geometric score ($\sim$0.76) is obtained for $\alpha_{\rm quasar}$$\geq$0.7, $\alpha_{\rm galaxy}$=0.6 and $\alpha_{\rm star}$=0.5. 
This configuration can be interpreted as boosting \texttt{Allosmod} predictions for the quasar and galaxy classes, while keeping the contribution from both classifiers for the stellar classes equal.
A similar evaluation of different $\alpha_{k}$ combinations of \texttt{Allosmod} and \texttt{Specmod} using the variable prior, where we identify the best configuration with the highest geometric score ($\sim$0.74) for $\alpha_{\rm quasar}$$\geq$0.5, $\alpha_{\rm galaxy}$=0.6 and $\alpha_{\rm star}$=0.5.
The best classifiers \texttt{Combmod}-${\alpha}$ correspond to $\alpha_{\rm glob}$=[1.0, 0.6, 0.5] using the global prior and $\alpha_{\rm vari}$=[0.6, 0.6, 0.5] using the variable prior.
{
The Appendix Figures \ref{fig:geoscoreSum}-\ref{fig:geoscoreSum_varprior} summarise the results obtained for different parametric combinations of \texttt{Allosmod} and \texttt{Specmod} using the global prior and the variable prior, respectively.
}
%

The overall performance evaluated on the {{cleaned}} validation data set (as described in Section \ref{subsec:cleanedvalset}) is summarised in Table \ref{table:tab_GDR3_and_newcombmod_summary} under the label `\texttt{Combmod}-$\alpha$'.
{Confusion matrices are provided in the Appendix Figures \ref{fig:confmat_updateddataset_globalprior_newcombmod2} and \ref{fig:confmat_updateddataset_variableprior_newcombmod2} for reference.}
%
The 2D performance of the \texttt{Combmod}-$\alpha$ classifier is compared to the GDR3 \texttt{Combmod} in Figure \ref{fig:2d_alldata_comparison}.
{To facilitate the comparison, we also compute 1D representations of the completenesses and purities as a function of Galactic latitude in Figures \ref{fig:1dsmooth_2_global}-\ref{fig:1dsmooth_2_variable}.}
%
Table \ref{table:tab_GDR3_and_newcombmod_2dsummary} summarises the 2D performance evaluated on the {{cleaned}} validation data set.
\texttt{Combmod}-$\alpha$ using the global prior achieves 
	average 2D purities for the quasar and galaxy classes of 79\% and 93\%,
	compared to 55\% and 88\% in the GDR3 \texttt{Combmod}.
	The average 2D completenesses for the quasar and galaxy classes are 82\% and of 93\%, respectively, 
	compared to 92\% and 95\% in the GDR3 \texttt{Combmod}.
Similarly, \texttt{Combmod}-$\alpha$ using the variable prior reports higher 2D completenesses for the quasars (an increase by 7 percentage points) but reduced 2D purities for the extragalactic classes (a drop by 3 percentage points).
In Table \ref{table:tab_GDR3_and_newcombmod_2dsummary_faint_bright}, the comparison at different magnitude limits shows that the highest increase in purity is obtained at fainter magnitudes.
For quasars at ${G}$$\geq$20, 
	\texttt{Combmod}-$\alpha$ using the global prior increases the average 2D purities by 42 percentage points but reduces the average 2D completenesses by 15 percentage points.
Similarly, \texttt{Combmod}-$\alpha$ using the variable prior increases the 2D purities by 31 percentage points but reduces the average 2D completenesses by 5 percentage points.
At magnitudes ${G}$<20, the new combined classifiers show a similar performance across all target classes, with a drop in the 2D completenesses by 2-5 percentage points and an increase in the 2D purities by 2-5 percentage points.

{
To summarise, the new combined classifier improves the purity of the extragalactic classes, particularly the quasar class (Table \ref{table:tab_GDR3_and_newcombmod_2dsummary}).
Compared to the GDR3 \texttt{Combmod}, \texttt{Combmod}-$\alpha$ increases the average 2D purities by 23 percentage points and 18 percentage points using the global prior and the variable prior, respectively.
In contrast, the new combined classifier reduces the average 2D completenesses by 10 percentage points and 3 percentage points using the global prior and the variable prior, respectively. 
Therefore, the new classifier achieves a significant improvement in purity for a small loss of completeness.
}

\begin{figure*}[htp!]
	\centering
	\begin{subfigure}{0.485\textwidth}
		\centering
		\includegraphics[scale=0.45, trim={0 0 0 0.8cm},clip]{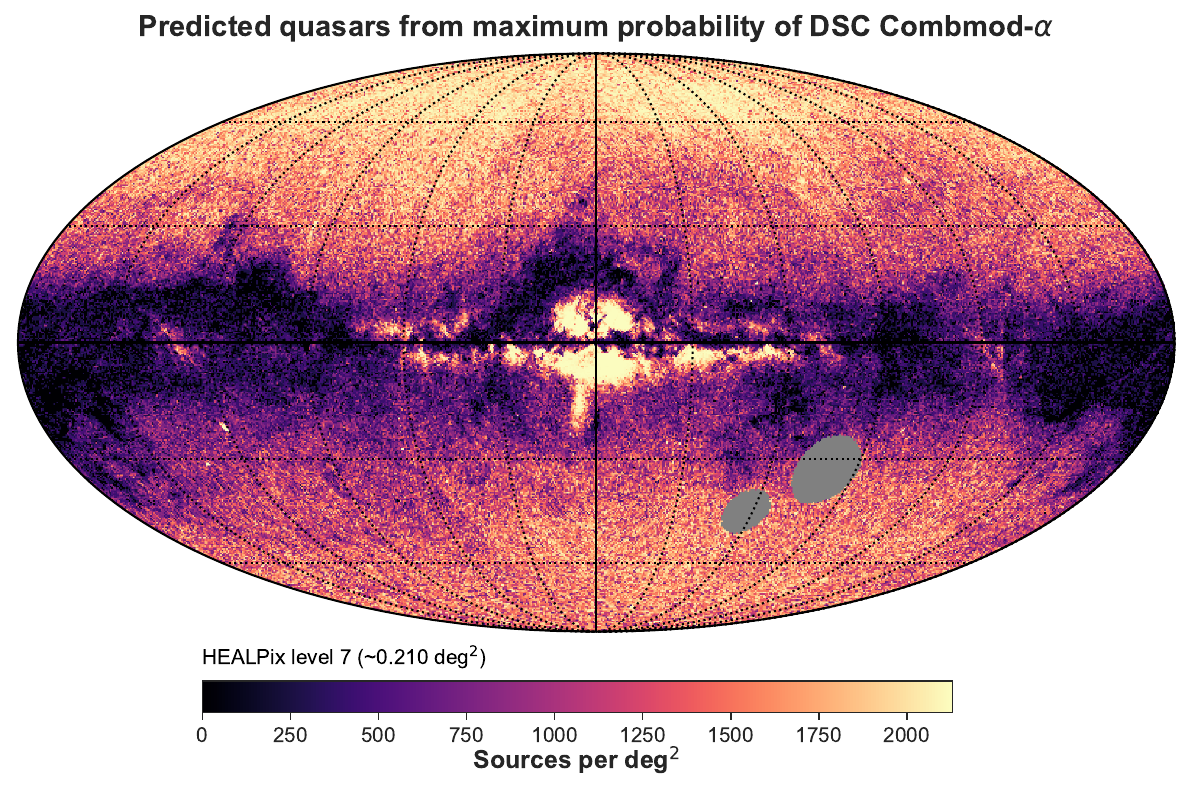}
		\caption{}
		\label{fig:gacs_combmod_combmodalpha_quasars_A}
	\end{subfigure}
	\hspace{0.2cm}
	\begin{subfigure}{0.485\textwidth}
		\centering
		\includegraphics[scale=0.45, trim={0 0 0 0.8cm},clip]{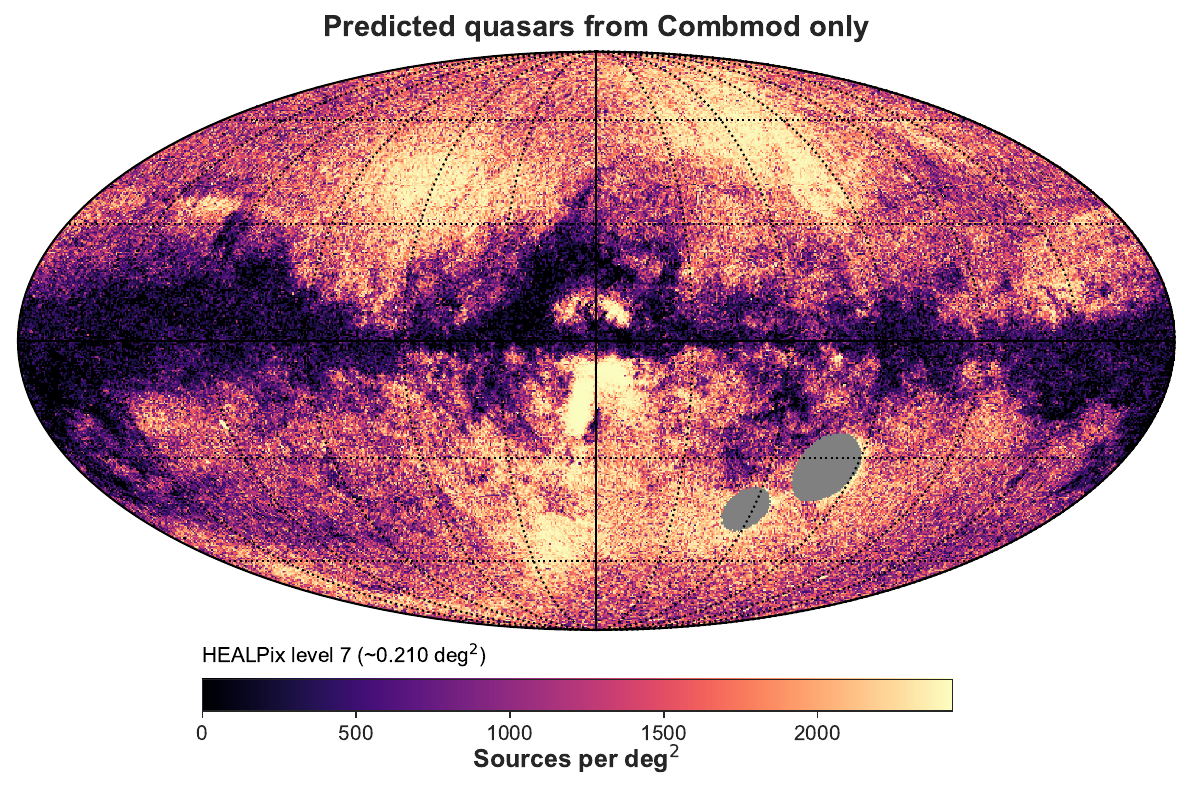}
		\caption{}
		\label{fig:gacs_combmod_combmodalpha_quasars_B}
	\end{subfigure}
	
	\begin{subfigure}{0.485\textwidth}
		\centering
		\includegraphics[scale=0.45, trim={0 0 0 0.8cm},clip]{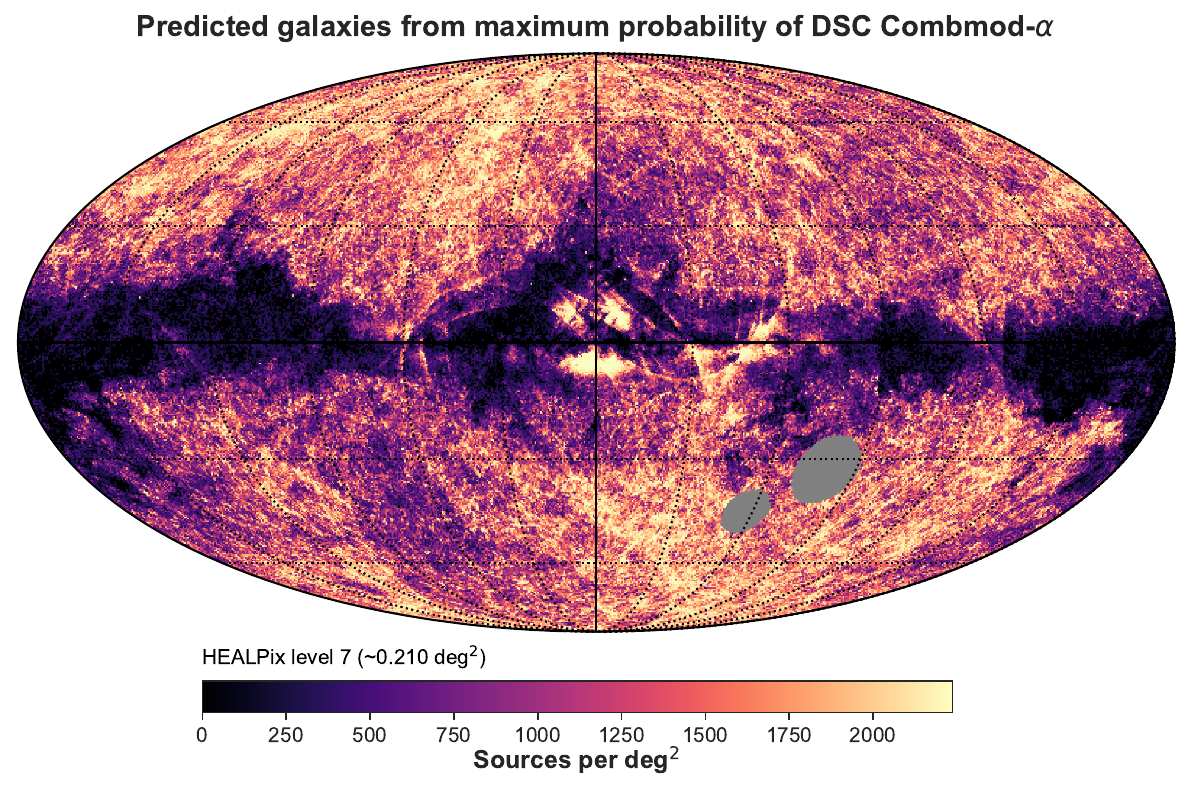}
		\caption{}
		\label{fig:gacs_combmod_combmodalpha_galaxies_A}
	\end{subfigure}
	\hspace{0.2cm}
	\begin{subfigure}{0.49\textwidth}
		\centering
		\includegraphics[scale=0.45, trim={0 0 0 0.8cm},clip]{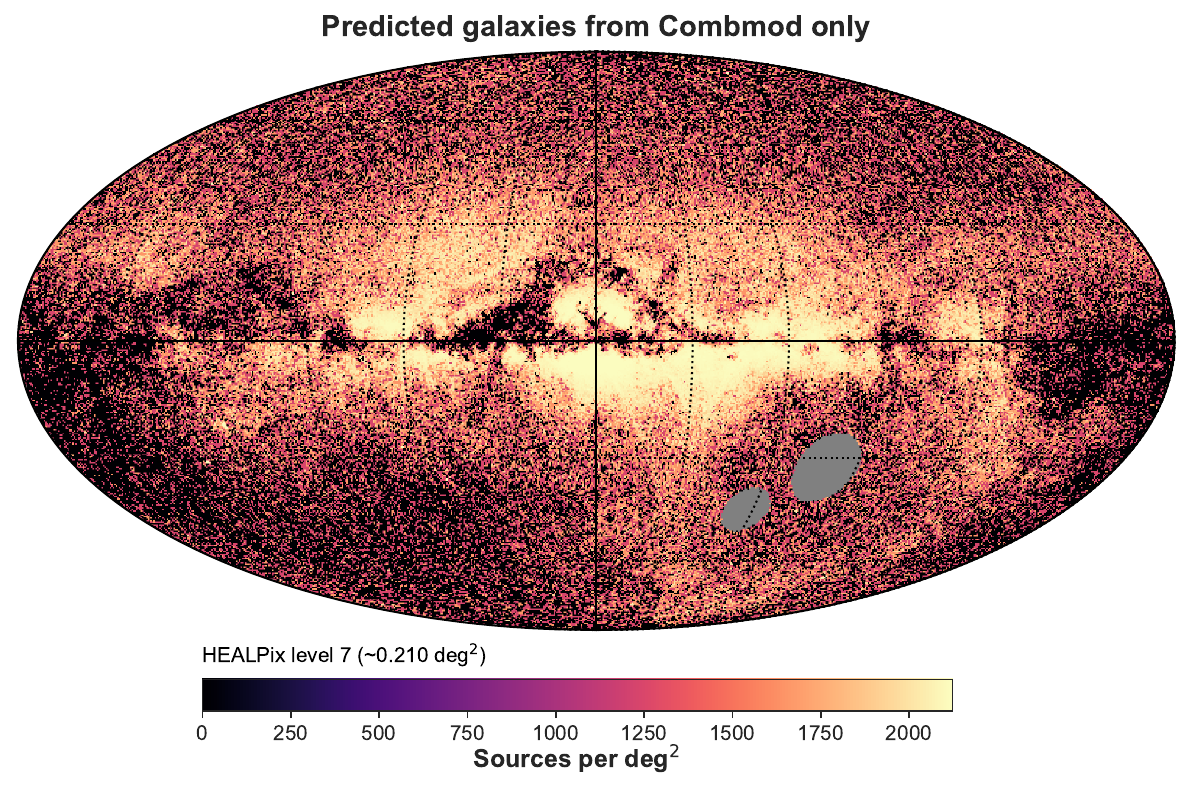}
		\caption{}
		\label{fig:gacs_combmod_combmodalpha_galaxies_B}
	\end{subfigure}
	
	\caption{
		Galactic sky distribution of sources classified from maximum probabilities 
		as quasars (top row) and galaxies (bottom row) 
		by \texttt{Combmod} and \texttt{Combmod}-$\alpha$ using the global prior 
    		at HEALpixel level 7 in Mollweide projection. 
		\textit{(a)} Quasar candidates identified by \texttt{Combmod}-$\alpha$ (1\,719\,638 sources).
		\textit{(b)} Quasar candidates identified only by \texttt{Combmod} (1\,230\,071 sources).
		\textit{(c)} Galaxy candidates identified by \texttt{Combmod}-$\alpha$ ({2\,971\,433} sources).
		\textit{(d)} Galaxy candidates identified only by \texttt{Combmod} (459\,835 sources).
		%
    		{The LMC and SMC regions are masked in grey.} 
	}
	\label{fig:gacs_combmod_combmodalpha_quasars_galaxy}
\end{figure*} 

\begin{figure*}[htp!]
\begin{minipage}[t]{1.\textwidth}  
	\centering
	\begin{subfigure}{1\textwidth}
	\centering
	\includegraphics[scale=0.40]{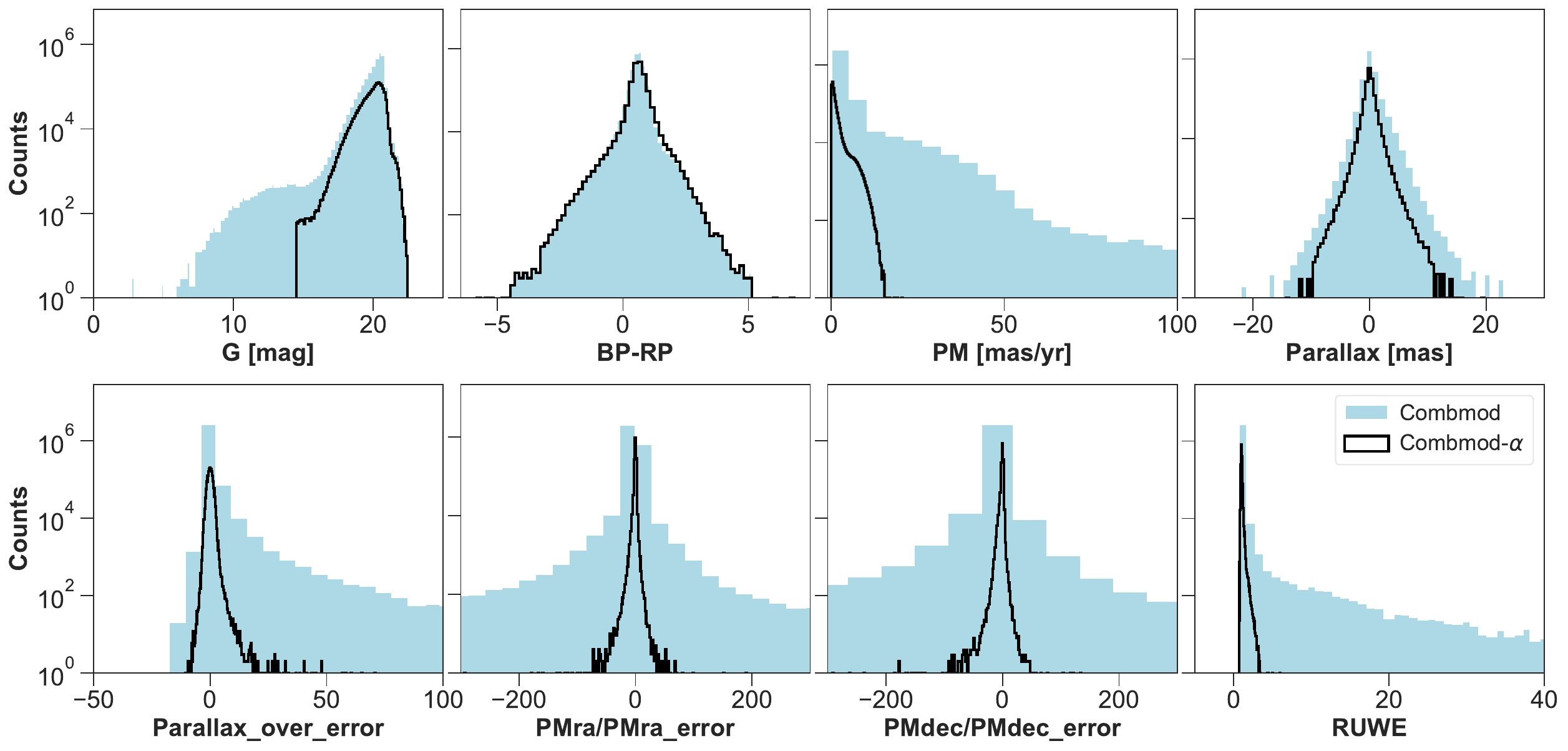}	
	\caption{}
	\end{subfigure}
	
	\vspace{0.5cm}
	\begin{subfigure}{1\textwidth}
	\centering
	\includegraphics[scale=0.40]{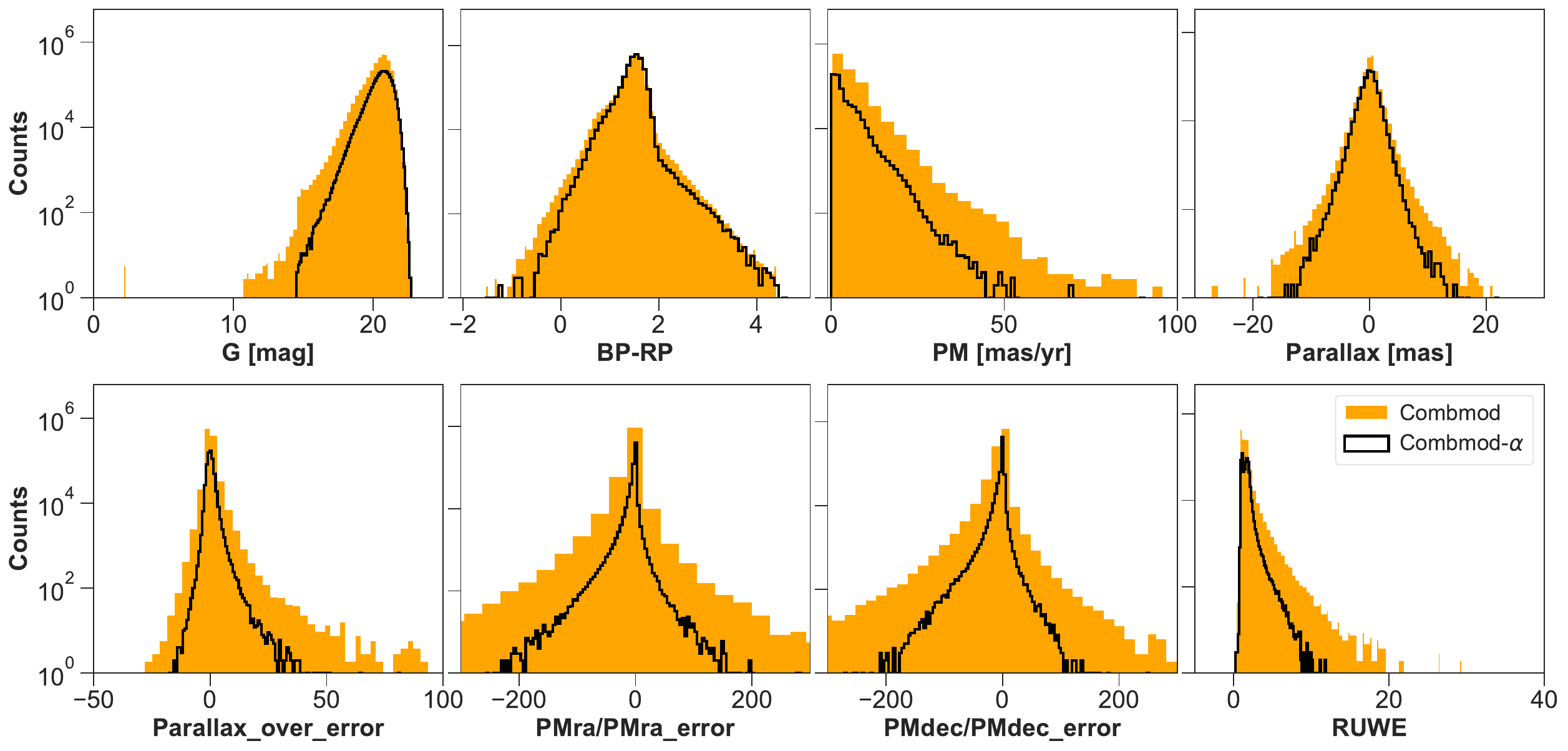}	
	\caption{}
	\end{subfigure}
	
	\caption{ 
	Features distributions of extragalactic candidates from maximum probabilities   
	classified by the GDR3 \texttt{Combmod} and \texttt{Combmod}-$\alpha$ using the global prior.
	\textit{(a)} Quasar candidates.
	\textit{(b)} Galaxy candidates.
	Excluding the LMC and SMC regions, class predictions from maximum probabilities of \texttt{Combmod}-$\alpha$ give 
		1\,719\,638 quasars and 2\,971\,433 galaxies
    		compared to 2\,909\,081 quasars and 3\,419\,626 galaxies classified by the GDR3 \texttt{Combmod}.
	{\texttt{Combmod} results are indicated in (a) blue for quasars and (b) orange for galaxies. 
	\texttt{Combmod}-$\alpha$ results are showed in black.}
    	}
	\label{fig:gacs_combmod_combmodalpha_quasars_galaxy_features}
	\end{minipage} 
\end{figure*} 

\subsubsection{Application to all of GDR3}\label{sec:combalpha_gdr3}

We now compute \texttt{Combmod}-$\alpha$ probabilities for {the $\sim$1.4 billion sources with} published\footnote{\url{https://gea.esac.esa.int/archive/}} GDR3 DSC \texttt{Specmod} and \texttt{Allosmod} probabilities.

{Table \ref{tab:ref_counts_gacs} reports the total numbers of the extragalactic candidates classified by \texttt{Combmod} and \texttt{Combmod}-$\alpha$ excluding the LMC and the SMC.}
Figures \ref{fig:gacs_combmod_combmodalpha_quasars_A} and \ref{fig:gacs_combmod_combmodalpha_galaxies_A} shows the Galactic sky distributions of the quasar and galaxy candidates obtained from maximum probabilities of \texttt{Combmod}-$\alpha$ using the global prior. 
%
%

\begin{table}[htp!]
   \caption{ 
   	Quasar and galaxy candidates classified by 
  	 GDR3 DSC \texttt{Combmod} and \texttt{Combmod}-$\alpha$ excluding the LMC and the SMC.}           
    \label{tab:ref_counts_gacs}  
    \small \centering    
    \setlength{\tabcolsep}{5pt}                        
    \begin{tabular}{l | r r }         
        \hline\hline                  
        		 \multirow{2}{6em}{\bf \centering Predicted by}  
		 & \multicolumn{1}{r}{\bf quasar} & \multicolumn{1}{r}{\bf galaxy} \\
		 &  \multicolumn{1}{r}{\bf candidates} & \multicolumn{1}{r}{\bf candidates} \\  
    	\hline\hline  
                	\texttt{Combmod} 					& 	2\,909\,081 	&	3\,419\,626 	 \\ 
                	\texttt{Combmod}-$\alpha $ 			& 	1\,719\,638 	&	2\,971\,433 	 \\ 
	\hline
                	common predictions  				& 	1\,679\,010 	&	2\,959\,791 	 \\ 
                	exclusive to \texttt{Combmod}  			& 	1\,230\,071 	&	459\,835 	 \\ 
                	exclusive to \texttt{Combmod}-$\alpha$  	& 	40\,628 		&	11\,642 	 	\\ 
	\hline 
    \end{tabular}
    \tablefoot{
    	 First two rows provide the total numbers of candidates from each classifier. 
  	 Last three rows summarise the common sources and the mismatched predictions between the classifiers.
	 }
\end{table}

Across the full sky, \texttt{Combmod}-$\alpha$ 
	predicts 2\,041\,979 quasars and 3\,031\,885 galaxies, 
	compared to 5\,254\,959 quasars and 3\,575\,099 galaxies from the GDR3 \texttt{Combmod}.
We exclude the LMC and SMC regions from this analysis due to the large stellar contamination, {and find}  
	1\,719\,638 quasars and 2\,971\,433 galaxies classified by \texttt{Combmod}-$\alpha$, 
	compared to 2\,909\,081 quasars and 3\,419\,626 galaxies classified by the GDR3 \texttt{Combmod}.
The difference in numbers in the classifications between the two classifiers is due, on one hand to the improved purity of \texttt{Combmod}-$\alpha$ rejecting far more stellar contaminants, and on the other hand to the bright magnitude limit in \texttt{Combmod}-$\alpha$.
In GDR3, DSC \texttt{Allosmod} assigne{s} a star label to all sources brighter than ${G}$=14.5. By design, \texttt{Combmod}-$\alpha$ inherits this {characteristic}.

The distributions of Gaia astrometry and photometry for the quasar and galaxy candidates are reported in Figure \ref{fig:gacs_combmod_combmodalpha_quasars_galaxy_features}. 
Compared to the GDR3 \texttt{Combmod}, the features distributions in the \texttt{Combmod}-$\alpha$ extragalactic candidates show a smaller scatter suggesting a lower level of stellar contaminants that is even more reduced for quasars.
In general, the distribution of normalised parallaxes and proper motions (PM) is expected to agree with noise for quasars and follow a normal distribution with a zero mean and a unit variance. 
For the \texttt{Combmod}-$\alpha$ quasars, the normalised parallaxes are closer to a normal distribution compared to the GDR3 \texttt{Combmod} quasars. 
The colour distribution shows no significant difference between the \texttt{Combmod} and \texttt{Combmod}-$\alpha$ quasars, only bluer galaxies (lower $G_{\rm BP}-G_{\rm RP}$) are filtered out by \texttt{Combmod}-$\alpha$. 
The G-band magnitude distribution of classified quasars and galaxies by \texttt{Combmod}-$\alpha$ is restricted to magnitudes fainter than 14.5, which corresponds to the magnitude limit set by \texttt{Allosmod} in GDR3. In the current configuration, \texttt{Combmod}-$\alpha$ uses exclusively the GDR3 \texttt{Allosmod} for quasars ($\alpha_{\rm quasar}$=1). 
Despite reducing the parameter $\alpha_{\rm quasar}$ from 1 to 0.9, so that \texttt{Specmod} contributes for the quasar class, we obtain similar results in the features distributions. This means the majority of bright sources at ${G}$=14.5 classified by \texttt{Specmod} as quasar or galaxy candidates are likely stellar contaminants now rejected by \texttt{Combmod}-$\alpha$.

{
The quasar and galaxy candidates classified as such only by \texttt{Combmod} and not by  \texttt{Combmod}-$\alpha$ (1\,230\,071 quasars and 459\,835 galaxies) are reported in Figures \ref{fig:gacs_combmod_combmodalpha_quasars_B}-\ref{fig:gacs_combmod_combmodalpha_galaxies_B}. Their sky distribution shows an excess of sources close to the Galactic plane, especially for galaxies. 
We also notice a higher density close to the LMC in the distribution of the quasar candidates of the GDR3 \texttt{Combmod}.
From the performance of the new combination discussed in Section \ref{seclperformance_alpha}, a large fraction of the $\sim$1.2 million sources classified exclusively by \texttt{Combmod} are likely stellar contaminants now rejected by \texttt{Combmod}-$\alpha$.
}

To summarise, compared to the GDR3 \texttt{Combmod}, the DSC extragalactic candidates from the new combination of the GDR3 \texttt{Specmod} and \texttt{Allosmod} constitute a purer sample, although incomplete, up to magnitudes of ${G}$=14.5.

\section{Conclusions}\label{sec:conclusions}

We have reassessed the performance of the DSC classification results in GDR3, and use a new combination of the \texttt{Specmod} and \texttt{Allosmod} classifiers to significantly improve the purity of the extragalactic classifications.

Classification results summarised through global metrics do not reflect how the performance of a classifier depends on the properties of the sources. We have therefore assessed the completeness and purity as a function of magnitude and Galactic latitude.
This reveals a significant variation in the performance allowing us, among other things, to identify regions most affected by stellar contamination. 
Unsurprisingly, completeness and purity are lowest at the faintest magnitudes.
%
In Table \ref{table:tab_GDR3_and_newcombmod_2dsummary}, we report the 2D completenesses and purities (i.e.\ averaged over magnitude and latitude bins), excluding the Magellanic Clouds where the contribution of the stellar contaminants heavily affects the purity.
Over all magnitudes and latitudes, the
	average 2D completenesses 
 for quasars, galaxies, and white dwarfs are 92\%, 95\%, and 43\%, and  
	the average 2D purities are 55\%, 88\%, and 94\%,
respectively.  This is a significant improvement over what was published and claimed in GDR3. This improvement comes primarily from the cleaned validation set. 
For ${G}$<20 the average 2D purities are 86\% for quasars and 92\% for galaxies; for ${G}$$\geq$20 the average 2D purities is 20\% for quasars and 83\% for galaxies (Table \ref{table:tab_GDR3_and_newcombmod_2dsummary_faint_bright}).

We have also examined the impact of a variable prior on the results. Rather than a single number reflecting the prior probability of a source being a quasar (for example), we modified this to be a function of $G$ magnitude and Galactic latitude. 
Compared to GDR3 \texttt{Combmod}, the application of the variable prior improves the 2D completenesses and purities of the quasars, galaxies, and white dwarfs only by a small amount (2-3 percentage points), mainly at fainter magnitudes.

More significantly, we introduced a new additive parametric combination of the GDR3 DSC baseline classifiers \texttt{Specmod} and \texttt{Allosmod}. 
The new combined classifier, which we call \texttt{Combmod}-$\alpha$, increases the overall purities by about 20 percentage points for the extragalactic classes, at the cost of only a small decrease in completeness (about 10 and 2 percentage points for the quasars and galaxies, respectively). 
\texttt{Combmod}-$\alpha$ achieves,
for the quasar and galaxy classes, respectively,
average 2D completenesses of 82\% and 93\%, and average 2D purities of 79\% and 93\% using the global prior. 
Using the variable prior, the average 2D completenesses are 89\% and 94\%, and the average 2D purities are 74\% and 90\%.  
Even for faint sources, the improvement with \texttt{Combmod}-$\alpha$ is significant: at ${G}$$\geq$20 the average 2D purities for the quasar class are 62\% using the global prior, and 51\% using the variable prior. This compares to the average 2D purities of just 20\% in the GDR3 \texttt{Combmod}. 
The newly introduced combination of the baseline classifiers is therefore able to produce much purer catalogues of extragalactic objects using the existing GDR3 results, without having to train or apply new classifiers. 
 
We plan to implement the presented approaches -- the continuous variable prior and a parametric combination of the baseline classifiers -- in future versions of DSC for the next Gaia Data Release. 
We expect that a new combined classifier would reach a similar performance to those presented here to provide higher purity catalogues of extragalactic candidates.
We also publish our software implementing the continuous variable prior and new \texttt{Combmod}-$\alpha$ classifier applied to the GDR3 DSC results as open source software.
 
\begin{acknowledgements}
We would like to thank René Andrae, Morgan Fouesneau, and Ruth Carballo for their constructive comments and helpful suggestions during the realisation of this work.
{We sincerely thank the referee for their detailed comments and suggestions that helped improve the manuscript.}
 
This work has made use of data from the European Space Agency (ESA) mission {\it Gaia} (\url{https://www.cosmos.esa.int/gaia}), processed by the {\it Gaia} Data Processing and Analysis Consortium (DPAC, \url{https://www.cosmos.esa.int/web/gaia/dpac/consortium}). 
Funding for the DPAC has been provided by national institutions, in particular the institutions participating in the {\it Gaia} Multilateral Agreement.

This work was funded in part by the DLR (German space agency) via grant 50 QG 2102.

\end{acknowledgements}

\bibliographystyle{aa} 
\bibliography{bib_1_abbrev_nopublisher} 

\begin{appendix} 

\section{Validation data set}
{
This section summarises the validation data set used in the study.
Table \ref{table:ref_xmatch} provides the list of the cross match catalogues used to identify mislabelled sources in the stellar classes (star, white dwarf and binary star classes) of the validation data set. The selection criteria are applied to the best match (minimal crossmatch radius of 1 arcsecond). 
Table \ref{table:ref_counts} reports a brief summary of the number of sources per target class after removal of the identified contaminants (i.e. mislabelled sources) and the Magellanic Clouds.
}
\begin{table}[htp!]
\begin{minipage}[t]{1\textwidth}  
   \caption{ 
   Summary of the cross-match catalogues and the criteria for contaminant removal from the stellar classes.}            
    \label{table:ref_xmatch}  
    \small \centering    
    \setlength{\tabcolsep}{10pt}                        
    \begin{tabular}{l | l }         
        \hline\hline                  
        		Catalogue & Selection criteria\\
    	\hline\hline  
                
                \multirow{2}{18em}{Milliquas v7.2-v8\\ \citep{flesch_half_2015, flesch_million_2023} }
                		& All candidates.  \\\\\\
		
                \multirow{2}{18em}{Assef R90 and C75 WISE AGN catalogues\\ \citep{assef_wise_2018} }
                		& All candidates. \\\\\\
		
                \multirow{2}{18em}{SDSS DR 17 photometry\\ \citep{abdurrouf_seventeenth_2022}} 
                		& Photometric classification \texttt{type=\{`GALAXY'\}}. \\\\\\
		
                \multirow{2}{18em}{LAMOST DR7\\ \citep{luo_vizier_2022}}
                		& Spectral type \texttt{Class=\{`QSO',`GALAXY',`UNKNOWN'\}}.  \\\\\\
		
                \multirow{2}{18em}{UKIDSS DR9\\ \citep{lawrence_ukirt_2007}}
                		&Source image profile classification \texttt{mergedClass=\{-3,0,1\}} \\
                		& referring to `probable galaxy', `noise', and `galaxy'.  \\\\\\
		
                \multirow{2}{18em}{DES DR2\\ \citep{abbott_dark_2021} }
                		& Classification from Sextractor measurements \texttt{ClsCoad=\{2,3\}} \\
                		&  and  from weighted average PSF magnitudes \texttt{ClWavg=\{2,3\}} \\
                		&  referring to `mostly galaxies' and `high-confidence galaxies'. \\\\\\
		
                \multirow{2}{16em}{SIMBAD\\ \citep{wenger_simbad_2000}}
                		& \texttt{main$\_$type} in \{`Galaxy',`Seyfert$\_$1',`Seyfert$\_$2',`BClG',\\
                    	& `ClG',`RadioG',`GinGroup',`GinCl',`GinPair',`EmG',`PairG',\\
                    	& `BlueCompG',`QSO',`QSO\_Candidate',`BLLac',`AGN',`HII',\\
                    	& `MolCld',`OpCl',`GlCl',`GlCl?$\_$Candidate',`NIR',`IR',`X',\\
                    	&`V*',`Planet',`Planet?$\_$Candidate',`Radio',`Radio(cm)', \\
                    	& `LensSystem$\_$Candidate',`multiple$\_$object',`Blue', \\
                    	& `PN?$\_$Candidate',`Inexistent',`Unknown$\_$Candidate'\}
                    \\\\
        \hline 
    \end{tabular}
    \end{minipage}
\end{table}

\begin{table}[htp!]
\noindent\hspace{.3\linewidth}
\begin{minipage}[c]{.7\textwidth}  
   \caption{ 
    Summary of the number of sources in the validation data set.}
    \label{table:ref_counts}  
    \small \centering    
    \setlength{\tabcolsep}{10pt}                        
    \begin{tabular}{l | r r }         
        \hline\hline                  
        		\multirow{2}{7em}{\centering Class} 
		 & \multicolumn{2}{c}{Validation data set}\\
		 \cline{2-3}
		 & \it original & \it cleaned\\
    	\hline\hline  
                	Quasar 			& 	308\,526 	&	308\,526 	 	\\ 
		Galaxy 			& 	  52\,359 	&	   51\,347 	 	\\
		Single star 		& 	717\,252 	&	702\,450 	 	\\
		White dwarf 		& 	  47\,245 	&	  45\,005 	 	\\
		Physical binary 		& 	 331\,526 & 	313\,827	 	\\
	\hline 
    \end{tabular}
    \tablefoot{The labels `original' and `cleaned' refer to the full validation data set and the {{cleaned}} validation data set, respectively. }
    \end{minipage}
\end{table}

\newpage ~\newpage
\section{DSC classification results in GDR3}
\begin{figure}[htp!]
\begin{minipage}[t]{1\textwidth}  
	\vspace{1cm}
	%
	\begin{subfigure}{.46\textwidth}
		\includegraphics[scale=0.40, trim={0 0 0 1.25cm},clip]{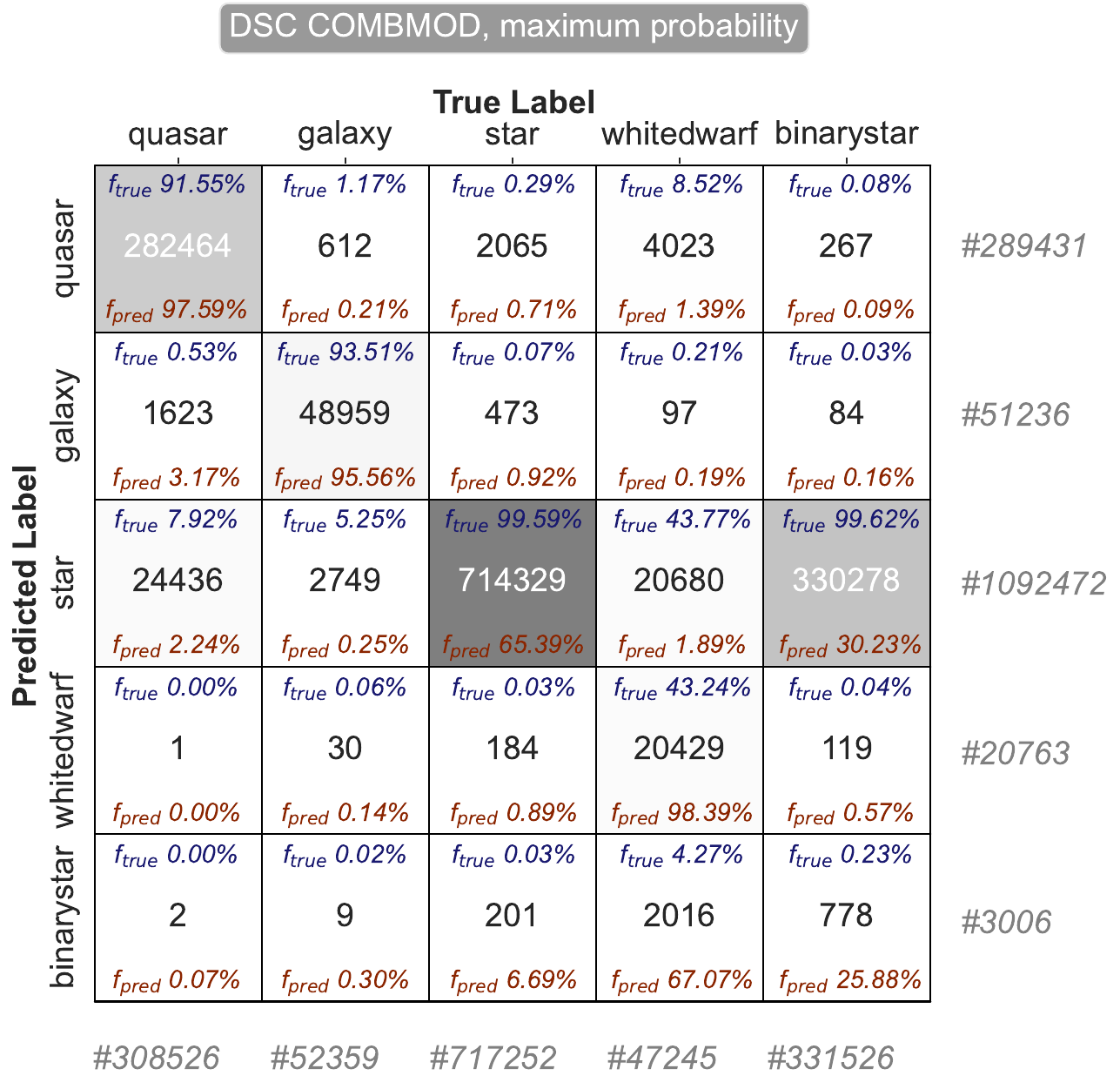}
		\caption{} 
		\label{fig:confmat_alldataset_globalprior}
	\end{subfigure}
	%
	\hspace{1cm} 
	\begin{subfigure}{.46\textwidth}
		\includegraphics[scale=0.40, trim={0 0 0 1.25cm},clip]{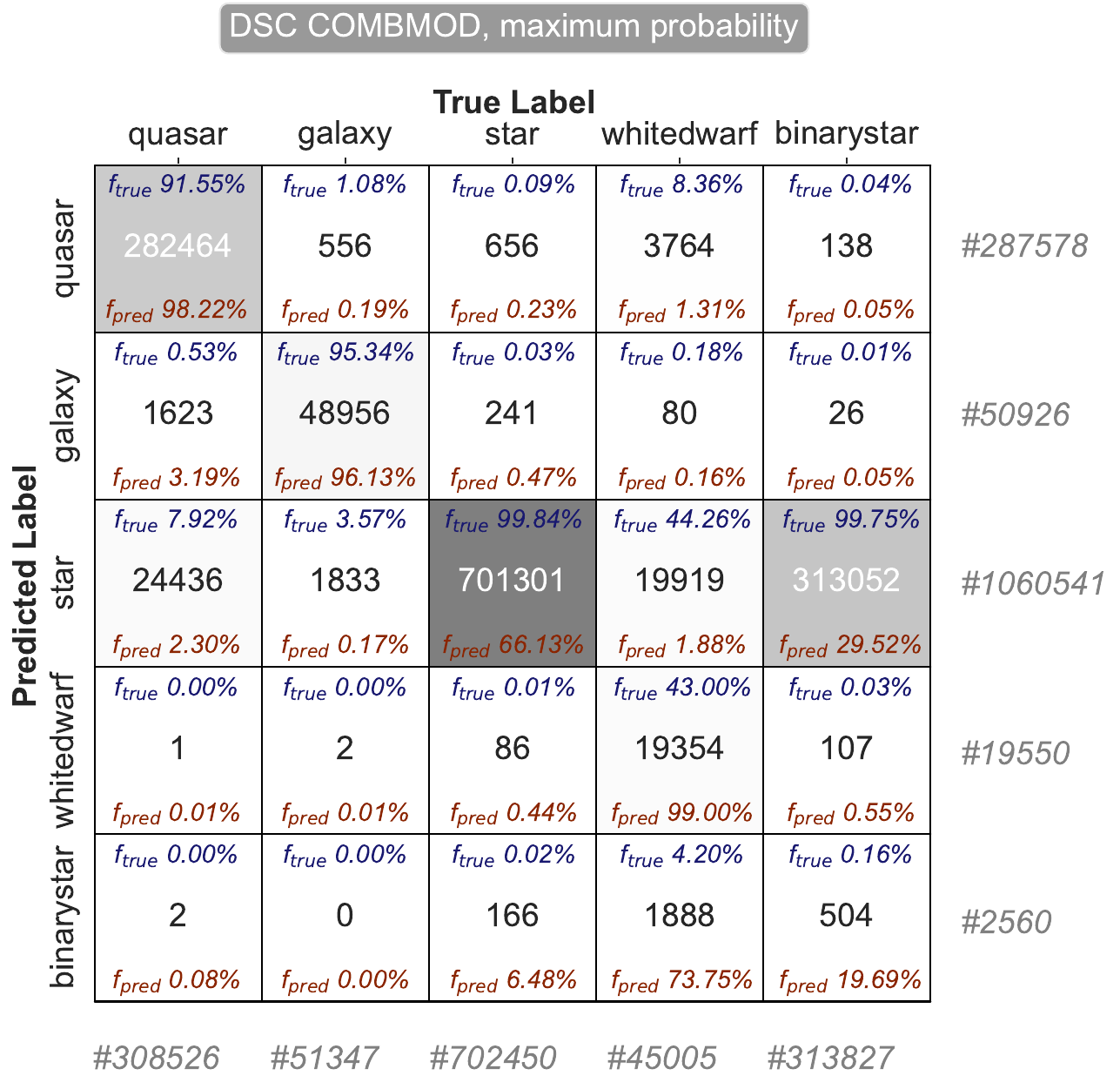}
		\caption{} 
		\label{fig:confmat_updateddataset_globalprior}	
	\end{subfigure}

	\caption{Unadjusted confusion matrices of DSC classifiers computed on the validation data set over the five classes. 
    	\textit{(a)} Confusion matrix computed over the full validation data set of 1\,456\,908 sources of DSC \texttt{Combmod} 
		using the global prior.
	\textit{(b)} Confusion matrix computed on the {{cleaned}} validation data set of 1\,421\,155 sources of DSC \texttt{Combmod}
		using the global prior.
	`Unadjusted' means that confusion matrices have not yet been adjusted to represent the expected class fractions.
	Indicated in blue colours are the fractions of the number of predictions per class normalised by the total number 
	of true labels in this class (last row in grey), 
	and in red colours the fractions of the predictions per class normalised by the total number of the predictions (last column in grey).
	The diagonal elements of the matrix in blue refer to the unadjusted completeness TP/(TP+FN) 
	and in red to the unadjusted purity TP/(TP+FP).
	{The colour scheme of the background is defined such that dark colours refer to higher number of sources and 
	white colours to lower values.}
	}
	\label{fig:confmats_1}
\end{minipage} 	
\end{figure}

\begin{figure}[htp!]
\begin{minipage}[t]{1\textwidth}  
	\centering	
	\hspace{1cm} 
	\begin{subfigure}{.46\textwidth}
		\includegraphics[scale=0.40, trim={0 0 0 1.25cm},clip]{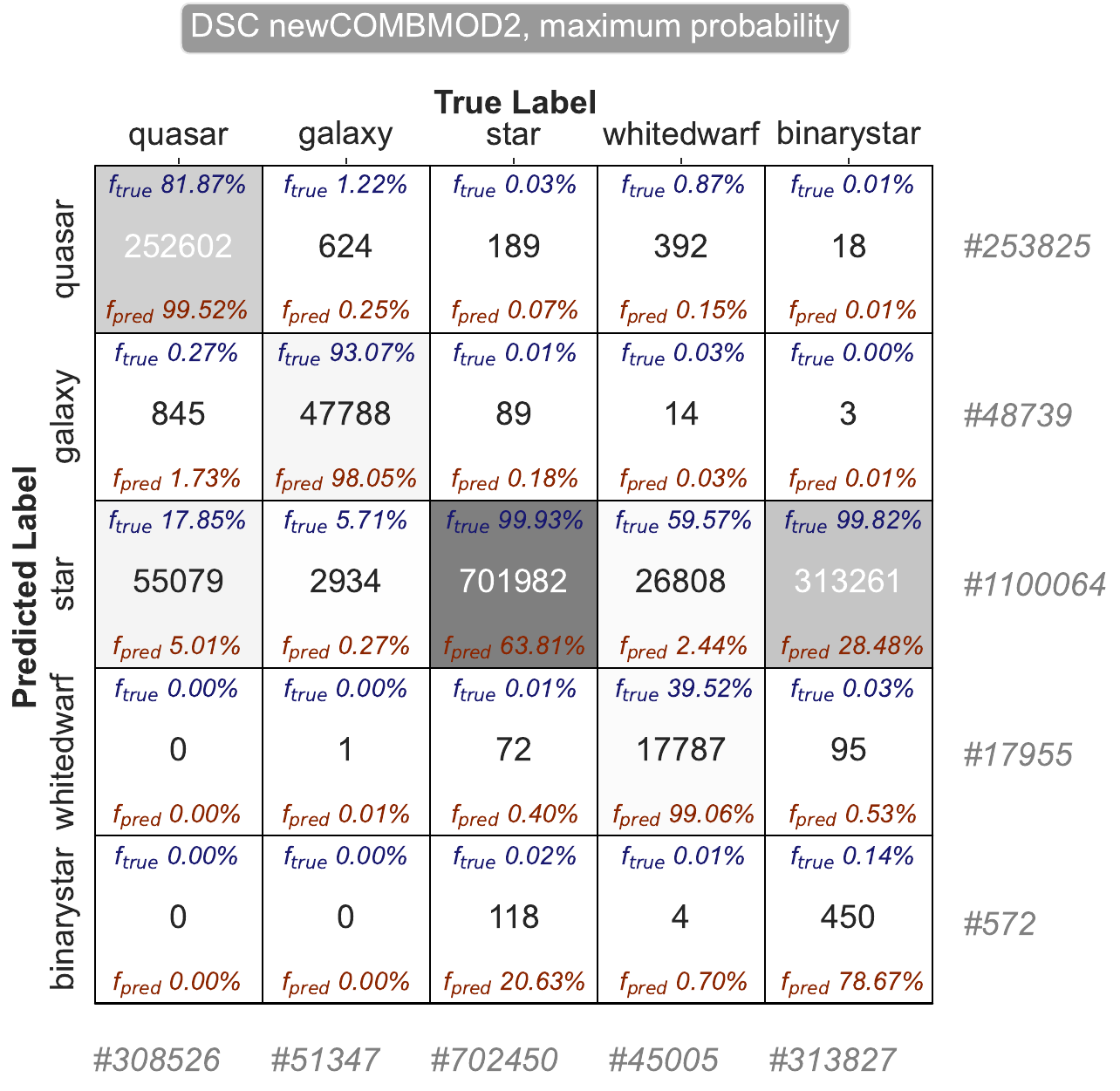}
		\caption{} 
		\label{fig:confmat_updateddataset_globalprior_newcombmod2}	
	\end{subfigure}
	
	\begin{subfigure}{.46\textwidth}
		\includegraphics[scale=0.40, trim={0 0 0 1.25cm},clip]{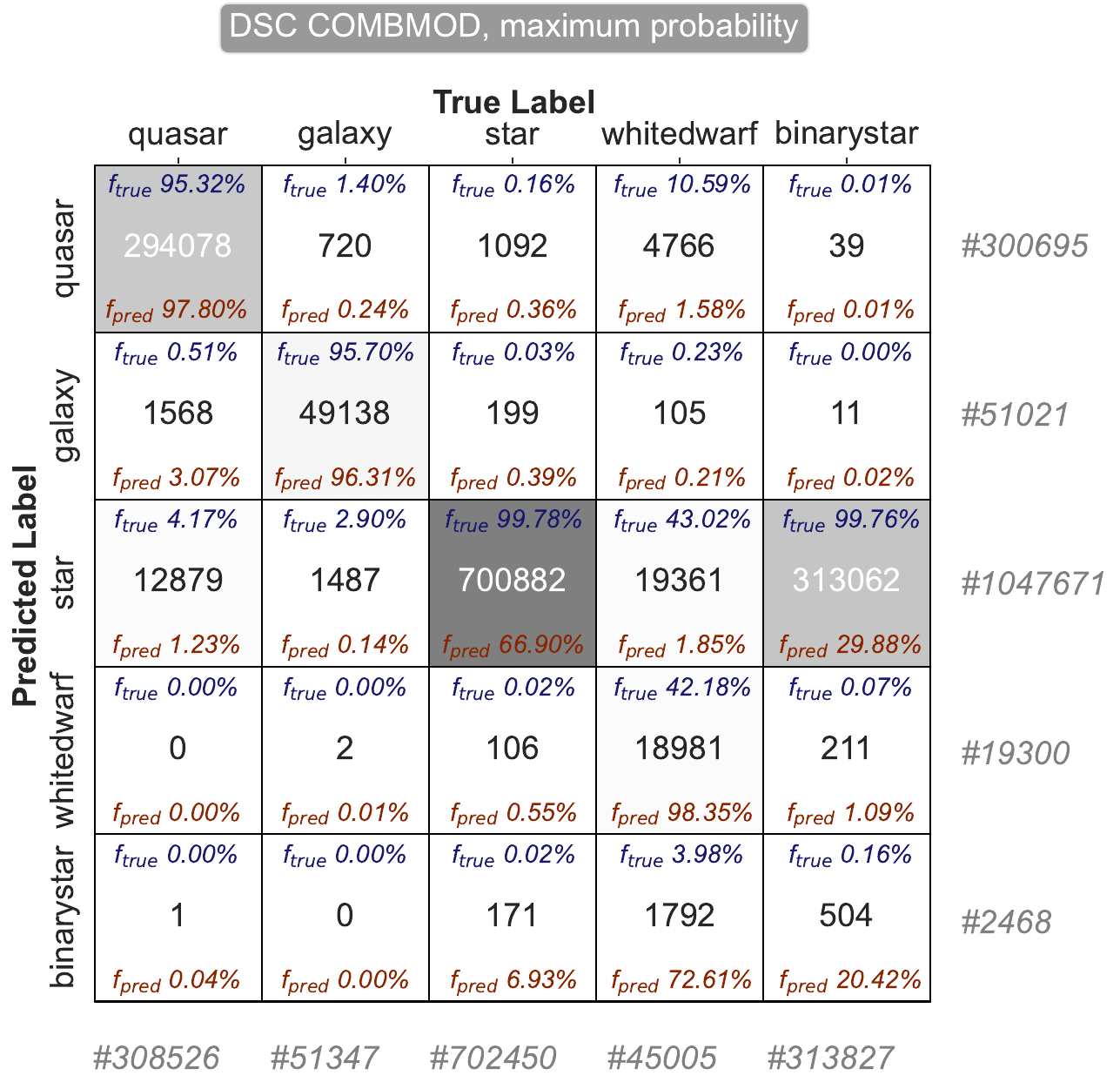}
		\caption{} 
		\label{fig:confmat_updateddataset_variableprior_combmod}	
	\end{subfigure}
	%
	\hspace{1cm} 
	\begin{subfigure}{0.46\textwidth}
		\includegraphics[scale=0.40, trim={0 0 0 1.25cm},clip]{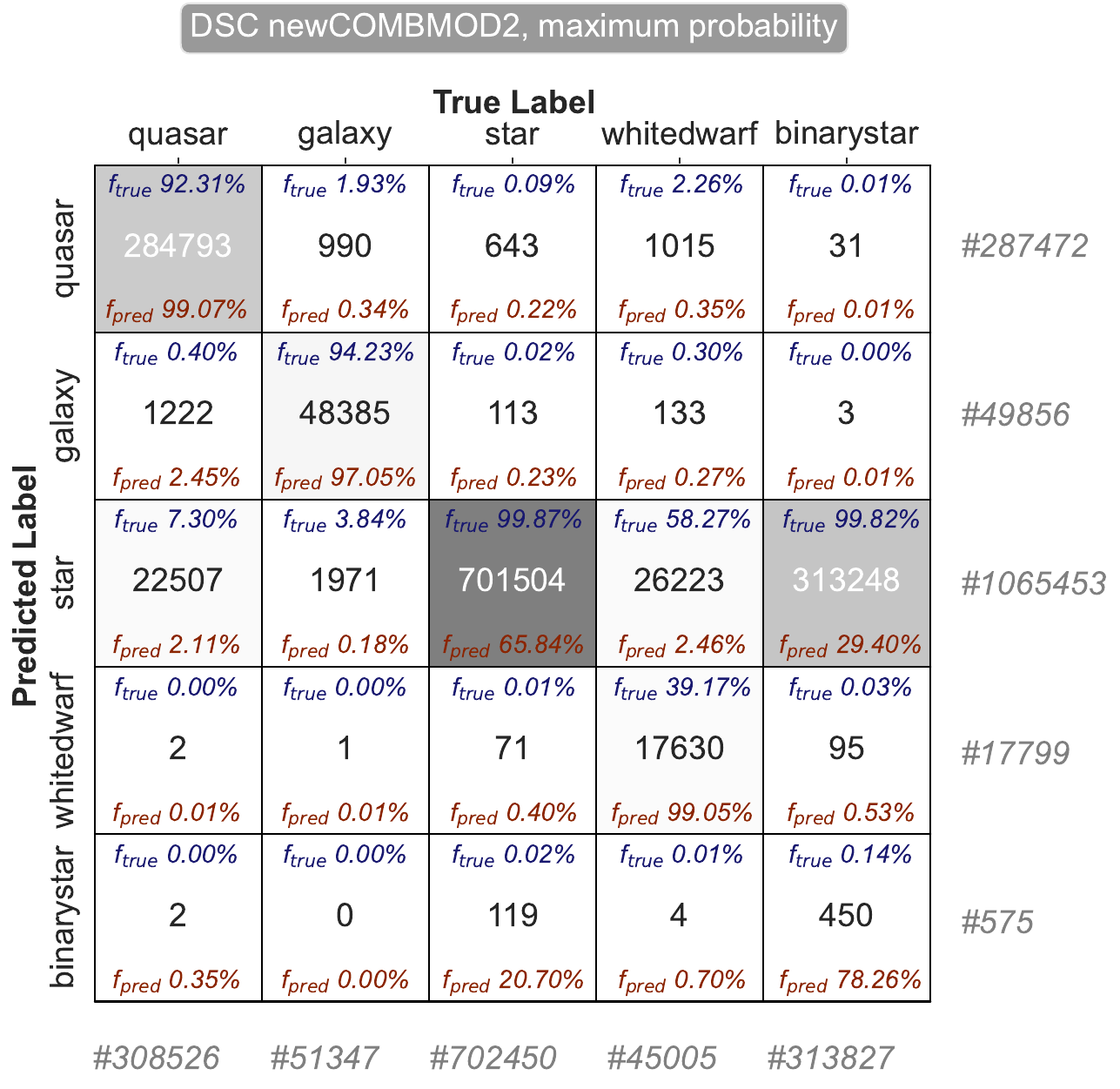}
		\caption{} 
		\label{fig:confmat_updateddataset_variableprior_newcombmod2}	
	\end{subfigure}

	\caption{
	As \ref{fig:confmats_1} but for \texttt{Combmod} using the variable prior and the new \texttt{Combmod}-$\alpha$ classifiers.
	\textit{(a)} Unadjusted confusion matrix computed on the {{cleaned}} validation data set of 1\,421\,155 sources for 
		\texttt{Combmod}-$\alpha$ with
		\{$\alpha_{\rm quasar}$=1.0, $\alpha_{\rm galaxy}$=0.6, $\alpha_{\rm star}$=0.5\}, using the global prior.
	\textit{(b)} Unadjusted confusion matrix computed on the {{cleaned}} validation data set of 1\,421\,155 sources for DSC \texttt{Combmod} 
		after application of the continuous variable prior.
	\textit{(c)} Unadjusted confusion matrix computed on the {{cleaned}} validation data set of 1\,421\,155 sources for 
		\texttt{Combmod}-$\alpha$ with
		\{$\alpha_{\rm quasar}$=0.6, $\alpha_{\rm galaxy}$=0.6, $\alpha_{\rm star}$=0.5\} after application of the continuous variable prior.
	}
	\label{fig:confmats_2}
\end{minipage} 
\end{figure}

{
To quantify the performance in multi-class classification, the purity (precision) and the completeness (recall) are computed from the confusion matrices.
In this section, we report the confusion matrices used to compute the overall metrics discussed throughout this study.
Figures \ref{fig:confmat_alldataset_globalprior} and \ref{fig:confmat_updateddataset_globalprior} refer to the confusion matrices of the DSC \texttt{Combmod} classifier using the global prior on the original validation data set and the cleaned validation data set, respectively.	
Figure \ref{fig:confmat_updateddataset_globalprior_newcombmod2} reports the confusion matrix of the new combined classifier \texttt{Combmod}-$\alpha$ using the global prior on the cleaned validation data set.
Figures \ref{fig:confmat_updateddataset_variableprior_combmod} and \ref{fig:confmat_updateddataset_variableprior_newcombmod2}  report the confusion matrices of the DSC \texttt{Combmod} and the new combined classifier \texttt{Combmod}-$\alpha$ using the variable prior on the cleaned validation data set.
}

{
The completeness and the purity before the adjustment to the class fractions (cf. Section \ref{subsec:DSC}) are listed in the leading diagonal of the confusion matrix, in blue and in red, respectively. 
For example, the unadjusted purity and completeness of the quasar class in \ref{fig:confmat_alldataset_globalprior} are $\sim$0.976  (= 282 464 / 289 431) and $\sim$0.916 (= 282 464 / 308 526), respectively.
After adjusting to account for the class fractions, the completenesses are unchanged while the purities are recomputed.
For the extragalactic classes, we note that the purities before the adjustment are $\sim$96\% and after the adjustment are decreased to $\sim$21\% to account for the reevaluated contribution of the stellar contaminants in the sky (Table \ref{table:tab_GDR3_and_newcombmod_summary}).
}

\newpage~\newpage

\newpage~\newpage

\section{Variable prior}
\begin{figure}[htp!]
\begin{minipage}[t]{1\textwidth}  
	\centering
	\includegraphics[scale=0.35]{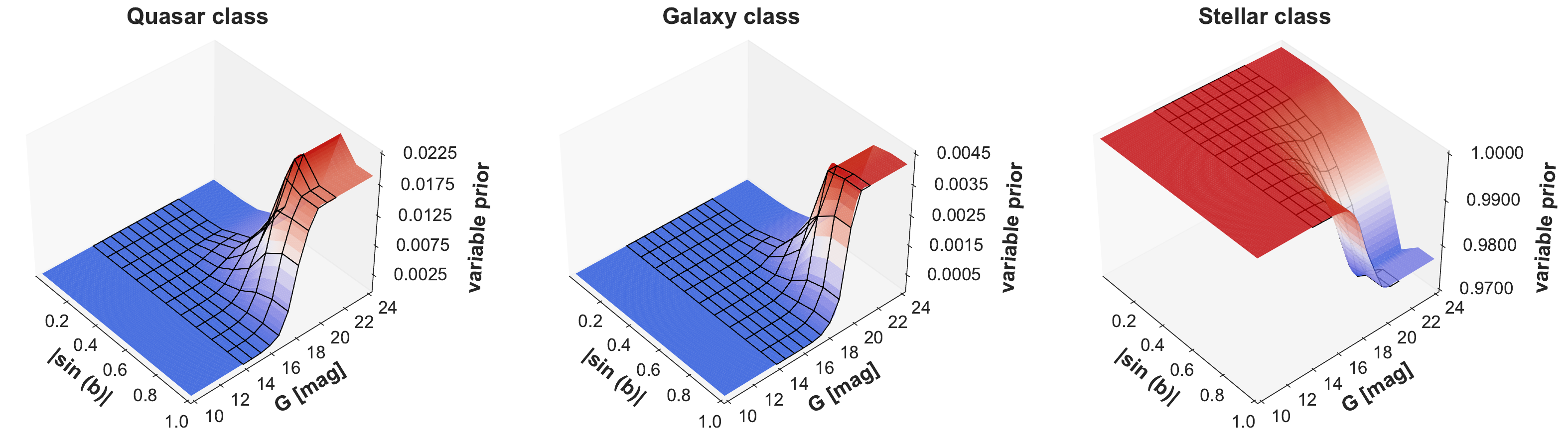}
	
	\caption{Representation of the variable prior for the quasar, galaxy, and stellar classes. 
        		For visualisation, the viewing angle is adapted to distinguish regions of interest in the mapping. 
       	 	The x-axis and y-axis refer to the Galactic latitude and brightness, and the z-axis reports to the prior.
		{The grid corresponds to the discretised 10$\times$10 definition of the prior. 
		The coloured surface represents the fitted spline expanding outside of the boundaries of the grid. 
		The colour scheme is defined such that red colours refer to higher values and  blue colours to lower values.}
		Left to right, are represented the priors for the quasar, galaxy, and star classes.
	}
	\label{fig:varprior}
	\end{minipage} 
\end{figure}

{
The section provides a representation of the variable prior defined in Section \ref{sec:varprior}.
Figure \ref{fig:varprior} shows the prior as a function of Galactic latitude and brightness for the quasar, galaxy, and stellar classes.
}


\section{Parametric combination of DSC classifiers}
\begin{figure*}[htp!]
\begin{minipage}[t]{1\textwidth} 
	\centering
	\begin{subfigure}{1\textwidth}
		\centering
		\includegraphics[scale=0.30]{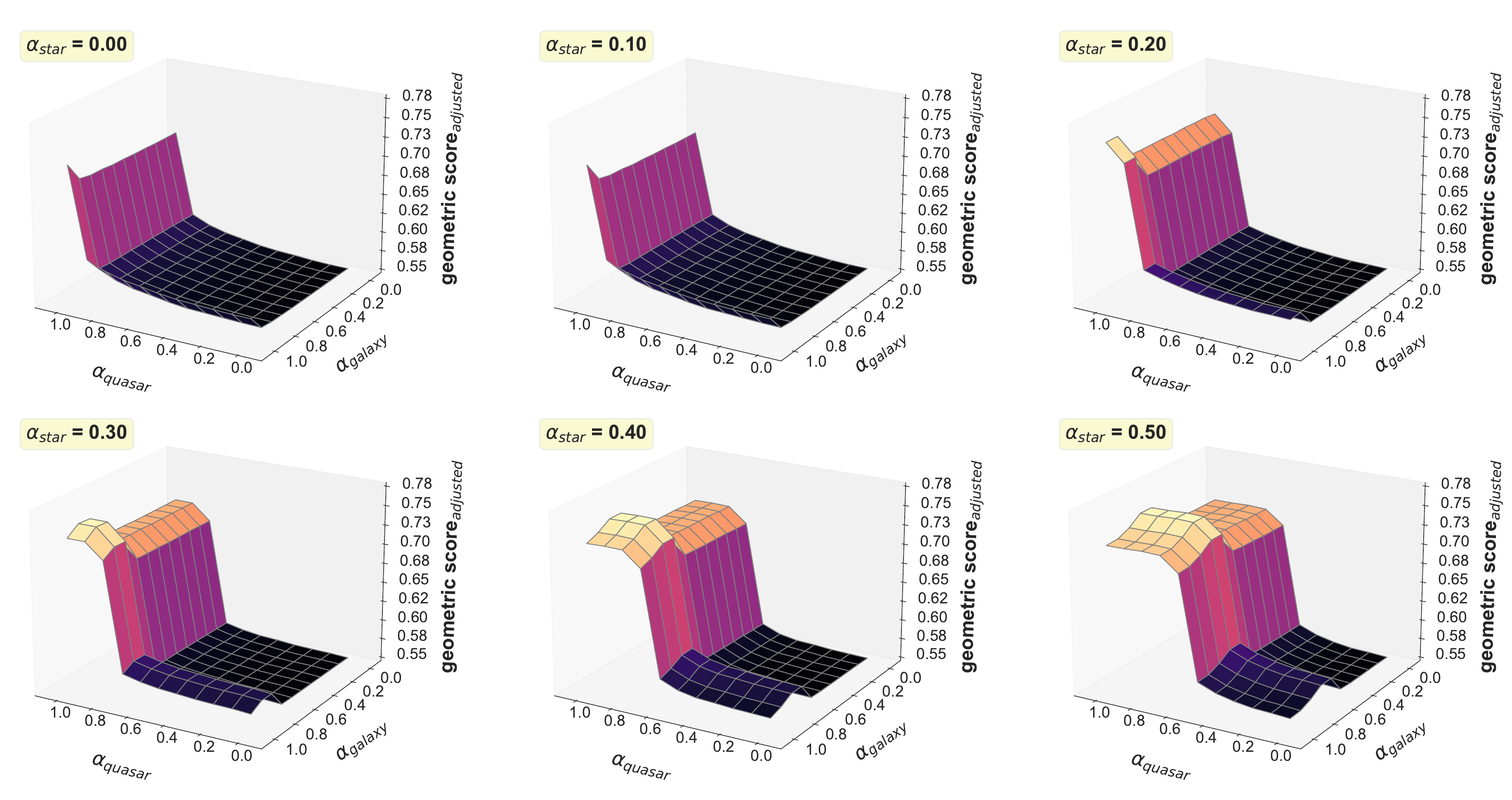}
		\caption{}
		\label{fig:geoscoreSum}
	\end{subfigure}

	\vspace{0.5cm}
	\begin{subfigure}{1\textwidth}
		\centering
		\includegraphics[scale=0.30]{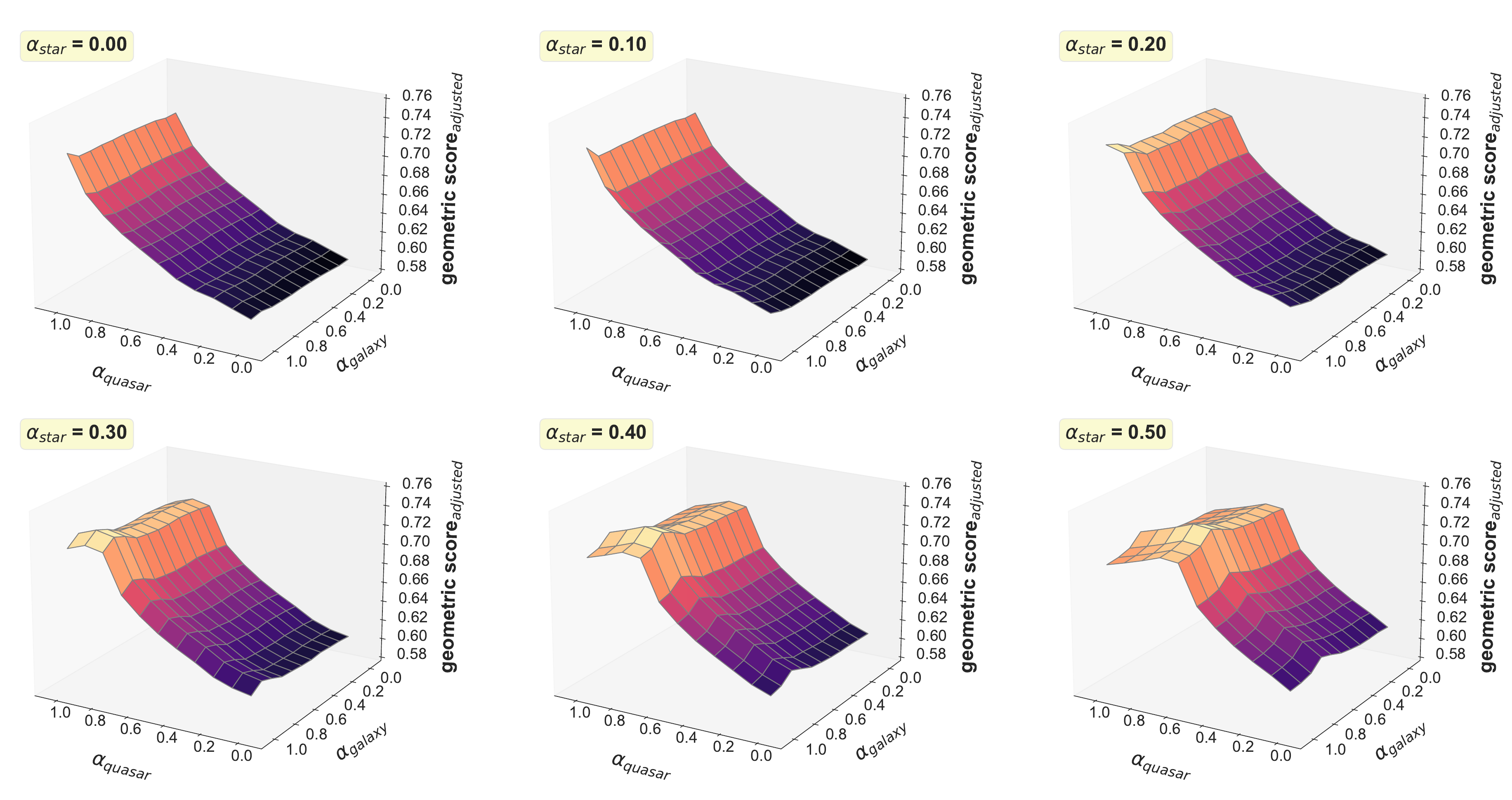}
		\caption{}
		\label{fig:geoscoreSum_varprior}
	\end{subfigure}
	
	\caption{ 
	Geometric score of the extragalactic classes for different parametric
	combinations of DSC \texttt{Specmod} and \texttt{Allosmod}.
	\textit{(a)} Results of all combinations using the global prior.
	\textit{(b)} As (a) but using the variable prior.
	The parameters $\alpha$=[$\alpha_{quasar}$, $\alpha_{galaxy}$, $\alpha_{star}$] are computed on a coarse grid.
	The geometric score of the extragalactic classes is computed at each combination.
	The evolving sequence can be followed from top to bottom, left to right, 
	where $\alpha_{\rm quasar}$ and $\alpha_{\rm galaxy}$ are set to vary between 0 and 1 
	while $\alpha_{\rm star}$ is fixed between 0 and 0.5. 
	Each panel reports the geometric score evaluated at each $\alpha_{\rm star}$ as a function of the other two parameters. 
	{The colour scheme of the mapping is defined such that bright colours refer to higher geometric score values and 
	dark colours to lower values.}
	}
	\end{minipage} 
	
\end{figure*}

{
This section reports the geometric score evaluated at different configurations of the parametric combination of DSC classifiers \texttt{Specmod} and \texttt{Allosmod}.
The parameters $\alpha$=[$\alpha_{quasar}$, $\alpha_{galaxy}$, $\alpha_{star}$] are computed on a coarse grid where $\alpha_{\rm quasar}$ and $\alpha_{\rm galaxy}$ are set to vary between 0 and 1, 
	while $\alpha_{\rm star}$ is fixed between 0 and 0.5.
For each combination of the $\alpha$ parameters, we compute the posterior probabilities and evaluate the classification metrics.
Figures \ref{fig:geoscoreSum} and \ref{fig:geoscoreSum_varprior} present the results obtained using the global prior and the variable prior, respectively.
}

\end{appendix}

\end{document}